\documentclass[twocolumn,trackchanges]{aastex701}

\usepackage{comment}
\usepackage{graphicx}
\usepackage{txfonts}

\begin{document}

\title{Interacting binaries on the Main Sequence as in-situ tracers of mass transfer efficiency and stability}

\author[orcid=0000-0002-8134-4854]{Koushik Sen}
\affiliation{Steward Observatory, Department of Astronomy, University of Arizona, 933 N. Cherry Ave., Tucson, AZ 85721, USA}
\email[show]{ksen@arizona.edu}  

\author[orcid=0000-0002-6718-9472]{Mathieu Renzo} 
\affiliation{Steward Observatory, Department of Astronomy, University of Arizona, 933 N. Cherry Ave., Tucson, AZ 85721, USA}
\email{mrenzo@arizona.edu}

\author[orcid=0000-0003-2946-9390]{Harim Jin}
\affiliation{Argelander-Institut f\"{u}r Astronomie, Auf dem Hügel 71, D-53121, Bonn, Germany}
\email{jin@MPA-Garching.MPG.DE}

\author[orcid=0000-0003-3026-0367]{Norbert Langer}
\affiliation{Argelander-Institut f\"{u}r Astronomie, Auf dem Hügel 71, D-53121, Bonn, Germany}
\email{nlanger@uni-bonn.de}

\author[orcid=0000-0002-2715-7484]{Abel Schootemeijer}
\affiliation{Argelander-Institut f\"{u}r Astronomie, Auf dem Hügel 71, D-53121, Bonn, Germany}
\email{aschoot@uni-bonn.de}

\author[orcid=0000-0002-7984-1675]{Jaime I. Villaseñor}
\affiliation{Max-Planck-Institut für Astronomie, Königstuhl 17, D-69117, Heidelberg, Germany}
\email{villasenor@mpia.de}

\author[orcid=0000-0003-0688-7987]{Laurent Mahy}
\affiliation{Royal Observatory of Belgium, Avenue Circulaire/Ringlaan 3, B1180 Brussels, Belgium}
\email{laurent.mahy@oma.be}

\author[orcid=0000-0002-2215-1841]{Aldana Grichener}
\affiliation{Steward Observatory, Department of Astronomy, University of Arizona, 933 N. Cherry Ave., Tucson, AZ 85721, USA}
\email{agrichener@arizona.edu}

\author[orcid=0000-0002-8465-8090]{Neev Shah}
\affiliation{Steward Observatory, Department of Astronomy, University of Arizona, 933 N. Cherry Ave., Tucson, AZ 85721, USA}
\email{neevshah@arizona.edu}

\author[orcid=0000-0002-0716-3801]{Chen Wang}
\affiliation{School of Astronomy and Space Science, Nanjing University, Nanjing, 210023, People's Republic of China}
\affiliation{Key Laboratory of Modern Astronomy and Astrophysics, Nanjing University, Ministry of Education, Nanjing, 210023, People's Republic of China}
\affiliation{Max Planck Institute for Astrophysics, Garching, Germany}
\email{chen_wang@nju.edu.cn}

\author[orcid=0000-0001-9565-9462]{Xiao-Tian Xu}
\affiliation{Tsung-Dao Lee Institute, Shanghai Jiao-Tong University, Shanghai, 201210, People’s Republic of China}
\email{xxu.astro@outlook.com}


\begin{abstract}

Understanding the transfer of mass and angular momentum in binary interactions is crucial for modelling the evolution of any interacting binary after the first mass transfer phase. Mass transfer physics assumptions shape the predictions for later stages of binary evolution, such as the immediate progenitors of stripped-envelope supernovae and gravitational wave mergers. We constrain the efficiency and stability of thermal timescale mass transfer in massive binary evolution using the observed population of 62 massive interacting binaries on the Main Sequence (`Algols') in the Milky Way, Large and Small Magellanic Clouds. 
We find that purely conservative or non-conservative mass transfer cannot explain the current mass ratio and orbital period of all massive Algols. Angular momentum conservation rules out conservative mass transfer in $\sim$28\,\% of massive Algols in the SMC. About three-quarters of all massive Algols are consistent with having undergone inefficient mass transfer ($\lesssim$\,50\,\%), while the remaining systems, mostly residing in the LMC and Milky Way, require mass transfer to have been more efficient than 25\%. For our fiducial assumption on the extent of envelope stripping, the current sample of massive Algols does not require mass transfer to be efficient at the shortest orbital periods ($\sim$2\,d) at any metallicity. We find evidence that mass transfer on the Main Sequence needs to be stable for initial accretor-to-donor mass ratios as unequal as $\sim 0.6$. 
Unless biased by observational selection effects, the massive Algols in the SMC seem to have undergone less efficient mass transfer than those in the LMC and Milky Way.

\end{abstract}



\section{Introduction}

Massive stars ($\geq8\,M_{\odot}$) are rarely found alone \citep{vanbeveren1998,Kobulnicky2007,Mason2009,sana2012,Sana2014,moe2017,Offner2023}. Surveys of O- and B-type stars have established that a majority reside in multiples \citep{Abt1990,Kiminki2008,Mahy2009,Mahy2013,Kobulnicky2012,kobulnicky2014,Sota2014,Dunstall2015,Sana2008,Sana2009,Sana2011,sana2013,Sana2014,Barba2017,Almeida2017,Trigueros2021,Villasenor2021,Villasenor2025,Shenar2022,Shenar2024,Banyard2022,Ritchie2022,Ramirez2024,Frost2025}. Most of them are so close that interactions through tides \citep{zahn1977,Goldreich1989,Marchant2016,mandel2016,deMink2016,Sun2023,Rosu2024,Rocha2025,Sen2025}, colliding winds \citep{Rauw2016,Kashi2020,Kashi2022,Naze2022}, Roche-lobe overflow \citep{Paczynski1971,philip1992,Vanbeveren1998b,langer2003}, and/or mergers \citep{podsiadlowski2006,schneider2019,wu2020,Menon2024,Frost2024,Patton2025} are inevitable. These interactions profoundly shape the evolutionary pathways of massive stars \citep{selma2014,Sen2023,Jin2025}, their feedback on the surrounding medium \citep{Eldridge2008,Stanway2016,Gotberg2017}, and their ultimate fates as progenitors of stripped-envelope supernovae \citep[SESNe,][]{Yoon2010,Zapartas2017,Sravan2020,laplace2020,Laplace2021,David2022a,David2022b,Ercolino2024,Ercolino2025} and compact object binaries that may later merge as gravitational-wave (GW) sources \citep{Belczynski2008,Eldridge2016,kruckow2018,Spera2019,Broekgaarden2022,vanSon2022b,Riley2022,Briel2023}. Understanding the physical processes at play in these systems is therefore essential not only for stellar astrophysics \citep{Marchant2024} but also for cosmology \citep{hopkins2014} and predicting the rates and properties of transient populations across the Universe \citep{Smith2011,Mandel2022}.

However, the treatment of mass transfer and angular momentum loss in massive binaries remains one of the most significant sources of uncertainty in stellar evolution theory \citep{Hjellming1987,Soberman1997,Woods2011,Pavlovskii2017,Ge2010,Ge2015,Ge2020,Temmink2023,Schurmann2024}. Whether the mass transfer is highly efficient or largely inefficient \citep{nelson2001,petrovic2005,selma2007,shao2016,Vinciguerra2020,Bouffanais2021,Sen2022,Romero2023,Nuijten2025,Lechien2025,Schurmann2025}, and how the lost mass carries away angular momentum \citep{Macleod2018a,Macleod2018b,Lu2022}, dramatically alters the orbital period and mass-ratio evolution of binaries \citep{Eldridge2011,renzo2019,Willcox2023,Wagg2025}. Population synthesis studies demonstrate that these choices can alter the diversity and rates of SESNe and GW mergers respectively \citep{Eldridge2013,Agrawal2022,Agrawal2023,Belczynski2022,Boesky2024,Rauf2024,Zapartas2025,Souropanis2025,Willcox2025}, while also reshaping the predicted orbital period distribution of compact object binaries, and masses and spin magnitudes of compact objects \citep{Belczynski2020,Olejak2021,Olejak2024,Tauris2022,Lau2024,Dorozsmai2024,Larsen2025}. Quantitatively constraining these processes is thus critical for linking present-day massive binaries with their end states \citep[e.g.][]{Mandel2025}. Without empirical constraints on early stages of binary evolution, parameter uncertainties and degeneracies in the predicted population of SESNe or GW mergers may not be resolved simply by an increase in their observed number in the coming decades.

An empirical approach is to study Algol-type binaries \citep{surkova2004,selma2007,mennekens2017,malkov2020,Sen2022}, which are interacting systems where mass transfer occurs while both stars are still on the Main Sequence \citep[`Case\,A',][]{Kippenhahn1967,pols1994,nelson2001}. These systems are especially valuable because they represent the earliest phase of mass transfer, prior to the additional complexities introduced by envelope expansion \citep{Romagnolo2023}, strong winds \citep{Smith2014,Renzo2017,Josiek2024}, or advanced nuclear burning \citep{Heger2002,Farmer2016,Grichener2025}. As such, they require the fewest assumptions regarding prior binary evolution, offering a relatively clean probe of the efficiency and stability of thermal-timescale mass transfer.

In Case A binaries, mass transfer begins on the donor’s thermal timescale (fast Case A) and transitions into a slower, nuclear-timescale phase as the donor expands during core hydrogen burning. This long-lived ‘slow Case A’ stage, which spans millions of years, yields observable semi-detached systems. Since all such binaries must have previously undergone an intense, fast Case A episode, comparing models with observed Algols constrains both phases. Moreover, Case A is especially relevant at high masses, where it is more common \citep{selma2007,Sen2023,Burt2025}, making massive Algols bright, numerous, and ideal testbeds for mass transfer physics. The demographics of observed Algols across different metallicities-Milky Way, Large Magellanic Cloud (LMC) and Small Magellanic Cloud (SMC)-therefore provide a powerful benchmark against which theoretical models can be tested \citep{selma2007,Sen2022}.

In this work, we study the mass transfer efficiency, angular momentum loss, and stability of mass transfer in the observed populations of 62 massive Algols in the Milky Way, LMC and SMC, and investigate any metallicity difference. We develop an analytical framework that links current observable binary properties—orbital period, donor mass, and accretor mass—to their initial parameters. By imposing physically motivated boundaries on the initial parameters, we derive constraints on the range of mass transfer efficiency and angular momentum loss during the mass transfer episode that can reproduce the observed population. We also determine limits on the stability of mass transfer required to obtain meaningful solutions for mass transfer efficiency. 

This paper is organised as follows. Section\,\ref{sec:method} describes the observational sample and our methodology for deriving initial binary parameters. Section\,\ref{sec:result} presents the range of mass transfer efficiency and angular momentum loss possible, for individual systems to overall trends across the Algol population. In Sect.\,\ref{sec:discussion}, we compare our results with the literature and evaluate the robustness of our assumptions. We summarise our main conclusions in Sect.\,\ref{sec:conclusion}.


\section{Methods}
\label{sec:method}
\subsection{Observed systems}

In this study, we include all the massive Algol binaries observed
in the Magellanic Clouds and the Milky Way. In the SMC, we use the
catalogue in \citet[][Table 1, 29 systems]{selma2007} curated from the observational
studies of the \citet{Harries2003} and \citet{hilditch2005}. We take
the massive Algols listed in Table\,1 and Table\,2 of \citet[][33 systems]{Sen2022}
in the LMC and the Milky Way, respectively (heterogeneous sample, see
references within for details on individual systems). We list the
current masses of the donors $M_{\rm d}$, accretors $M_{\rm a}$ and
binary orbital periods $P_{\rm orb}$ used in this study for all the
systems in Table\,\ref{table:all_algols_smc} and Table\,\ref{table:all_algols_lmc_mw}.

The binaries span an accretor mass range of $\sim 8$--$40\,M_\odot$,
accretor-to-donor mass ratios $q=M_{a}/M_{d}$ from near unity up to $\sim 3$, and orbital periods between
$\sim 1$ and $10$\,d (the observed systems in the SMC are limited by
a maximum orbital period of 5\,d, \citealp{Harries2003,hilditch2005}).
Circular orbits were assumed for most systems. This sample of short-period,
massive binaries thus enables a systematic comparison with grids of
binary evolution tracks to probe the efficiency of Case\,A mass transfer across a large range of metallicities (near Solar to $\sim$one-fifth of Solar).

\subsection{Initial stellar and binary parameters}
\label{sect:assumptions}
Detailed binary evolution models predict that the fast thermal
timescale mass transfer on the Main Sequence strips the outer
envelope of the mass-donating star approximately up to its initial convective
core at Zero-Age Main Sequence (Fig.\,2 of \citealp{Schurmann2024b},
see also Fig.\,1 of \citealp{Sen2023}). We assume that the
present masses of the donors in the massive Algols correspond
to their initial convective core masses that include the overshooting region (we discuss the caveats
in this assumption in Sect.\,\ref{sect:envelope_stripping}). Therefore, the initial
mass of the donor star can be estimated from its initial convective core mass
$M_{\rm ccd,i}$. At the range of donor
masses relevant for our study (2-20\,$M_{\odot}$), the slow nuclear
timescale mass transfer (the `Algol' phase) removes $\leq$\,1$M_{\odot}$ 
mass from the donor \citep{pols1994,selma2007,Sen2022}. Using
grids of detailed binary evolution models (Sect.\,\ref{sec:detailed_models}), we estimate the initial donor mass
$M_{\rm d,i}$ of each massive Algol binary (listed in
Table\,\ref{table:all_algols_smc}) and Table\,\ref{table:all_algols_lmc_mw}.

For each value of initial donor mass, there exists a maximum
initial orbital period $P_{\rm orb,i,max}$ up to which mass
transfer can initiate on the Main Sequence \citep{nelson2001,
selma2007,Sen2022,Pauli2022}. This maximum initial orbital
period is weakly dependent on the initial mass ratio
of the binary, but sensitive to stellar radii and the physics
that determines them (opacity primarily, and at second order
mass loss and inflation, e.g., \citealt{sanyal2015,Xin2022}). For each massive Algol, we estimate
the maximum initial orbital period $P_{\rm orb,i,max}$ that the binary could have
started with, using the binary models described in Sect.\,\ref{sec:detailed_models}
(listed in Table\,\ref{table:all_algols_smc}) and Table\,\ref{table:all_algols_lmc_mw}.

The difference between the initial donor mass and the current
donor mass in each Algol donor gives the mass lost by the donor
$\Delta M_{\rm d}$ and is equal to the maximum amount of
mass that the mass accreting star could have accreted during
the fast thermal timescale mass transfer phase. Subtracting
the above difference from the present mass of the accretor
gives the lowest initial mass of the accretor $M_{\rm a,i,min}$
at its Zero-Age Main Sequence, if the accretor could accrete
all the transferred mass (conservative mass-transfer limit).
Here, we limit the lowest possible initial accretor mass to be
0.08\,$M_{\odot}$, required to initiate core hydrogen burning in its
core.

If the accretor does not accrete any of the mass lost by the
donor, the initial mass of the accretor is equal to its observed
current mass (assuming negligible mass lost via stellar winds
on the Main Sequence). Here, the maximum initial accretor mass
$M_{\rm a,i,max}$ is limited to the initial donor mass estimated
above, so as not to invert the direction of mass transfer at its
initiation. The above procedure gives a range of possible initial
accretor masses for each system analysed in this study (Table\,\ref{table:all_algols_smc} and Table\,\ref{table:all_algols_lmc_mw}).

\subsection{Mass transfer efficiency and specific angular momentum loss}

We define the mass transfer efficiency $\varepsilon$
as the ratio of the mass accreted by the accretor to the mass lost
by the donor. In terms of initial and current masses,
\begin{equation}
    \varepsilon = -\frac{\Delta M_{\rm a}}{\Delta M_{\rm d}} = \frac{M_{\rm a} - M_{\rm a,i}}{M_{\rm d,i} - M_{\rm d}},
    \label{eq:mte}
\end{equation}
where $M_{\rm a,i} \in[M_{\rm a,i,max}, M_{\rm a,i,min}]$ is
the initial mass of the accretor. For the corresponding range
of initial accretor masses [$M_{\rm a,i,max}$, $M_{\rm a,i,min}$],
we calculate the range of possible mass transfer efficiencies for
each system [$\varepsilon_{\rm min}$,$\varepsilon_{\rm max}$].

Equation\,(\ref{eq:mte}) can be written in terms of the total
mass $M_{\rm T} = M_{\rm a} + M_{\rm d}$ and mass ratio
$q = M_{\rm a}/M_{\rm d}$ of the binary system as \citep{Soberman1997,Nuijten2025}
\begin{equation}
    \varepsilon = \left(\frac{1+q}{1+q_{\rm i}}-\frac{M_{\rm T}}{M_{\rm T,i}}\right) \left(q\frac{M_{\rm T}}{M_{\rm T,i}} - q_{\rm i}\frac{1+q}{1+q_{\rm i}}\right)^{-1},
\end{equation}
where $M_{\rm T,i} = M_{\rm a,i} + M_{\rm d,i}$ and
$q_{\rm i}=M_{\rm a,i}/M_{\rm d,i}$ are the initial
total mass and initial mass ratio of the binary, respectively.

For conservative mass transfer $\varepsilon = 1$, there is no angular
 momentum lost from the binary (neglecting stellar winds; moreover, winds tap into the spin angular momentum of the stars, which is much smaller than the orbital angular momentum), and the initial orbital period $P_{\rm orb,i,cons}$ is given by \citep[e.g.][]{Tauris2023}
\begin{equation}
    \ln\left(\frac{P_{\rm orb,i,cons}}{P_{\rm orb}}\right) = 3\ln\left(\frac{M_{\rm d} M_{\rm a}}{M_{\rm d,i} M_{\rm a,i}}\right).
    \label{eq:conservative}
\end{equation}

For each possible value of mass transfer efficiency $\varepsilon
\in [\varepsilon_{\rm min},\varepsilon_{\rm max}] \neq1$, a range
of initial orbital periods can lead to the present orbital period,
depending on the assumed angular momentum lost $\Delta J$ per unit
mass lost $\Delta M$ from the binary
\begin{equation}
    j_{\rm loss} = \frac{\Delta J}{\Delta M} = \gamma \frac{J_{\rm orb}}{M_{\rm T}} = \gamma j_{\rm orb},
\end{equation}
where $\gamma\,(>0)$ is a multiplicative factor to capture different
modes of mass loss \citep{PolsMarinus1994}, $J_{\rm orb}$ is
the total orbital angular momentum of the binary, $j_{\rm orb}$ is
the specific orbital angular momentum of the binary and
$j_{\rm loss}$ is the specific angular momentum lost from
the binary.

In reality, $\gamma$ depend on how the mass flows away from
the binary \citep[e.g.][]{Soberman1997,Lu2022}, and thus
could vary as the stars evolve through mass transfer. As a
first approximation, we consider $\gamma$ as constant to
relate analytically the initial and present-day period.
The initial orbital period $P_{\rm orb,i}$ can be related
to the orbital period and masses of the binary components
\citep[see Eq.\,(A2) of][]{PolsMarinus1994}
\begin{equation}
    \ln\left(\frac{P_{\rm orb,i}}{P_{\rm orb}}\right) = 3\ln\left(\frac{M_{\rm d} M_{\rm a}}{M_{\rm d,i} M_{\rm a,i}}\right) + \left(3\gamma+\frac{3}{2}\right)\ln\left(\frac{M_{\rm T,i}}{M_{\rm T}}\right).
\end{equation}
We calculate the initial orbital period values for $\gamma$ =
{0, 1, 2, 3}. Values of $\gamma>3$ are expected to be
unlikely \citep[see discussion in Sect.\,2.4 of][]{Nuijten2025}.
Magnetic braking \citep{Bour2025} or L2 mass loss \citep{Lu2022}
can lead to high values of $\gamma$, but most massive stars
are not found to host detectable dipolar magnetic fields
\citep{Wade2016}. The strong differential rotation during
the accretion phase, however, may generate magnetic fields
\citep{Braithwaite2017} and have been recently detected 
in a few semi-detached and contact binaries \citep{Hubrig2026}. 
For isotropic re-emission from the surface of the accretor,
$\gamma = M_{\rm d}/M_{\rm a}$. We discuss the possibility of
$\gamma>3$ in Sect.\,\ref{sec:discussion}. The initial orbital
period $P_{\rm orb,i}$ is limited by the maximum initial orbital
period $P_{\rm orb,i,max}$ up to which mass transfer can
initiate on the Main Sequence. The initial orbital period
(for $\varepsilon$ constant over time, Eq.\,(A4-A5) of
\citealp{PolsMarinus1994}) is given by
\begin{equation}
    \ln\left(\frac{P_{\rm orb,i}}{P_{\rm orb}}\right) = 3\ln\left(\frac{M_{\rm d}}{M_{\rm d,i}}\right) + \frac{3}{\varepsilon}\ln\left(\frac{ M_{\rm a}}{ M_{\rm a,i}}\right) + \frac{3}{2}\ln\left(\frac{M_{\rm T}}{M_{\rm T,i}}\right), 0 < \varepsilon \leq 1
    \label{eq:inefficient}
\end{equation}
and
\begin{equation}
    \ln\left(\frac{P_{\rm orb,i}}{P_{\rm orb}}\right) = 3\ln\left(\frac{M_{\rm d}}{M_{\rm d,i}}\right) + \frac{3}{2}\ln\left(\frac{M_{\rm T}}{M_{\rm T,i}}\right) + 3\frac{M_{\rm d,i} - M_{\rm d}}{M_{\rm a}},\rm{for\,\,} \varepsilon = 0.
    \label{eq:epsilon0}
\end{equation}


\begin{table*}
\caption{Summary of grid parameters. \textbf{Notes.} $\Delta$ = grid spacing; $N_{\rm Case\,A}$ = number of models undergoing Case\,A mass transfer. Models \href{https://wwwmpa.mpa-garching.mpg.de/stellgrid/}{here}.}
\centering
\begin{tabular}{lrrrrrr}
\hline\hline
Galaxy & metallicity & initial donor mass $M_{\rm d,i}$ & initial orbital period $P_{\rm i}$ & initial mass ratio $q_{\rm i}$ & $j_{\rm loss}$ & $N_{\rm Case\,A}$ \\
\hline
LMC & 0.004831 &  5-40 $M_{\odot}$ ($\Delta$ log $M_{\rm d,i}$ = 0.05) & 1.4 - 3162\,d ($\Delta$logPi = 0.025 days) & 0.025 - 0.975 ($\Delta q_{\rm i}$ = 0.025) & $j_{\rm acc,orb}$ & $\sim$15000 \\
SMC & 0.002174 & 5-100 $M_{\odot}$ ($\Delta$ log $M_{\rm d,i}$ = 0.05) & 1.0 - 3162\,d ($\Delta$logPi = 0.025 days) & 0.300 - 0.975 ($\Delta q_{\rm i}$ = 0.025) & $j_{\rm acc,orb}$ & $\sim$19000 \\
\hline
\end{tabular}
\label{table:grid_details}
\end{table*}

\subsection{Detailed binary evolution models}
\label{sec:detailed_models}
We calculate the cumulative mass transfer efficiency in
detailed binary evolution models introduced in \citet[][for
the SMC]{wang2020} and \citet[][for the LMC]{pablothesis,Sen2022}.
At any timestep during the evolution of the binary, the
cumulative mass transfer efficiency of the model is defined as in
Eq.\,(\ref{eq:mte}), where $M_{\rm d}$ and $M_{\rm a}$
denote the mass of the donor and accretor at that timestep,
and $M_{\rm d,i}$ and $M_{\rm a,i}$ denote their initial
masses at the Zero-Age Main Sequence. The contribution
from each model is weighted by the initial mass function
\citep{salpeter1955}, the orbital period and mass ratio
distribution \citep{sana2012,sana2013}, and the timestep
during the Algol configuration (see Eq.\,(4) of \citealp[]{Sen2022}).

While the detailed binary models include wind mass loss,
we do not remove the contribution from the wind mass loss
in calculating the mass transfer efficiency in the models.
We discuss the effects of ignoring wind mass loss on our conclusions
in Sect.\,\ref{sec:wind_mass_loss}. From our synthetic Algol
 population, we removed timesteps where mass transfer occurs from the initially less massive accretor to the donor, that is, when inverse mass transfer occurs. We also remove the timesteps where both binary components fill its Roche lobe \citep[contact configuration,][]{Menon2021,Fabry2022,Fabry2023,Henneco2024,Vrancken2024}

The stellar and binary physics assumptions in the models
are described in detail in \citet{wang2020,Sen2022}. During
a mass transfer phase, mass accretion onto the accretor
is limited by the spin of the accretor star. Critically
rotating accretors are assumed not to be able to accrete
any more mass \citep{Langer1998}. The excess mass transferred
is lost as winds from the accretor, carrying the specific
orbital angular momentum of the accretor \citep{langer2003,
petrovic2005}. Mass transfer is assumed to be stable as
long as the excess mass lost can be driven by the combined
luminosity of both the binary components (see Eq.\,(2) of
\citealp{Sen2022}). Strong tides (\citealp[][]{zahn1977},
implemented as in \citealp{detmers2008}) in short-period
($\leq$5\,d) systems can halt the accretor spin-up and
lead to efficient mass accretion; longer period systems
undergo inefficient mass transfer \citep{Sen2022}.
This leads to an orbital period-dependent mass transfer
efficiency in the models (see Fig.\,F2 of 
\citealp{Sen2022} for quantitative numbers). When the 
combined luminosity of both stars is insufficient to drive 
the required excess mass loss, the binary is assumed to merge.

\subsection{Mass transfer stability}

Mass transfer on the Main Sequence is not stable down to
arbitrarily low values of initial mass ratios, owing to
the increasingly divergent thermal timescales of the donor
and the accretor \citep{Ge2020}. The mass transfer rate
is typically set by the evolutionary timescale of the donor star
(thermal initially and later nuclear),
while the accretor can readjust its structure on its longer
thermal timescale as compared to the donor. As such, most
binaries with initial mass ratios $<$ 0.65 may merge during
the thermal timescale mass transfer on the Main Sequence
\citep{selma2013,Ge2020}.

In the models of \citet[][Fig.\,1,
green models]{Sen2022}, there exist two threshold values
of initial mass ratios, such that 1) above the higher value
$q_{\rm max}$ (bottom left of green region above the black
shaded region), all binary models survive the
thermal timescale mass transfer; 2) below the lower value
$q_{\rm min}$ (top left of green region below the Case\,A
boundary) of the initial mass ratio, all
binary models are predicted to merge. The value of $q_{\rm
max}$ and $q_{\rm min}$ are a function of the initial donor
mass of the binary \citep[][Fig.\,1]{Sen2023}. Between the
two thresholds, the fate of the binary depends on its
initial orbital period. Below the mass ratio $q_{\rm max}$,
all the accretors spin up to critical rotation soon after
the onset of thermal timescale mass transfer \citep{packet1981},
and the mass transfer efficiency in such models is less than
10\% (top panels of Fig.\,F2 of \citealp{Sen2022}).

As a conservative estimate and for the sake of simplicity,
we assume that all binaries may undergo arbitrarily efficient stable
mass transfer up to an initial mass ratio given by $q_{\rm max}$ in the rotationally limited mass accretion scheme. 
Below $q_{\rm max}$, a higher mass transfer efficiency may be
supported by the formation of a disk, where the coupling of the
disk to the accretor can aid the accretion of mass while removing
angular momentum from the accretor \citep{Paczynski1991,Popham1991}.
\citet{Schurmann2024} investigate the response of non-rotating
accretor stars for constant mass transfer efficiencies. We
parametrise their results for the onset of L2 overflow, as a
function of the initial donor mass and mass transfer efficiencies,
to derive the minimum mass ratio for stable mass transfer in the
presence of a disk (Fig.\,8 of \citealp{Schurmann2024}).

Table\,\ref{table:1} lists the values of $M_{\rm ccd,i}$,
$q_{\rm min}$, $q_{\rm max}$ and $P_{\rm orb,i,max}$, as described above, 
for different values of initial donor mass in the LMC and SMC.
We linearly interpolate between the values and assume that
the stellar and binary parameters for LMC metallicity are
applicable for the Milky Way systems. The difference in
values of $M_{\rm ccd,i}$ are within 0.1\,$M_{\odot}$ between the
LMC and SMC for the same initial donor mass. The values of
$q_{\rm min}$, $q_{\rm max}$ for the two metallicities are separated by at most 0.05. The maximum
initial orbital period $P_{\rm orb,i,max}$ is a significant
function of metallicity, and we
discuss in Sect.\,\ref{sec:outliers} that our conclusions
on angular momentum loss are not strongly affected by metallicity.

\begin{figure*}
    \centering
    \includegraphics[width=0.49\linewidth]{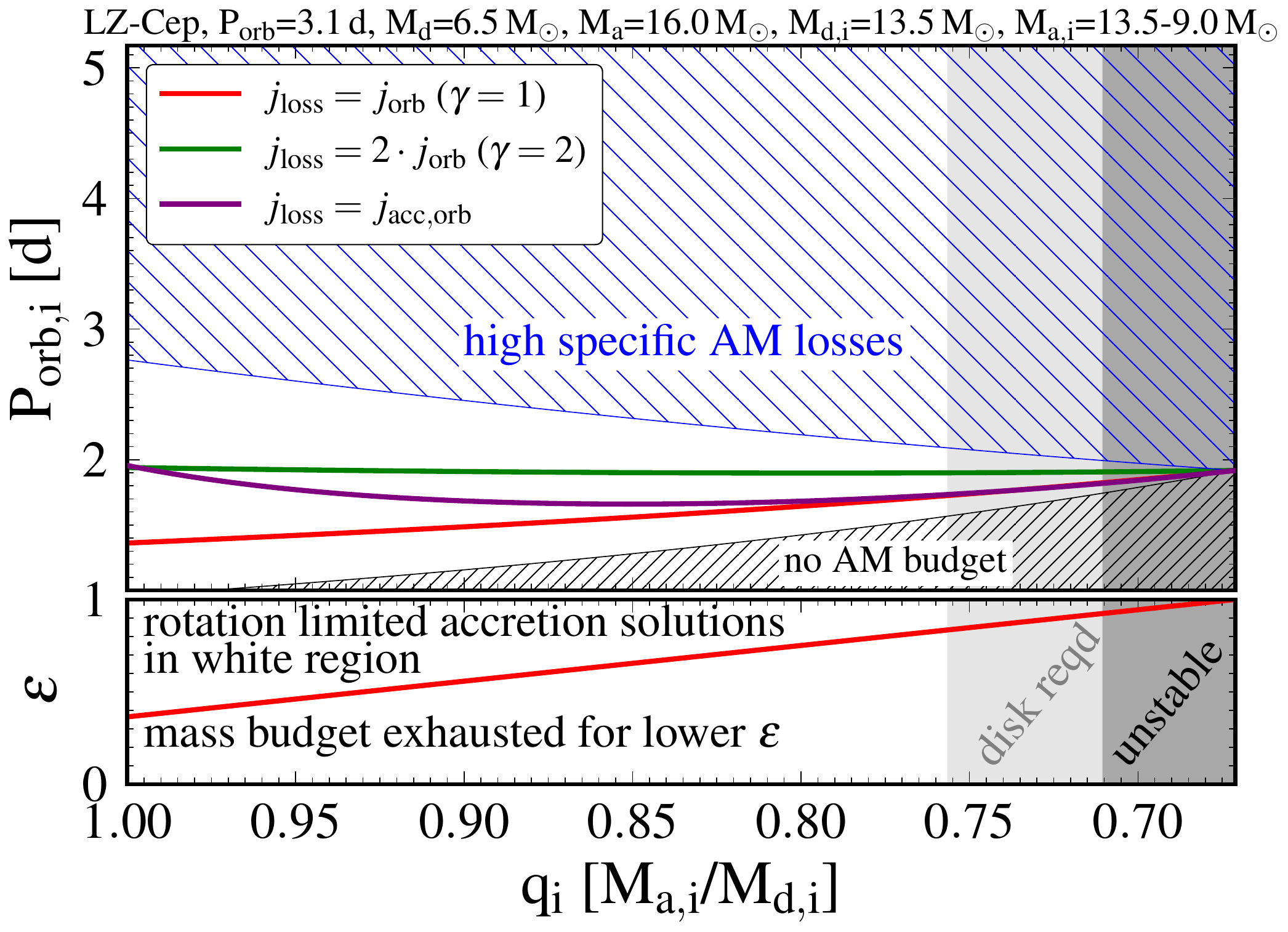}
    \includegraphics[width=0.49\linewidth]{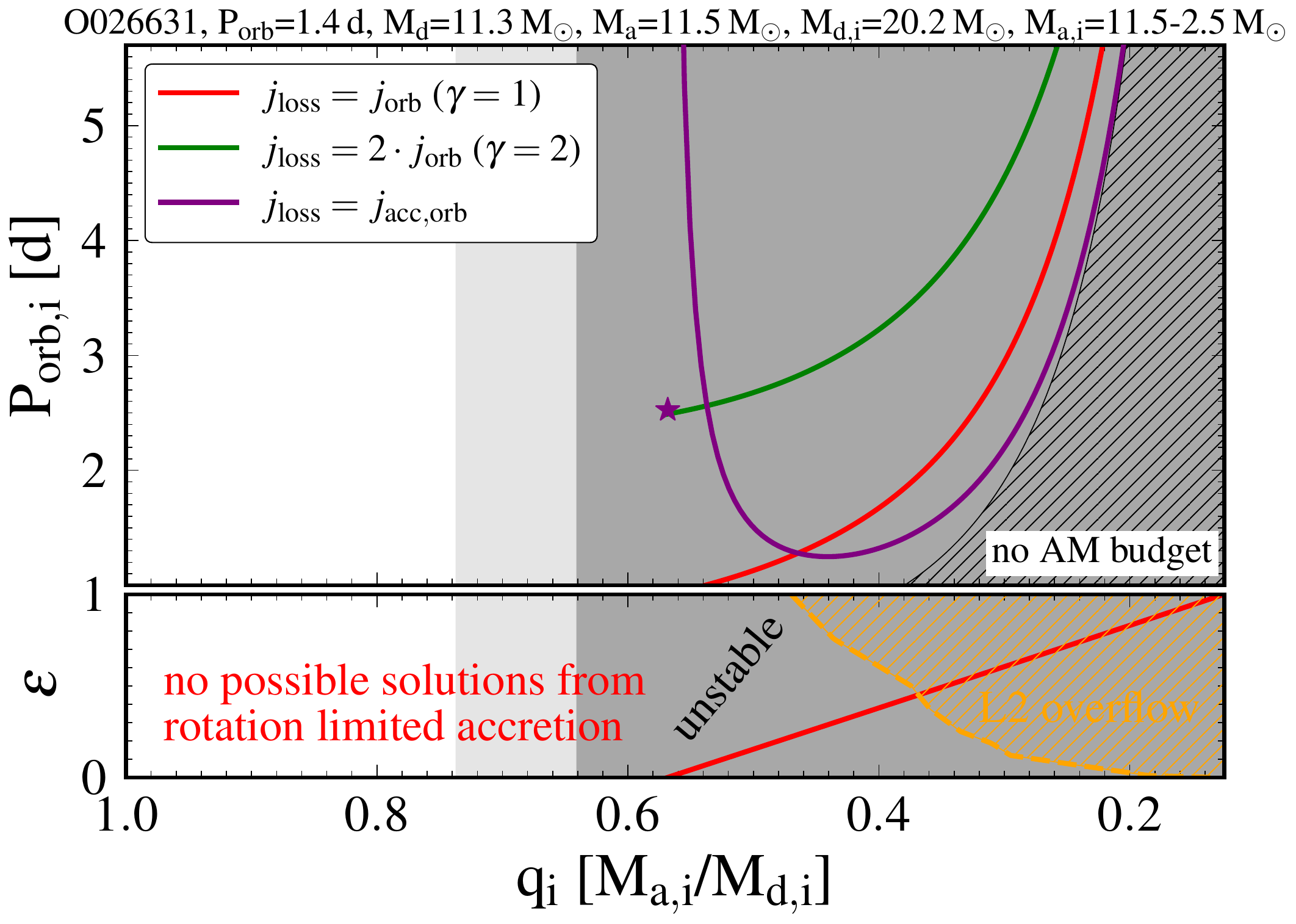}
    \caption{The parameter space of mass transfer efficiency
$\varepsilon$ (\textit{bottom panels}) and initial orbital period $P_{\rm orb,i}$ (\textit{top panels}) as a function of the decreasing initial mass ratio $q_{\rm i}$ for LZ\,Cep (\textit{left panel}, representing a typical case) and O026631 (\textit{right panel}, representing an outlier system). The white regions in both panels indicates the most likely values of mass transfer efficiency and specific angular momentum loss from the rotation-limited accretion model. Different colours correspond to curves of $P_{\rm orb,i}$ for alternative assumptions of $\gamma$=1 (red), 2 (green) and $M_{\rm d}/M_{\rm a}$ (purple, isotropic remission from the surface of the accretor). In the right panel, the purple star shows the initial orbital period for $\varepsilon=0$ (Eq.\,(\ref{eq:epsilon0})). The black and blue hatched regions correspond to $\gamma\leq0$ and $\gamma\geq3$, respectively. The dark grey colour shows the region where all detailed binary evolution models merge. All detailed binary evolution models survive the thermal timescale mass transfer to the left of the light grey region (the highest value of $q_{\rm i}$ in the light grey region is the $q_{\rm max}$ limit). Models in the light grey region merge or survive depending on the binary orbital period; the shortest-period models merge, and vice versa  (the lowest value of $q_{\rm i}$ in this region is the $q_{\rm min}$ limit). The orange dashed line shows the limiting mass transfer efficiency as a function of the initial mass ratio for which non-rotating models of \citet{Schurmann2024} undergo L2 overflow. The orange shaded region, therefore, denotes the parameter space where an accretion disk-mediated mass transfer onto the accretor may not be stable. The upper limit to the top panels is given by $P_{\rm orb,i,max}$ calculated for each system. }
    \label{fig:individual_examples}
\end{figure*}


\section{Results}
\label{sec:result}
In this section, we evaluate the possible range of mass
transfer efficiency $\varepsilon$ and angular momentum
loss factor $\gamma$ for each massive Algol binary in
Table\,\ref{table:all_algols_smc} and Table\,\ref{table:all_algols_lmc_mw}. We describe in detail our
analysis for one representative binary (Sect.\,\ref{sec:example}),
discuss the outliers (Sect.\,\ref{sec:outliers}), and then
show the population properties in Sect.\,\ref{sec:pop}
and derive empirical constraints on mass transfer efficiency
during the thermal timescale mass transfer phase on the
Main Sequence.

\subsection{Example analysis}
\label{sec:example}

We describe our analysis in detail for the massive Algol in the Milky Way, LZ\,Cep (Fig.\,\ref{fig:individual_examples}, left panel). The binary has a Roche-lobe filling donor star of mass 6.5\,$M_{\odot}$ in a 3.07\,d orbit with a 16\,$M_{\odot}$ companion \citep{mahy2011}. For the current donor mass $M_{\rm d}=6.5$\,$M_{\odot}$, we find the initial donor mass $M_{\rm d,i}=13.46$\,$M_{\odot}$ and maximum possible initial orbital period $P_{\rm orb,i,max}$ of 5.62\,d (see Table\,\ref{table:1}). This implies that the donor has lost 6.96\,\,$M_{\odot}$ during the fast Case\,A mass transfer phase. For the case of conservative mass transfer, the accretor could have accreted all the mass lost by the donor (neglecting wind mass loss), and the minimum initial mass of the accretor could be 9.04\,$M_{\odot}$. Since the initial mass of the accretor cannot be higher than that of the donor, the maximum initial accretor mass can be 13.46\,$M_{\odot}$. Hence, the initial mass of the accretor can range between 9.04-13.46\,$M_{\odot}$.

The accretor $M_{\rm a}=16$\,$M_{\odot}$ must have accreted at least 2.54\,$M_{\odot}$ of matter from the donor, up to the maximum of 6.96\,$M_{\odot}$. So, the mass transfer efficiency for this binary ranges from $\sim$0.36 to 1 (Fig.\,\ref{fig:individual_examples}, left bottom panel). However, since binaries with low initial mass ratios are not expected to survive an arbitrarily efficient thermal timescale mass transfer on the Main Sequence (grey regions), the higher values of mass transfer efficiency are less likely. The white area in Fig.\,\ref{fig:individual_examples} shows the permitted range of mass transfer efficiency and specific angular momentum loss for LZ\,Cep. Hence, we expect that the mass transfer efficiency in this system may lie in the range $\sim0.36-0.85$ (bottom panel), where we exclude the values of $\varepsilon$ in the grey regions.

For conservative mass transfer ($\varepsilon=1$), we can calculate the initial orbital period the binary must have had to reach its current orbital period of 3.07\,d (Eq.\,(\ref{eq:conservative}), top left panel of Fig.\,\ref{fig:individual_examples}). The upper limit to the initial orbital period is given by the maximum initial orbital period for which mass transfer can occur on the Main Sequence. For LZ\,Cep, we see that angular momentum conservation allows for conservative mass transfer, although mass transfer stability arguments make it unlikely. For each value of inefficient mass transfer possible for LZ\,Cep ($0.36\leq\varepsilon<1$), we can calculate a range of values of initial orbital period, depending on the assumption of specific angular momentum loss from the binary (Eq.\,(\ref{eq:inefficient}), different colored lines in the upper panels of Fig.\,\ref{fig:individual_examples}).

We can derive the plausible range of specific angular momentum loss factor $\gamma$ that can lead to the current orbital period $P_{\rm orb} = 3.07$\,d of the binary, for each value of $\varepsilon\neq1$. The black hatched region is excluded because $\gamma<0$ implies the binary gains angular momentum on mass loss. We deem the blue shaded region unlikely because the system has to lose a high amount of angular momentum ($\gamma>3$, \citealp{Nuijten2025}). To the left of the grey regions, we find the plausible range of the initial orbital period $P_{\rm orb,i}$, and in turn $\gamma$\footnote{The analytical equations for gamma (Eq.\,\ref{eq:inefficient}\,\&\,\ref{eq:epsilon0}) are valid only when epsilon is constant throughout the mass transfer phase. Since the mass transfer efficiency in the detailed binary models is a function of time, the analytical line showing $j_{\rm loss} = j_{\rm acc-orb}$ does not directly correspond to the detailed binary evolution model’s specific angular momentum loss.}. We set the lower limit to the initial orbital period to 1\,d as most binaries having shorter orbital periods are expected to enter into a nuclear timescale contact phase shortly after the onset of mass transfer on the Main Sequence \citep{Menon2021,Abdul-Masih2021,Abdul-Masih2022,Fabry2022,Fabry2023,Rickard2023,Henneco2024,Vandersnickt2025,Gull2025}. The allowed range of $\gamma$ for this system is $\sim0.5-3.0$. We also note that the binary is less likely to have originated from the most extended initial orbital periods due to considerations of angular momentum loss (blue hatched regions have $\gamma>3$).

\begin{figure*}
    \centering
    \includegraphics[width=0.49\linewidth]{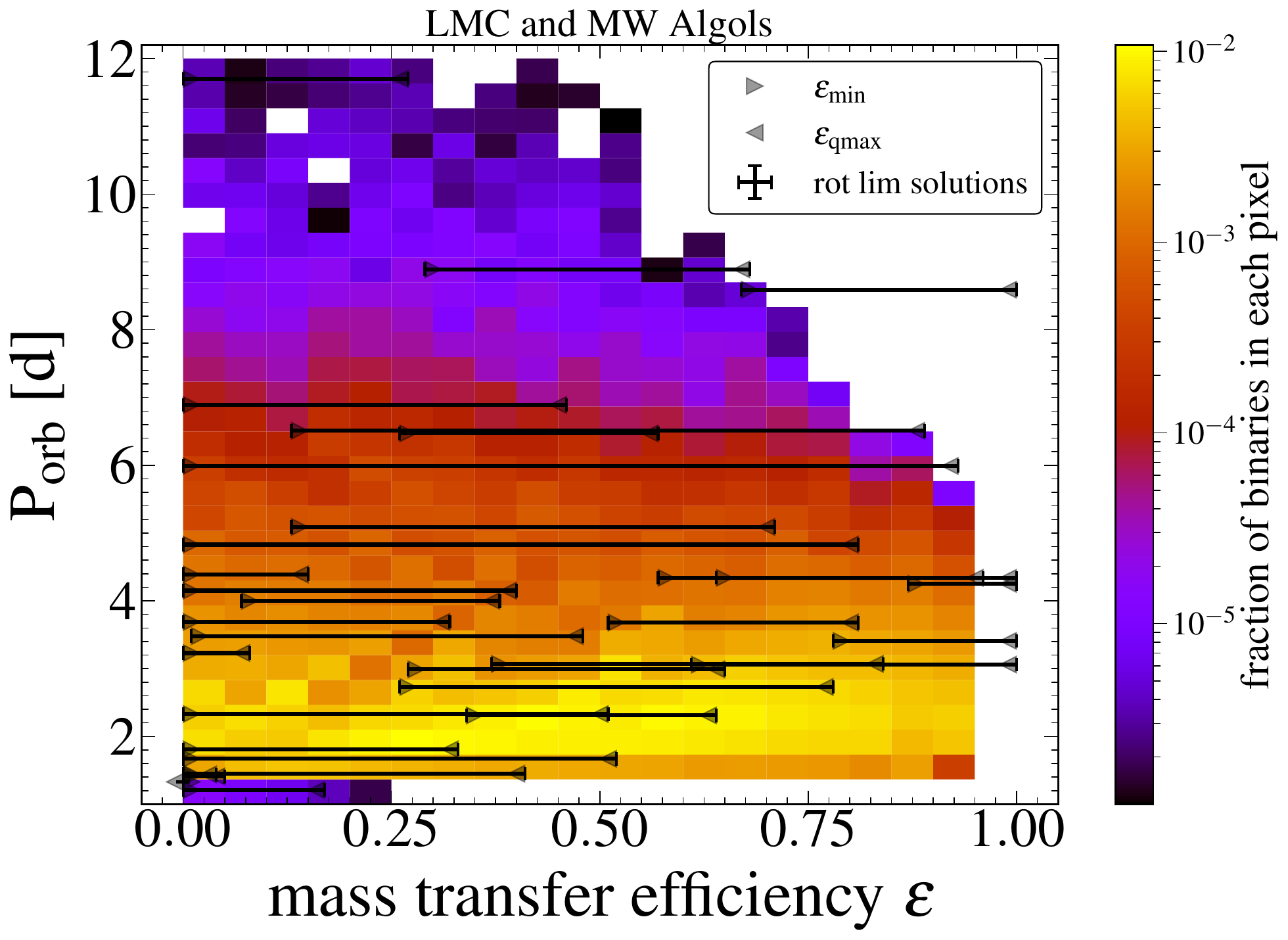}
    \includegraphics[width=0.49\linewidth]{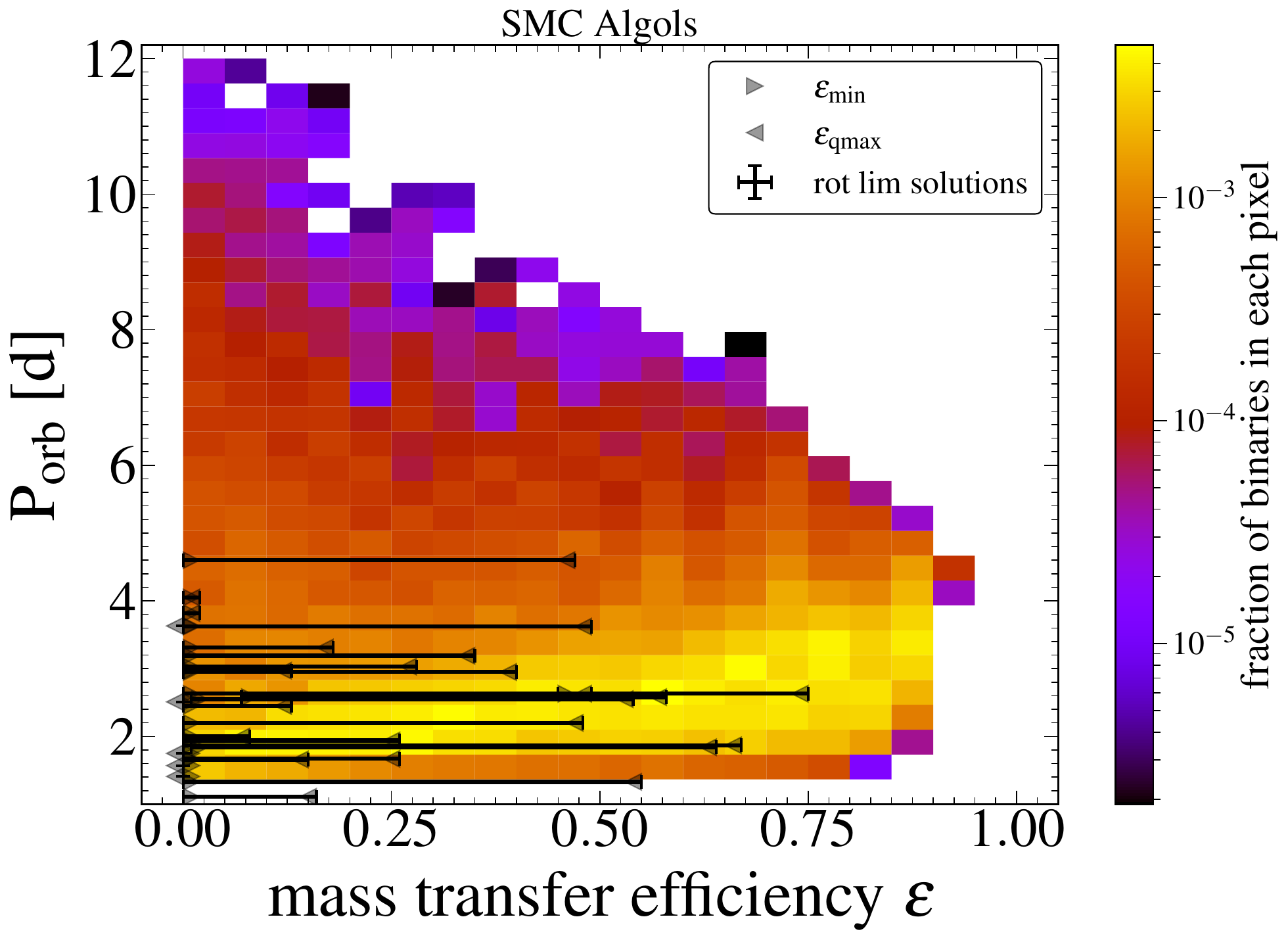}
    \caption{Inferred range of mass transfer efficiency (black lines) as a function of the orbital period of massive Algols in the LMC and Milky Way (left panel) and the SMC (right panel). Triangles indicate the minimum and maximum mass transfer efficiency possible for each system, in the rotationally limited mass accretion scheme. The 2D histogram displays the calculated mass transfer efficiency in detailed binary evolution models that also incorporate the rotationally limited mass accretion prescription (note the log scale on the colour bar). The total probability is normalised such that the integrated sum over the entire area is 1. }
    \label{fig:porb_beta}
\end{figure*}

\subsection{Outliers}
\label{sec:outliers}

The right panel of Fig.\,\ref{fig:individual_examples} shows that the mass transfer stability consideration (exclusion of grey regions) can lead to cases (e.g. O026631 in the SMC) where we find no solutions for mass transfer efficiency or angular momentum loss factor. We see that our assumption on the mass transfer stability has to be relaxed at least to $q_{\rm max} \approx 0.56$ for this system to have a non-conservative solution ($\varepsilon=0$), and below 0.5 to have increasingly more efficient mass transfer ($\varepsilon>0$). A similar case in the Milky Way is TT\,Aur (Fig.\,\ref{fig:appendix_lmc_mw}), where the mass transfer stability criterion needs to be relaxed to at least $q_{\rm min} = 0.68$ at the lowest initial donor mass to have a $\varepsilon=0$ solution.

Detailed binary evolution models predict that the mass transfer leads to mergers at increasing initial mass ratios for decreasing initial donor masses (Table\,\ref{table:1}). We find that our analytical approach leads to no or highly inefficient solutions in most of the least massive Algols, both in the Milky Way and the SMC (see the top few rows of Fig.\,\ref{fig:appendix_smc} and Fig.\,\ref{fig:appendix_lmc_mw}). The least massive Algols have the strongest constraints from the mass transfer stability criterion ($q_{\rm min}\sim0.75$). Hence, we conclude that the least massive Algols give evidence that mass transfer must be stable up to $q_{\rm min}\sim0.6$ at initial donor masses $M_{\rm d,i}\approx10-15$\,$M_{\odot}$. This implies that the difference in stability of mass transfer when using detailed models compared to when using semi-analytic fits from \citet{hurley2002} may impact rates of astrophysical events of interest, including neutron star X-ray binaries and binary neutron star mergers (\citealp[]{Xu2025}, see also \citealp{Eldridge2017,Agrawal2023,Fragos2023,Chattaraj2025}).

The allowed range of mass transfer efficiency $\varepsilon$ is more strongly constrained from the mass transfer stability criterion than from angular momentum conservation (c.f. grey region vs black hatched area in the right panel of Fig.\,\ref{fig:individual_examples}). We also find that high values of $\gamma$ are forbidden by the maximum limit on initial orbital period for Case\,A mass transfer (absence of blue hatched regions in the top right panel of Fig.\,\ref{fig:individual_examples}). Therefore, the upper limit to the maximum initial orbital period for mass transfer to initiate on the Main Sequence $P_{\rm orb,i,max}$ plays a less restrictive role in deriving our conclusions on mass transfer efficiency and stability in this system. The above case is similar for the least massive Algols in the Milky Way as well (c.f. first few rows of Fig.\,\ref{fig:appendix_smc} and Fig.\,\ref{fig:appendix_lmc_mw}). Hence, even though we assume the same upper limit to initial orbital periods for the LMC and Milky Way binaries, we do not expect that the range of mass transfer efficiency inferred from our analysis will change by increasing $P_{\rm orb,i,max}$. The allowed range of $\gamma$ will increase if $P_{\rm orb,i,max}$ is increased for the Milky Way systems, although the highest values of initial orbital period will remain less likely.

To obtain a non-zero solution to the mass transfer efficiency, our detailed binary models will require a mechanism to remove angular momentum from the accretor while continually accreting mass, in the form of a decretion disk or magnetic braking. The orange dashed line shows the minimum mass ratio as a function of mass transfer efficiency such that non-rotating accretors do not undergo L2 overflow \citep{Schurmann2024}. The intersection of the orange dashed line with the black line represents the maximum mass transfer efficiency that the system could achieve in the presence of a disk to siphon angular momentum out of the accretor. For O026631, the maximum mass transfer efficiency $\varepsilon_{\rm max}^{\rm disk}$\,$\sim$\,0.45. We also note that rotation, however, can reach critical values in the accretor's outer layers and have a large effect on the radius; therefore, we consider the models of \citet{Schurmann2024} conservative in the sense that nature may reach L2 overflow at much higher initial mass ratios than inferred from non-rotating models. Moreover, it has been recently shown that L2 mass loss can occur when the decretion disk itself extends beyond the L2 point, significantly shrinking the orbit and leading to possible merger scenarios well before the non-rotating limit above \citep{Lu2022}.

\subsection{Population properties}
\label{sec:pop}

In this section, we derive the possible range of mass transfer efficiencies if each of the observed systems had undergone rotationally limited mass transfer only (Sect.\,\ref{sect:rot}), or for the case of a disk-mediated mass transfer that can siphon angular momentum from the accretor to the orbit while the accretor continually accretes mass (Sect.\,\ref{sect:disk}). We calculate the maximum value of mass transfer efficiency possible in the rotationally limited mass accretion case for each system at $q_{\rm max}$, and tabulate our results as $\varepsilon_{\rm qmax}$ in Table\,\ref{table:all_algols_smc} and Table\,\ref{table:all_algols_lmc_mw}. We also show the minimum and maximum values of the angular momentum loss factor, tabulated as $\gamma_{\rm min,qmax}$ and $\gamma_{\rm max,qmax}$ in Table\,\ref{table:all_algols_smc} and Table\,\ref{table:all_algols_lmc_mw}. We also show the maximum value of mass transfer efficiency possible for a disk-supported mass transfer $\varepsilon_{\max}^{\rm disk}$ in Table\,\ref{table:all_algols_smc} and Table\,\ref{table:all_algols_lmc_mw}.

For about $\sim$50\% of the massive Algols in the LMC (e.g. HV\,2241, VFTS\,652, SC1\,105) and Milky Way (e.g. XX\,Cas, AB\,Cru), we find that non-conservative mass transfer ($\varepsilon=0$) is ruled out because the allowed range of initial accretor masses has to be lower than the initial donor mass (mass budget). In these systems, the current accretor masses are higher than the calculated initial donor masses, such that the accretors must have accreted mass from the donor to reach the current configuration. However, only two out of 29 systems in the SMC necessarily require such non-zero solutions to mass transfer efficiency.

\begin{figure*}
    \centering
    \includegraphics[width=0.49\linewidth]{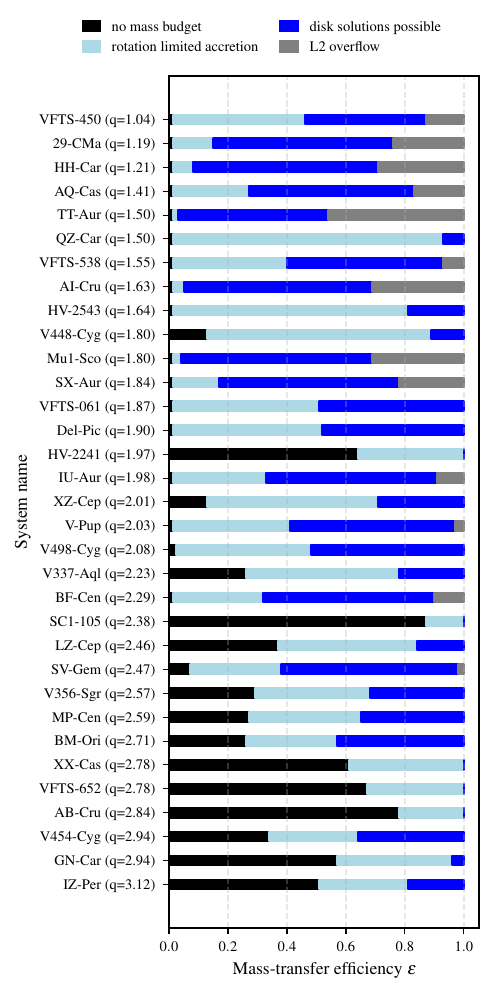}
    \includegraphics[width=0.49\linewidth]{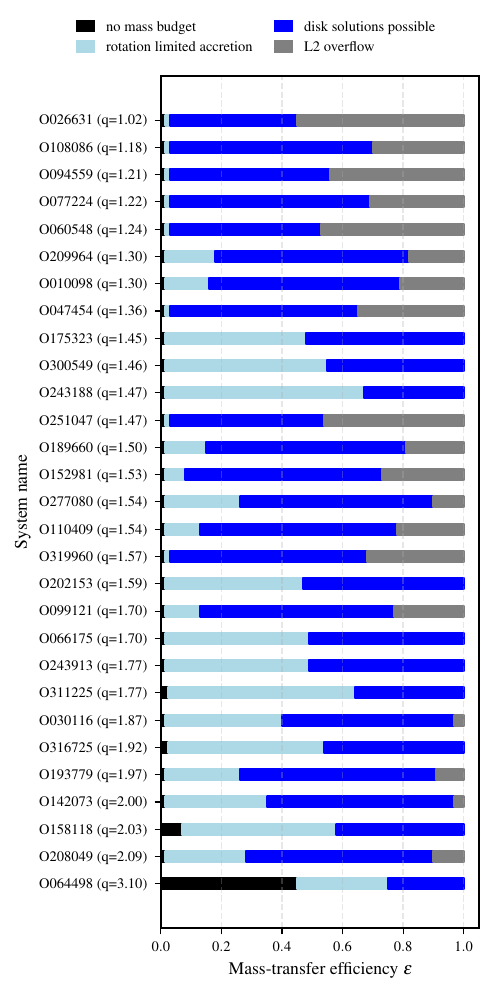}
    \caption{Disallowed range (black bar - initial accretor mass becomes larger than initial donor mass; grey bar - L2 overflow of the accretor) and allowed range (light blue - rotationally limited mass accretion scheme; dark blue - disk supported mass transfer scheme) of mass transfer efficiencies for each system in the LMC \& Milky Way (\textit{left panel}) and SMC (\textit{right panel}), arranged in increasing order of observed mass ratio. }
    \label{fig:bar}
\end{figure*}

In some systems (e.g. V356\,Sgr, AQ\,Cas, see Fig.\,\ref{fig:appendix_lmc_mw}), conservative mass transfer ($\varepsilon=1$) is also excluded from angular momentum loss constraints. In such cases, the initial orbital period required for conservative mass transfer by the binary is much higher than the maximum initial orbital period where mass transfer can occur on the Main Sequence. Angular momentum conservation rules out conservative mass transfer in $\sim$28\,\% of massive Algols in the SMC and $\sim$\,11\% of massive Algols in the Milky Way. However, we find that the constraint on mass transfer efficiency from angular momentum loss consideration ($\gamma < 0$) is less restrictive than the mass transfer stability consideration ($q_{\rm max}$ and L2 overflow limits) in all the analysed massive Algols (see Fig.\,\ref{fig:appendix_smc} and Fig.\,\ref{fig:appendix_lmc_mw}).

\subsubsection{Rotationally limited mass accretion scheme}
\label{sect:rot}

Figure\,\ref{fig:porb_beta} shows the allowed range of mass transfer efficiencies as a function of the current orbital period of each massive Algol. We observe that the current population of massive Algols does not require mass transfer to be very efficient at the shortest periods, unlike what is theoretically expected for rotationally limited accretion in the presence of strong tides. We observe evidence for an increase in the upper limit to the mass transfer efficiency as a function of the binary orbital period for systems in the Milky Way and the LMC. However, the mass transfer efficiency of the SMC systems shows no correlation with the orbital period (c.f. \citealp{selma2007}). Detailed binary evolution models predict that the majority of massive Algols at the shortest orbital period undergo fairly conservative mass transfer. However, the observed population of massive Algols do not populate the highest probability regions marked by the yellow background area in the $P_{\rm orb}-\varepsilon$ plane.

Comparing the top and middle panels of Fig.\,3 in \citet{Sen2022}, we see that most models that undergo highly efficient thermal timescale mass transfer ($\varepsilon>0.8$) merge during the Main Sequence after entering into a contact configuration. Most of the predicted population of massive Algols that do not merge on the Main Sequence undergoes inefficient mass transfer ($\varepsilon<0.5$). Therefore, the observed population of massive Algols may only represent the predicted population of Algols that do not merge on the Main Sequence. We speculate that highly efficient mass transfer at the shortest periods may lead to a contact phase on the Main Sequence (due to radial expansion of the accretor) at a much faster timescale than predicted by current binary evolution models \citep[see also][]{Menon2021,Henneco2024}.

In the orbital period range 3-5\,d, the observed systems in the Milky Way and LMC form two separate sub-populations, one requiring $\varepsilon<0.5$ and vice versa. In detailed binary evolution models, systems with initial mass ratios above 0.8 undergo fairly efficient mass transfer and vice versa (Fig.\,F2 of \citealp{Sen2022}). Therefore, the observed sub-populations of Algols in the Milky Way and the LMC may suggest a mass ratio dependence in mass transfer efficiency, which could also be an outcome of thermal timescale-limited mass transfer efficiency \citep[e.g.][]{Belczynski2008,Schneider2015}.

At low metallicity, accretors spin down more slowly on the Main Sequence due to weaker stellar winds. For rotationally limited mass accretion, as in the detailed binary evolution models, the accretors spin up faster due to mass transfer at lower metallicity, because stars are more compact and hence have a smaller moment of inertia. Tides are also less effective in halting the spin-up of the accretor because the radius of the accretor is smaller. Starting from the same initial rotation velocity, orbital period of the binary and mass of the accretor, the above three effects combine to predict a less efficient mass transfer episode at lower metallicity. The maximum mass transfer efficiency of the models in the right panel is lower compared to the left panel of Fig.\,\ref{fig:porb_beta}. Likewise, we find that the observed Algols in the SMC seem to have undergone less efficient mass transfer on average than those in the LMC and Milky Way.

\subsubsection{Possible disk solutions}
\label{sect:disk}

Here, we assume that the presence of a disk during the thermal timescale mass transfer phase can support arbitrarily efficient mass transfer, unless the accretor's radial response to the mass accretion leads to L2 overflow in the binary. Therefore, mass transfer efficiency solutions up to $\varepsilon_{\rm max}^{\rm disk}$ may be possible for each system in the presence of a disk. 

Fig.\,\ref{fig:bar} shows the range of mass transfer efficiency solutions possible for the different models of mass transfer- rotationally limited (light blue) and disk-mediated (dark blue, but also includes the light blue region by construction). We observe a strong mass ratio dependence in the allowed range of mass transfer efficiencies. Systems with a mass ratio less than $\sim$2 are consistent with them having undergone largely inefficient mass transfer and vice versa. We see that L2 Overflow rules out conservative mass transfer in the smallest mass ratio systems, while the mass budget constraints rule out non-conservative mass transfer in the highest mass ratio systems. 

Most of the current population of observed massive Algols in the SMC has mass ratios less than 2, whereas half of the Algol population in the LMC and Milky Way has mass ratios above 2. This may be an observational selection effect where it is harder to detect unequal mass ratio binaries in the SMC than the Milky Way (note that there are only seven observed massive Algols in the LMC in the current sample, and four of them are from the recent multi-epoch high-resolution VFTS-TMBM survey). However, there is a system with mass ratio above 3, and it is one of the least massive systems in our sample (Table\,\ref{table:all_algols_smc}, second row). If not biased by observational selection effects, our results strongly suggest that there may be a metallicity dependence in the mass transfer efficiency. A future multi-epoch spectroscopic survey of the remaining eclipsing binary sample from \citet{Harries2003} and \citet{hilditch2005}, can extend the observed sample and shed light on the lack of observed unequal mass ratio binaries in the SMC. 


\section{Discussion}
\label{sec:discussion}

\subsection{Mass transfer efficiency and stability}

In \citet{Sen2022}, we found that the orbital period and mass ratio distribution of massive Algols in the LMC and Milky Way give evidence for efficient mass transfer in some systems, while others have undergone inefficient mass transfer. We also found that the least massive systems are not well reproduced by their accretor spin-dependent mass transfer prescription. In this work, we show that the mass transfer stability criterion that requires the combined luminosity of two stars to be able to drive the excess mass loss via winds during an inefficient mass transfer episode is too strict to provide acceptable solutions for mass transfer efficiency in the least massive Algols (Fig.\,\ref{fig:appendix_lmc_mw}). \citet{selma2007} reported a similar lack of acceptable solutions for the least massive Algols in the SMC (see their Table\,3, $\chi^2>10$), which may again be due to the lack of models that underwent stable mass transfer at low initial mass ratios. Mass loss from the L2 Lagrangian point \citep{Lu2022} may provide an additional mechanism to remove the excess mass. However, L2 mass loss inherently implies a high value for $\gamma$, which may be ruled out by the upper limit on the maximum orbital period for Case\,A mass transfer (see blue regions, or the lack thereof, in the top four rows of the Fig.\,\ref{fig:appendix_smc} and Fig.\,\ref{fig:appendix_lmc_mw}).

In the SMC, \citet{selma2007} found a weak correlation between the mass transfer efficiency and initial orbital period of their best-fit binary models to the observed population of massive Algols. In our work, we also see a weak dependence of the allowed range of mass transfer efficiency on the current orbital period of the same sample, where the upper limit to the mass transfer efficiency from the rotationally limited accretion model decreases for increasing orbital period (Fig.\,\ref{fig:porb_beta}). The detailed binary model grids of \citet{selma2007,wang2020,Sen2022} assume that the mass lost due to inefficient mass transfer carries away the specific orbital angular momentum of the accretor. If $\varepsilon$ is assumed to be constant during the entire mass transfer episode, purple curves in Fig.\,\ref{fig:individual_examples} represent the initial orbital periods for various mass transfer efficiencies. However, the mass transfer efficiency in the detailed binary evolution models is not constant during the mass transfer phase \citep[e.g.][]{Renzo2021}.

From our analytical study, we observe that the assumption of higher mass transfer efficiencies necessitates that the binary models begin with longer orbital periods (see black shaded region in the lower right of the left panel of Fig.\,\ref{fig:individual_examples}). In some of the least massive Algols in the SMC (first four rows in Fig.\,\ref{fig:appendix_smc}), strongly efficient mass transfer ($\varepsilon>0.8$) is also ruled out from angular momentum conservation arguments alone: the initial orbital period required for $\varepsilon>0.8$ is higher than the maximum initial orbital period for mass transfer to commence on the Main Sequence in the SMC.

The observed population of short-period stripped star+OB star binaries is expected to be the evolved counterparts of massive Algol binaries \citep{wellstein2001,Sen2022}. At the high masses, these constitute the Wolf-Rayet+O star (WR+O) binaries. Most of the WR+O binaries have been found to require highly inefficient mass transfer, both in the Milky Way \citep{petrovic2005,shao2016,Nuijten2025} and the SMC \citep{Schurmann2024}. This is in agreement with our finding that a large majority of the massive Algols may have undergone inefficient mass transfer. On the other hand, low- and intermediate-mass Algols require a wide range of efficiencies to explain the observed populations \citep{nelson2001,vanRensbergen2006,VanRensbergen2010,Deschamps2013,deschamps2015,mennekens2017,Dervisoglu2018,negu2018,vanRensbergen2021}. Evidence for a mass dependence of the accretion efficiency just in the massive star regime is discussed in \citet{Schurmann2025}.

Recent studies on Be+sdOB binaries found that they require mass transfer to be significantly more efficient than in the case of massive Algols and WR+O binaries (\citealp{Lechien2025,Bao2025}, see also \citealp[]{schootemeijer2018,Wang2021,Picco2025}). Most of these systems are predicted to have undergone mass transfer after the main-sequence evolution of the donor star (Fig.\,4 of \citealp{Lechien2025}), that is, they are not the direct evolutionary counterparts to Algol binaries. These systems strongly disfavour the accretion spin-up dependent mass transfer efficiency model (see also, \citealp{Dervisoglu2010}), as tides in long-period systems are less effective in tidally locking the accretor to the orbital angular velocity of the binary. The presence of a viscous decretion disk around the Be star in these binaries may allow for the loss of angular momentum via the boundary layer coupling between the star and the disk while continually accreting mass \citep{Paczynski1991,Popham1991,Colpi1991,Bisnovatyi-Kogan1993,selma2013,Martin2025}.

To reduce the predicted binary black hole merger rate in gravitational wave population synthesis models, \citet{Romero2023} proposed conservative Case\,B mass transfer to increase the orbital periods of the resulting compact object binaries such that they do not merge within Hubble time. Similarly, previous studies propose that the mass transfer efficiency should be as high as 50\% to explain the observed population of long-period Be X-ray binaries \citep{Shao2014,Vinciguerra2020,Schurmann2025}, where such binaries are expected to have undergone Case\,B mass transfer. On the other hand, the observed population of short-period WR+O binaries in the local Universe seems to have undergone relatively inefficient mass transfer \citep{petrovic2005, shao2016, Nuijten2025}. Likewise, \citet{Lau2024} discuss that the observed population of gravitational wave sources may have undergone very inefficient mass transfer (see also, \citealp{Bouffanais2021,vanSon2022b,Picco2024}). Recently, \citet{Zapartas2025} interpreted that the current sample of binary companions to stripped-envelope supernovae also favours inefficient mass accretion.

\citet[][Fig. 8]{Schurmann2024} showed that the lowest mass ratio for which L2 Overflow is avoided is a function of the mass transfer efficiency, and it is always greater than 0.5 for conservative mass transfer. While their models are at SMC metallicity, the accretors swell up even more at higher metallicities, and the parameter space for contact formation and L2 overflow increases to even higher initial mass ratios. This is consistent with our finding that the least massive Algols require inefficient mass transfer. 

\subsection{Extent of envelope stripping}
\label{sect:envelope_stripping}
\begin{figure}
    \centering
    \includegraphics[width=\linewidth]{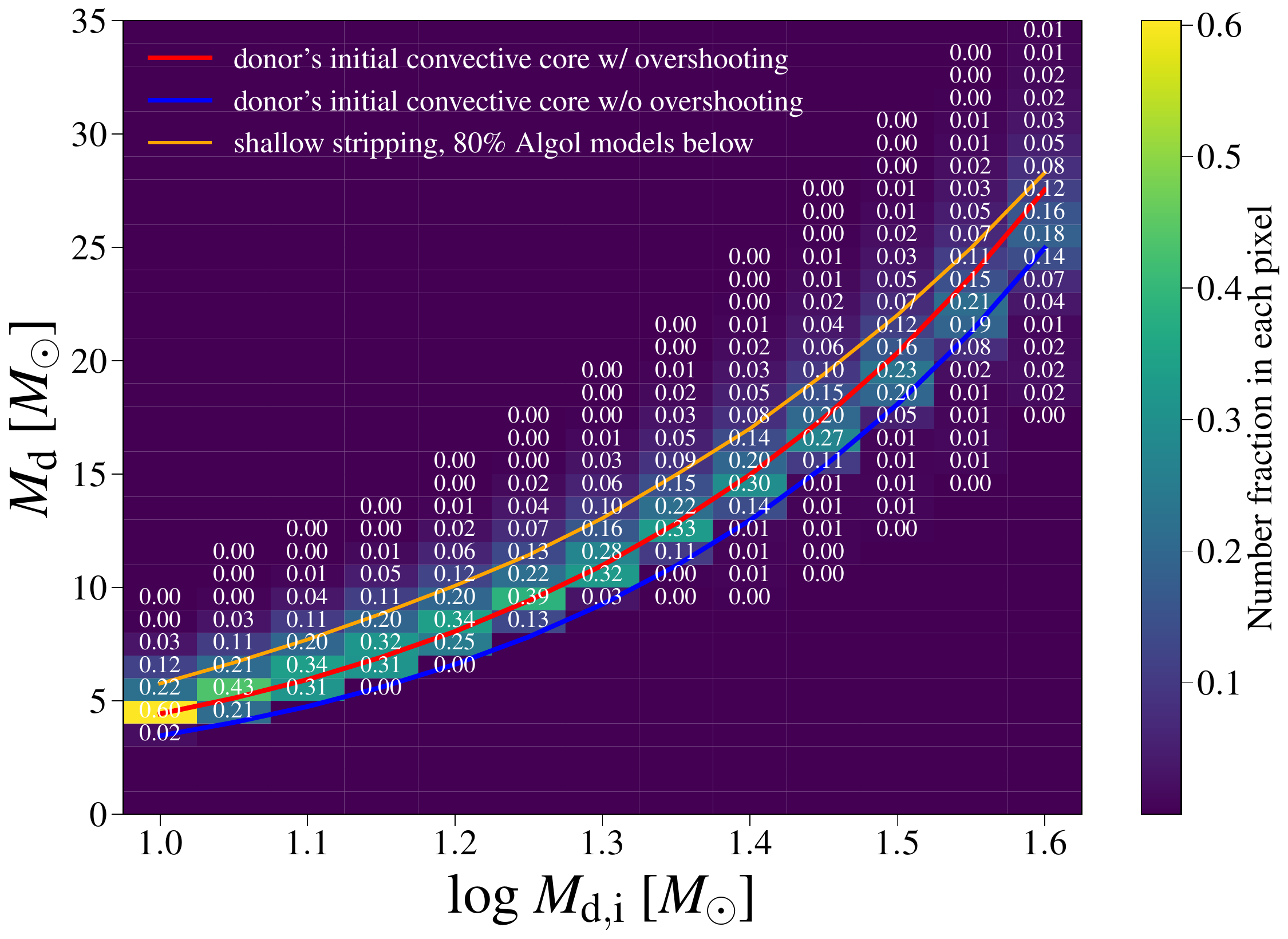}
    \caption{2D-histogram showing the distribution of the donor masses as a function of the logarithm of initial donor mass during the Algol phase, for the detailed binary evolution models with LMC metallicity \citep{Sen2022}.  The total probability is normalised such that the sum over all pixels for each logarithmic initial mass bin is unity (overplotted numbers show contribution from each pixel up to two significant digits). The red and blue curves show the convective core mass of the donor at Zero-Age Main Sequence with and without overshooting, respectively. The orange curve indicates that 80\% of donor models during the Algol phase are stripped below the curve. }
    \label{fig:ccdi}
\end{figure}

Figure\,\ref{fig:ccdi} shows the distribution of donor masses during the Algol phase. During the thermal timescale mass transfer on the Main Sequence, the donor may get stripped up to its initial convective core mass, which includes the overshooting region (e.g. Fig.\,2 of \citealp{Schurmann2024b}). The above is represented by the red curve in the 10-40\,$M_{\odot}$ range of initial donor masses, where the majority of the donor masses at each initial donor mass bin reside during the Algol phase. During the nuclear timescale mass transfer phase, the donors lose further mass. We observe that the blue curve, representing the donor's initial convective core mass excluding the overshooting region, provides an adequate approximation to the deepest extent of envelope stripping due to mass transfer on the Main Sequence (including both fast- and slow-Case A, e.g., \citealt{klencki:25}).

Physically, the response of the (inner layers of the) donor to the mass lost is primarily determined by its thermal state, that is the temperature gradient. Therefore, whether to include or not the convective boundary mixing region should be decided based on its gradient: for convective penetration \citep[e.g.,][]{Anders2022}, the gradient is adiabatic and no distinction with the convective core should be expected. However, other processes are known to contribute to convective boundary mixing, and the thermal gradient they establish on timescales that are long compared to the convective turnover is not always clear. In our study, we used the red curve (including convective boundary mixing) to determine the initial donor masses of the observed Algols (Sect.\,\ref{sec:method}) as our fiducial extent of envelope stripping during the fast Case\,A mass transfer. Assuming a deeper extent of envelope stripping to estimate the initial donor mass in our analysis would lead to the inference of more inefficient mass transfer occurring in the observed systems. 

We see that a non-negligible fraction of Algol donor models lie above the red curve. If our models accurately capture the extent of envelope stripping during the Algol phase, some of the observed Algols may have donors that have not yet stripped down to their initial convective core mass. If we assume a shallower stripping, such that all the observed donors get stripped up to the orange curve, the maximum mass transfer efficiency that the observed systems may have undergone shifts to higher values (Fig.\,\ref{fig:shallow_stripping_porb_beta}), and many observed Algols may have undergone fairly efficient mass transfer. Inefficient mass transfer solutions also become possible for the least massive Algols without having to relax the mass transfer stability criterion for the least massive Algols. However, there are six systems where the minimum initial accretor mass for conservative mass transfer exceeds the initial donor mass derived from the shallow stripping curve (see black triangles at $\varepsilon=1$); hence, these are unphysical solutions and indeed give evidence that not all observed Algols justify the shallow stripping assumption. We also see that a handful of Algols in the SMC still require inefficient mass transfer ($\varepsilon < 0.5$). Even for the shallower stripping assumption, the observed Algols in the SMC appear to have undergone less efficient mass transfer, on average, than those in the LMC and Milky Way systems. Surface nitrogen abundance measurements of the observed population of massive Algols can be another independent diagnostic to constrain the efficiency of mass transfer in these systems.

Finally, the extent of envelope stripping may non-trivially depend on the mass transfer efficiency itself, where the mass and angular momentum exchange between the donor and the accretor determines the closest orbit during the thermal timescale of mass transfer. The mass transfer efficiency in our models is dependent on both the initial orbital period and the initial mass ratio for a given initial donor mass. The detailed models cover the full range of mass transfer efficiencies (Fig.\,F2 of \citealp{Sen2022}). While quantifying the effect of mass transfer efficiency on the extent of envelope stripping is beyond the scope of this work, we note that the occurrence of both extremes in mass transfer efficiencies in our models also implies that the predicted spread of donor masses during the Algol phase in Fig.\,\ref{fig:ccdi} includes the effect of mass transfer efficiency.

Our detailed binary evolution models suggest that slow Case\,A mass transfer, occurring at the nuclear timescale, is highly efficient except at the longest initial orbital periods ($\varepsilon>0.6$, see Fig.\,F2 of \citealp{Sen2022}). The estimates for mass transfer efficiency derived in this work give the time-averaged mass transfer efficiency for each system \citep{PolsMarinus1994}. Starting from the same initial donor mass, if the current donor mass is an outcome of mass lost during both fast and slow Case\,A, we expect that the prior thermal timescale mass transfer phase may have been slightly less efficient than our derived range of values for each observed massive Algol.

\subsection{Convective core mass}

We calculate the initial mass of the donor from the initial mass of the convective core. The Ledoux criterion sets the initial convective core mass in our models for convection. Assuming the Schwarzschild criterion \citep[e.g. in][]{selma2007} for convection will lead to higher estimates of initial convective core masses for the same initial mass. Recent studies of gravity-mode asteroseismology of B stars \citep{Pedersen2021} and analyses of eclipsing binaries \citep{Claret2019,Tkachenko2020} suggest that the convective cores of stars are larger than assumed in many stellar evolution models. Building on this observational evidence, theoretical work has demonstrated that convective penetration in O- and B-type stars can enlarge the convective core by about 10–30\% of the pressure scale height at the core boundary \citep{Anders2022,Jermyn2022,Johnston2024}. Larger convective cores in donors imply smaller envelopes for the same total initial mass. For systems where the current accretor mass is larger than the initial donor mass, the lower limit on mass transfer efficiency shifts to higher values. The initial mass ratio required for conservative mass transfer also relaxes to higher values.

The detailed binary evolution models treat convective boundary mixing with parametric algorithms, which assume a radiative gradient in this region. The models of \citet{wang2020,Sen2022} have a step overshooting parameter equal to 0.335 times the local pressure scale height and a radiative gradient in the overshooting region, calibrated to rotational velocity measurements from the FLAMES Survey of Massive Stars at $\sim$15\,$M_{\odot}$ (\citealp{Hunter2008,brott2011}, see also the recent theoretical study by \citealp{Andrassy2024}). \citet{Castro2014,Castro2018} later confirmed this calibration for stars of similar mass in the Milky Way and the SMC, but noted that stars of lower or higher mass appear to require correspondingly smaller or larger overshooting values \citep[see also][]{Baraffe2023,Anders2023,Johnston2024}. Moreover, \citet{Johnston2021} also find evidence of scatter in the convective core masses for the same total mass of an observed star, up to stellar masses $\leq$24\,$M_{\odot}$. This scatter in convective core masses for the same total stellar mass may also be responsible for the separate sub-populations of massive Algols in the LMC and the Milky Way, which require either efficient or inefficient mass transfer at the same orbital period.

\subsection{Wind mass loss}
\label{sec:wind_mass_loss}
The wind mass loss rate in the detailed binary evolution models are
set as in \citet{brott2011}. The mass loss rate on the Main Sequence
is low compared to that of evolved systems and also decreases with decreasing metallicity \citep{Vink2001,mokiem2007,Renzo2017,Bjorklund2021,Hawcroft2021,Brands2022,Gormaz-Matamala2023,Krticka2025}.
The donors in the binary models typically lose $\leq1$\,$M_{\odot}$ via
stellar winds before the thermal timescale mass transfer phase
initiates on the Main Sequence \citep{brott2011,Sen2022}.
Uncertainties in the wind mass-loss rates can also impact these
predicted core masses, but this effect is small for initial masses below
$\simeq30M_{\odot}$: at solar metallicity, the corresponding systematic error on Helium core masses is
$\lesssim0.4M_{\odot}$ \citep{Renzo2017}.  The amount of mass lost via winds is lower for SMC metallicity.
The mass transfer rate is orders of magnitude higher than the wind mass loss rate, such that our conclusions on mass transfer efficiency are not significantly affected by ignoring the mass lost by the donor and accretor via winds during the mass transfer phases.

\subsection{Observational sample}

The current sample of massive Algol binaries is highly heterogeneous.
The component masses for the Galactic Algols were taken from individual
studies, and the data analysis techniques vary from one study to another.
Moreover, most of the observational data were taken several decades ago,
without recent spectroscopic follow-ups. The mass estimates are often
without error bars. Only four out of the seven massive Algols in the
LMC have estimates on surface abundances and rotational velocities of
the individual components. Of the eight massive Algols in the LMC and
the Milky Way that have surface rotation measurements, only two show
donors that are rotating synchronously with the orbit \citep{Sen2022},
while our current binary evolution models predict that all Roche-lobe
filling donors should be synchronised to the orbit.

None of the SMC systems has surface abundance or rotational velocity
measurements either. The effective temperature measurements of the
primary components of the massive Algols in the SMC were based on
spectral type classification and the effective temperature of the
binary companion was derived from the I-band flux ratios \citep{
Harries2003,hilditch2005}, which lie in the Rayleigh-Jeans tail and
are not the most sensitive probes of the effective temperatures of
the O- and B-type stars. Therefore, modern spectroscopic surveys of
massive Algols binaries are essential to obtain accurate information
on most of the existing literature, beyond the component masses and
the orbital periods. Moreover, a larger number of observational
constraints, such as rotational velocities and surface abundances, can be crucial to understand the extent of envelope
stripping and the mode of mass transfer (disk or no disk) occurring
in these massive Algols. Like the OGLE survey, the ASAS-SN survey
\citep{Kochanek2017} can be an excellent dataset to mine eclipsing
semi-detached binaries for follow-up homogeneous spectroscopic
monitoring campaigns in the Milky Way and the Magellanic Clouds.


\section{Conclusion}
\label{sec:conclusion}

We investigated the efficiency, angular momentum loss, and stability of mass transfer in massive interacting binaries on the Main Sequence (Algols) across the Milky Way, Large Magellanic Cloud, and Small Magellanic Cloud. Using an analytical framework that connects present-day binary properties (donor mass, accretor mass, and orbital period) to their initial parameters, and combining these with physical constraints, we derived the ranges of mass transfer efficiency and angular momentum loss consistent with the observed populations. Our main conclusions are:
\begin{itemize}
\item Conservative ($\varepsilon=1$) or non-conservative ($\varepsilon=0$) mass transfer cannot reproduce the observed properties of all massive Algols at any metallicity. A majority of systems favour inefficient transfer ($\varepsilon \leq 0.5$), and we find no necessity for highly efficient mass transfer even at the shortest orbital periods ($\sim$2\,d) for our fiducial assumption on envelope stripping (Fig.\,\ref{fig:porb_beta}). 

\item The current set of massive Algols requires that mass transfer remain stable down to initial mass ratios of $\sim$0.6, lower than typically assumed in the literature \citep{selma2013,Ge2020} as well as in our detailed binary evolution models \citep{wang2020,Sen2022}. The least massive Algols ($M_{\rm d,i} < 20\,M_{\odot}$) provide the most substantial evidence for these relaxed stability criteria. 

\item Our analysis relies on the assumption (Sect.\,\ref{sect:assumptions} and Sect.\,\ref{sect:envelope_stripping}) that the fast thermal-timescale initial mass-transfer strips the donor down to its Zero Age Main Sequence convective core mass. Some massive Algols cannot be reproduced by the detailed models (e.g. right panel of Fig.\,\ref{fig:individual_examples}), which assume rotationally limited accretion efficiency and efficient tides. This implies that the mass-transfer efficiency may not entirely be regulated by rotational processes. Observations of massive Algols in the earliest post-mass-transfer phase may require improving the physical picture of mass transfer in detailed stellar evolution models.

\item We find a correlation between efficiency and stability: more efficient mass transfer requires its stability at initial mass ratios significantly below standard limits. Constraints from angular momentum loss exclude conservative mass-transfer solutions in a quarter of systems, particularly in the SMC, but are generally less restrictive than the stability criterion.

\item The current population of massive Algols in the SMC may have undergone less efficient mass transfer than their counterparts in the LMC and the Milky Way. However, we cannot rule out an observational bias in the current population of Algols in the SMC, as all except one have mass ratios below $\sim$2.1, while there are a lot of observed Algols in the Milky Way and LMC with more unequal mass ratios (Fig\,\ref{fig:bar}). 

\item Our results show that massive Algols serve as powerful in-situ probes of mass transfer physics across a range of metallicities. Follow-up observational studies on Algols in the SMC can provide further constraints on the possibility of a metallicity dependence in mass transfer efficiency.
\end{itemize}

Multi-epoch spectroscopic surveys of massive stars and binaries (e.g. the OWN survey, \citealp[][]{Barba2017}; the IACOB project, \citealp[][]{Simon-Diaz2014}; the VFTS survey, \citealp[][]{Evans2011}; the BLOeM survey, \citealp[][]{Shenar2024}), especially with constraints from rotational velocities and surface abundances, are crucial for further disentangling the degeneracy between efficiency and stability, and for anchoring population synthesis predictions of stripped-envelope supernovae and gravitational-wave sources. Simultaneous improvements in multi-dimensional modelling of the thermal timescale mass transfer \citep{Bisikalo1998,Dessart2003,Lu2022,Cehula2023,Ryu2025,Scherbak2025b,Scherbak2025a} and improved orbital modelling in 1D codes \citep{Rocha2025,Parkosidis2025} will help advance the field, enabling more accurate modelling of the early stages of mass transfer.


\begin{acknowledgements}
    M.R. acknowledges support from NASA (ATP: 80NSSC24K0932). JIV acknowledges support from the European Research Council for the ERC Advanced Grant 101054731. This research was supported in part by grant NSF PHY-2309135 to the Kavli Institute for Theoretical Physics (KITP). Part of this work was initiated at the "Stable mass transfer 2.0" workshop held at the Center for Computational Astrophysics of the Flatiron Institute, which is supported by the Simons Foundation. We thank Pablo Marchant, Tom Maccarone, Hugues Sana and the anonymous referee for helpful comments on the manuscript. 
\end{acknowledgements}

\appendix

\section{Additional plots and tables}

\begin{figure*}
    \centering
    \includegraphics[width=0.24\linewidth]{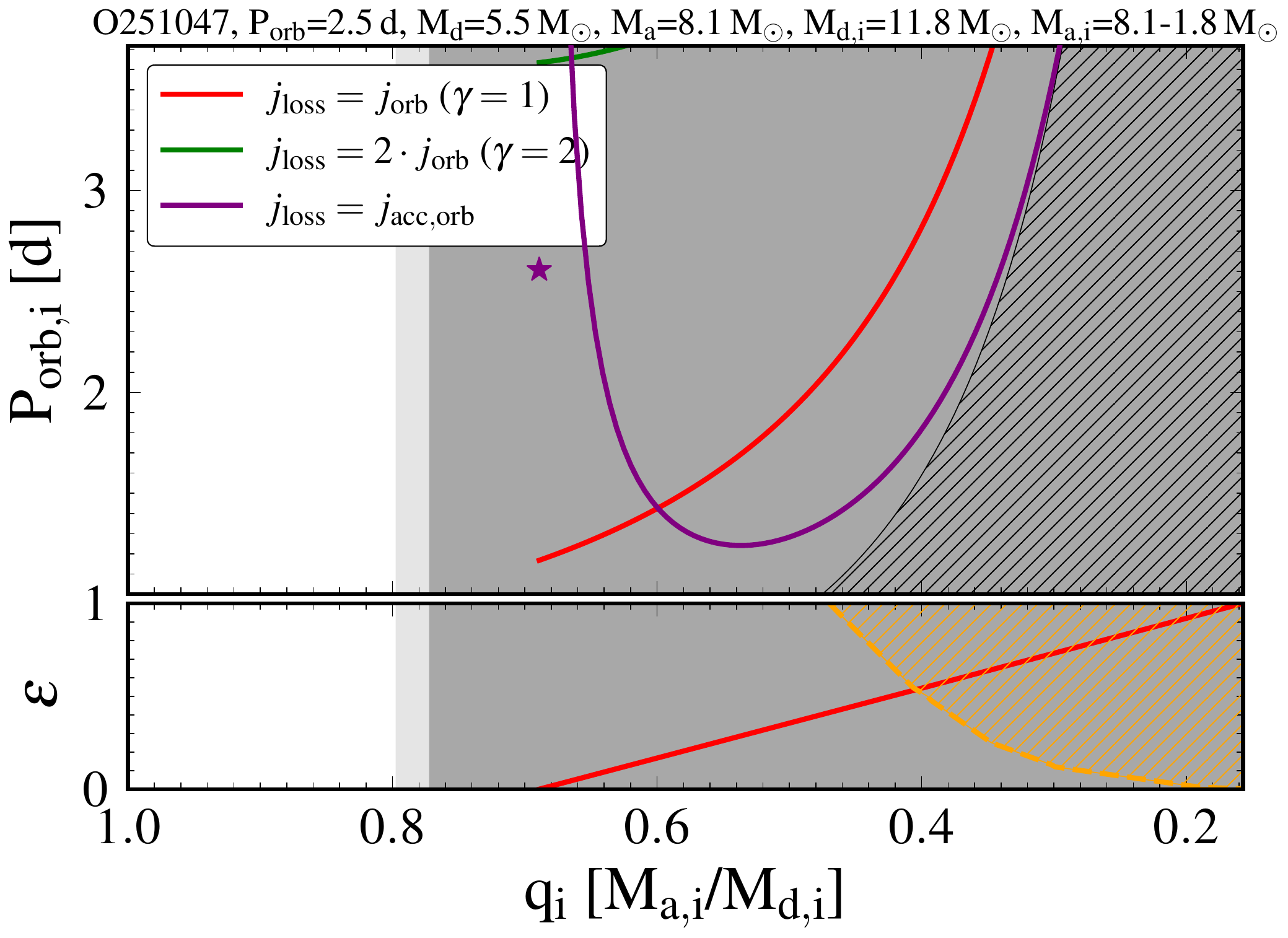}
    \includegraphics[width=0.24\linewidth]{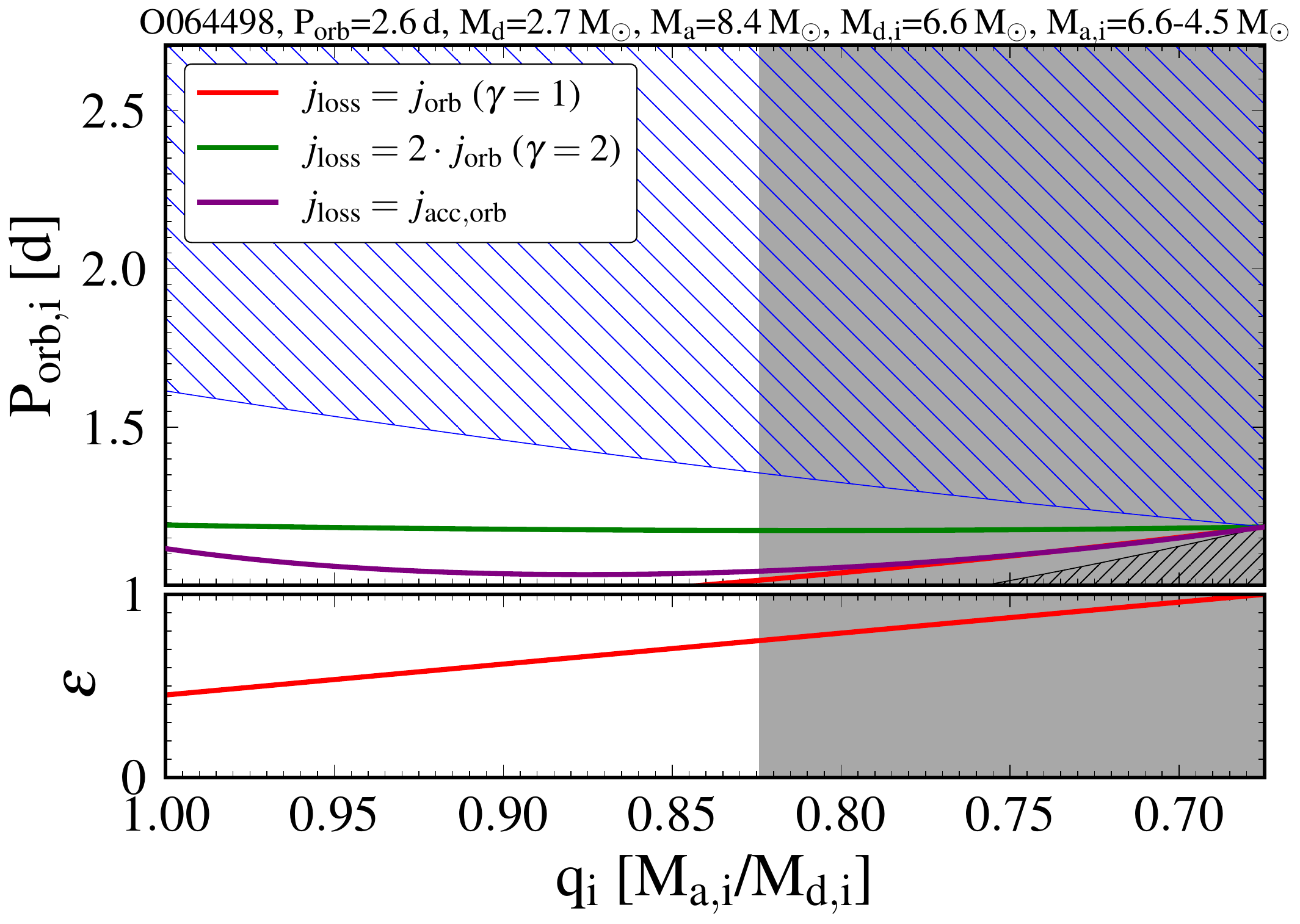}
    \includegraphics[width=0.24\linewidth]{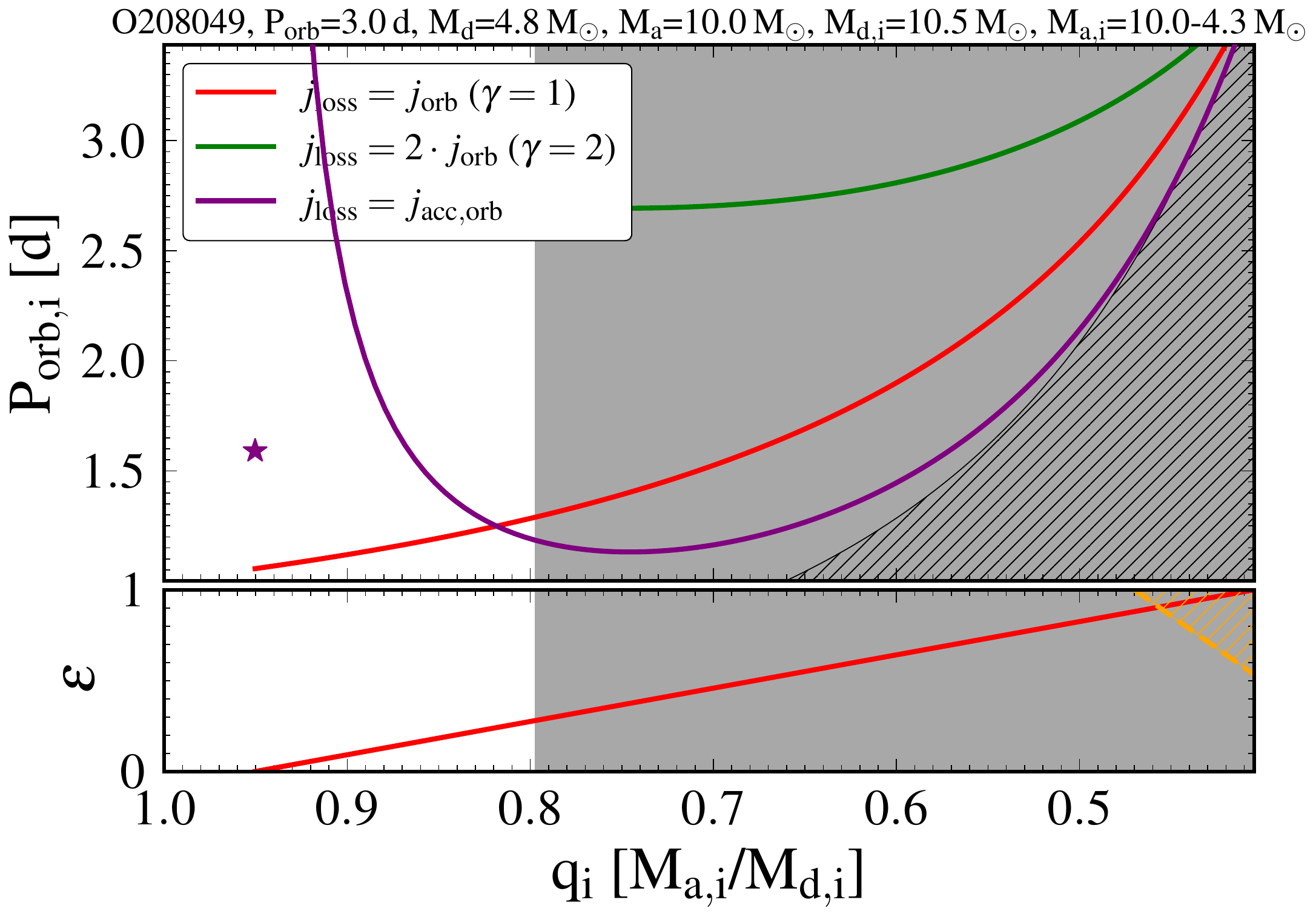}
    \includegraphics[width=0.24\linewidth]{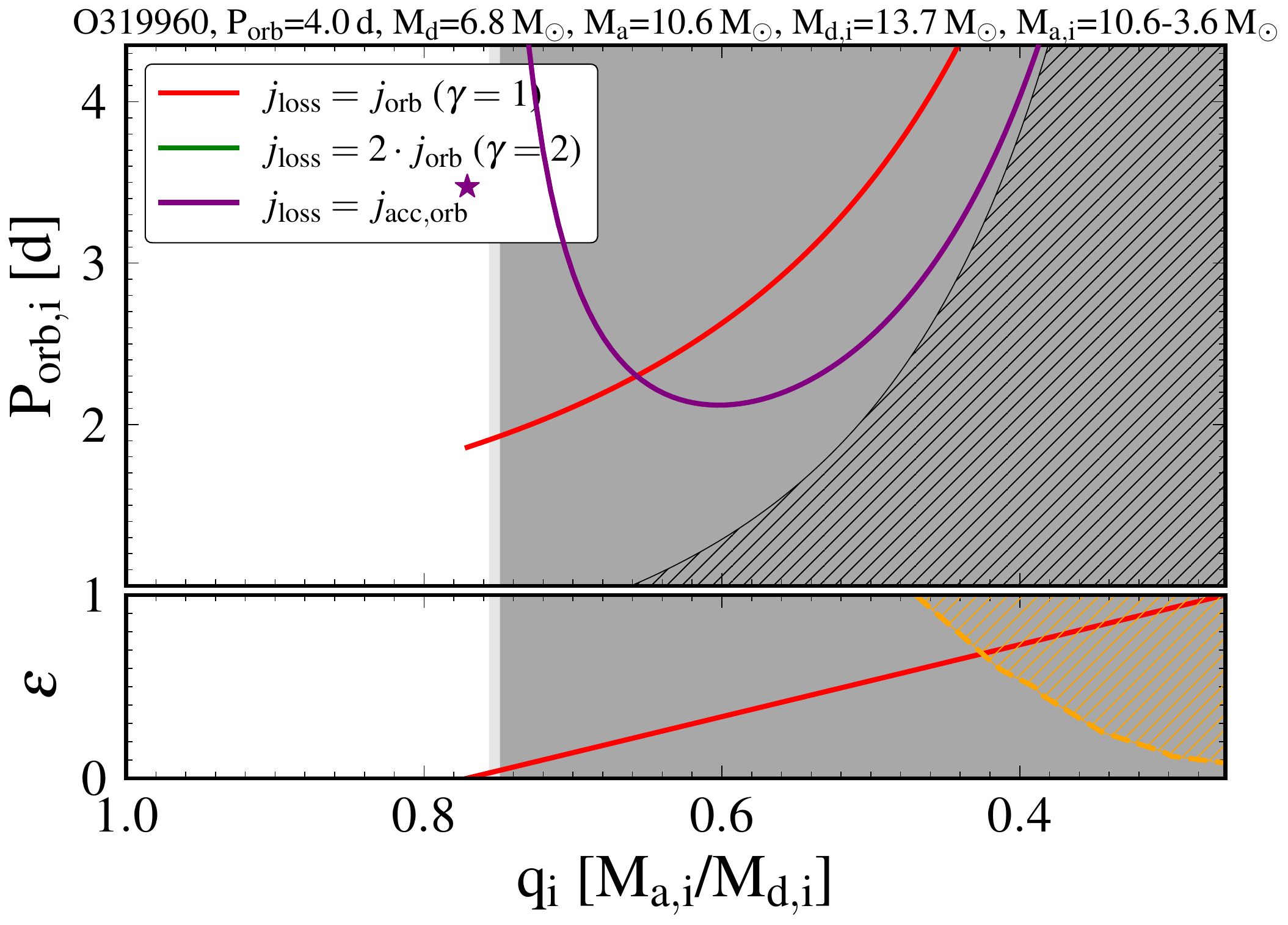}
    \includegraphics[width=0.24\linewidth]{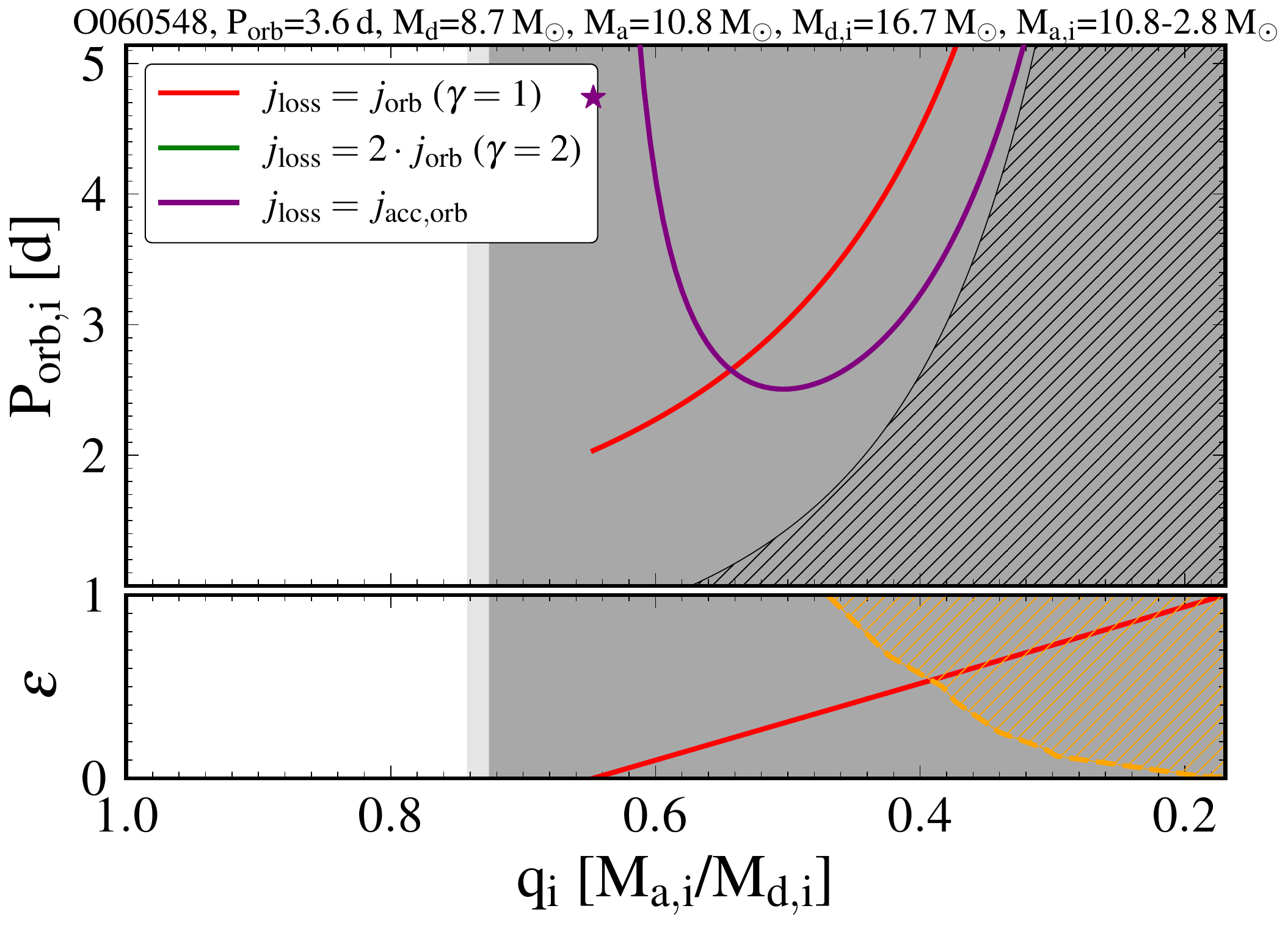}
    \includegraphics[width=0.24\linewidth]{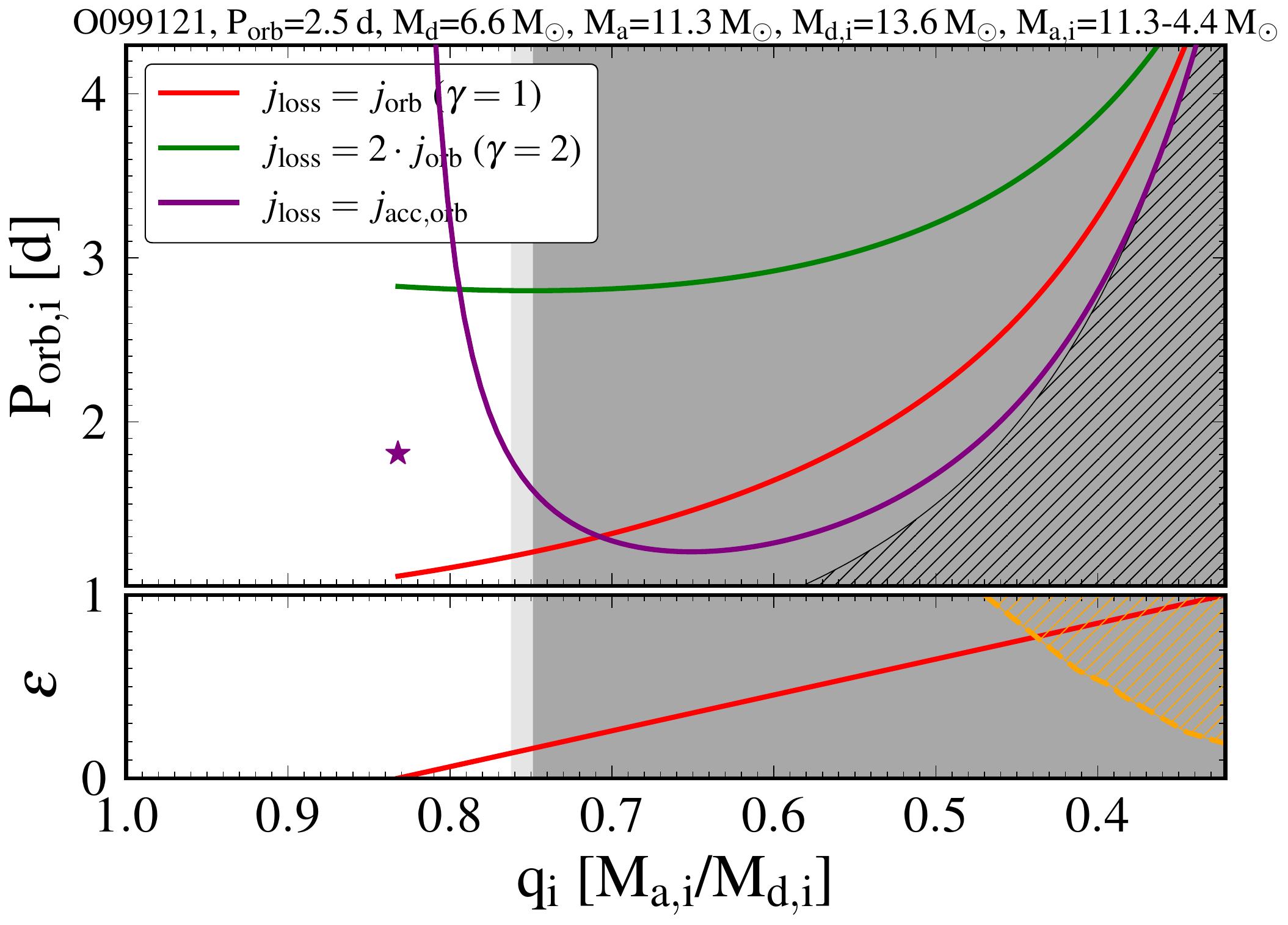}
    \includegraphics[width=0.24\linewidth]{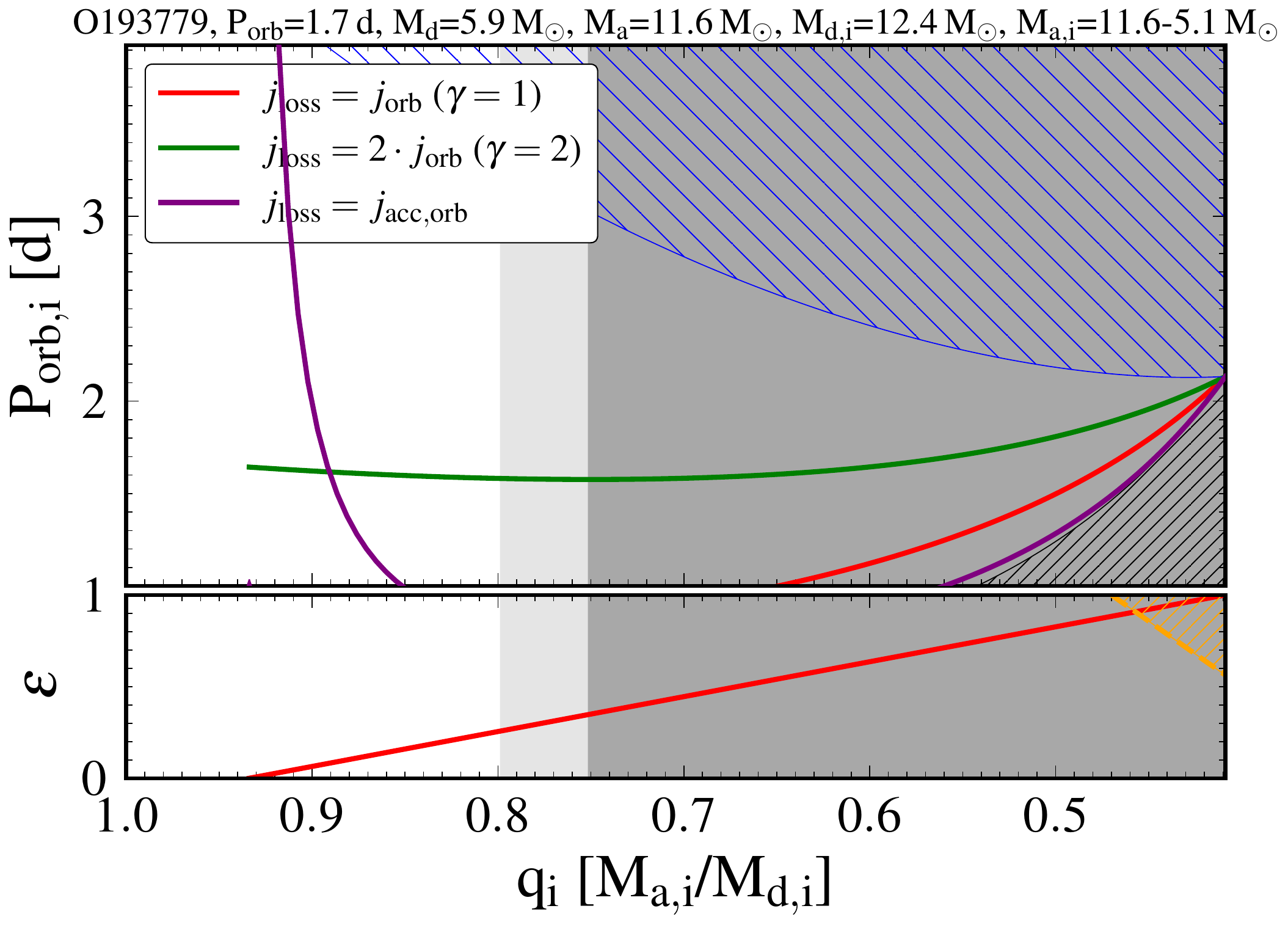}
    \includegraphics[width=0.24\linewidth]{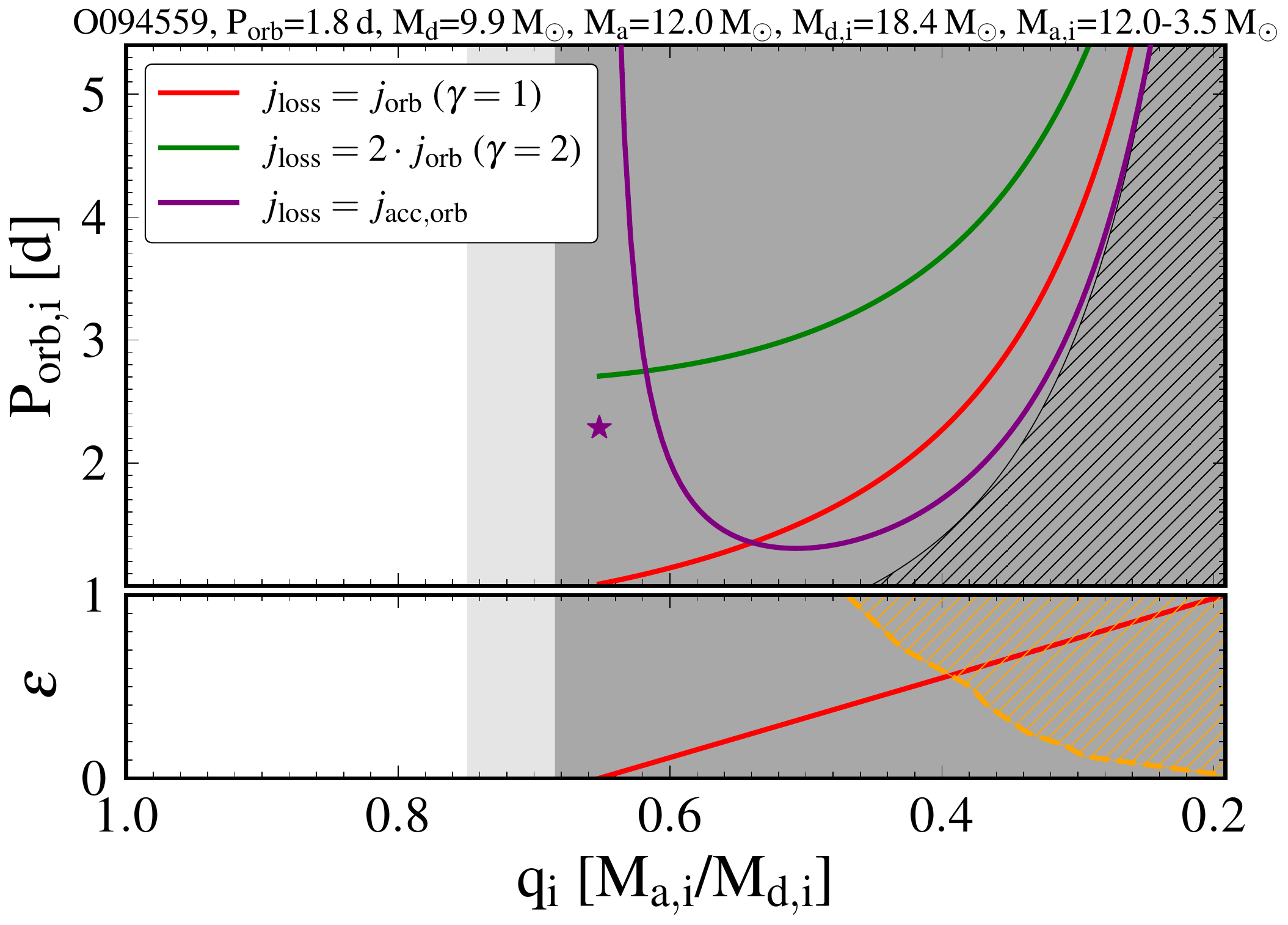}
    \includegraphics[width=0.24\linewidth]{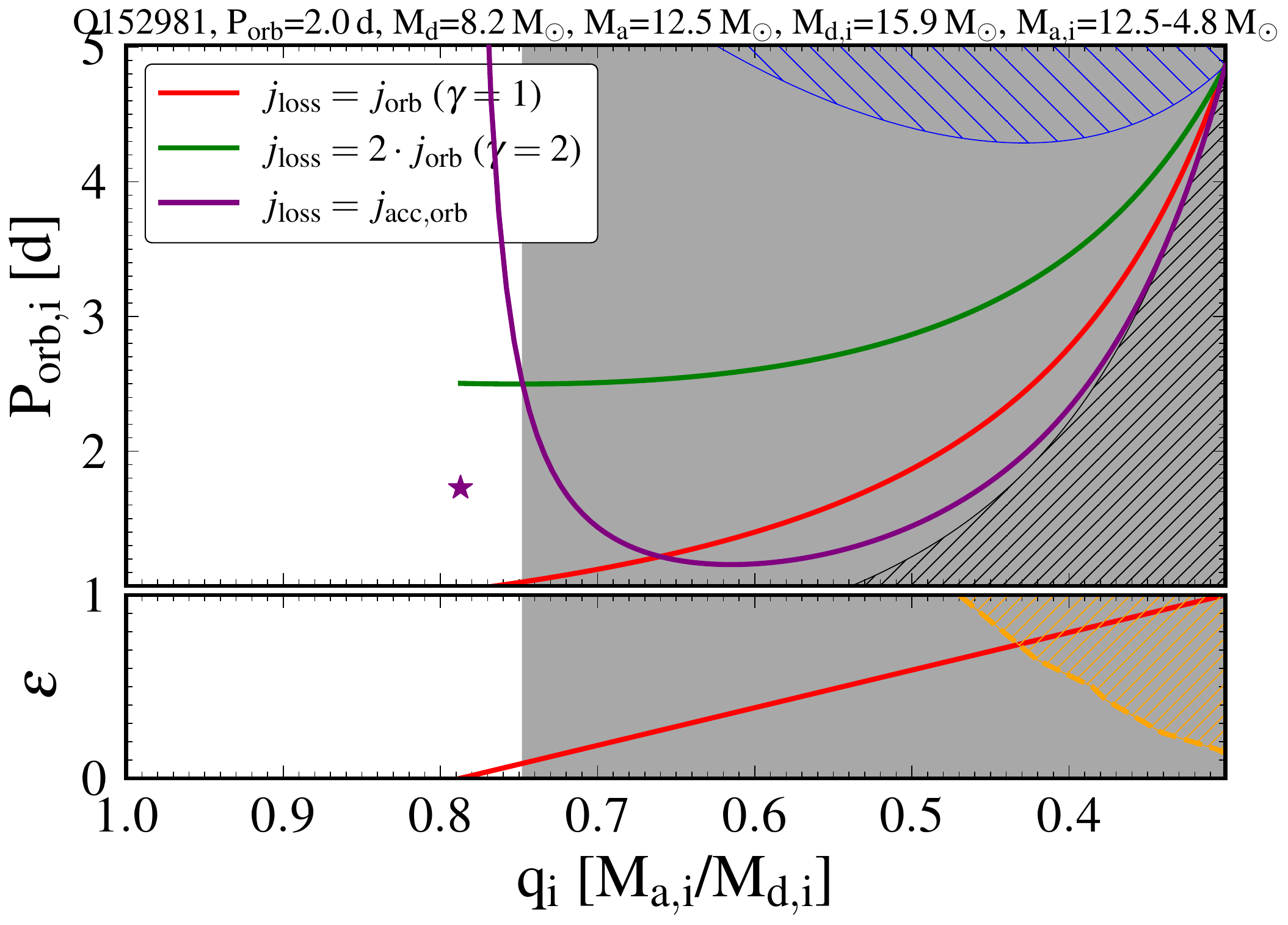}
    \includegraphics[width=0.24\linewidth]{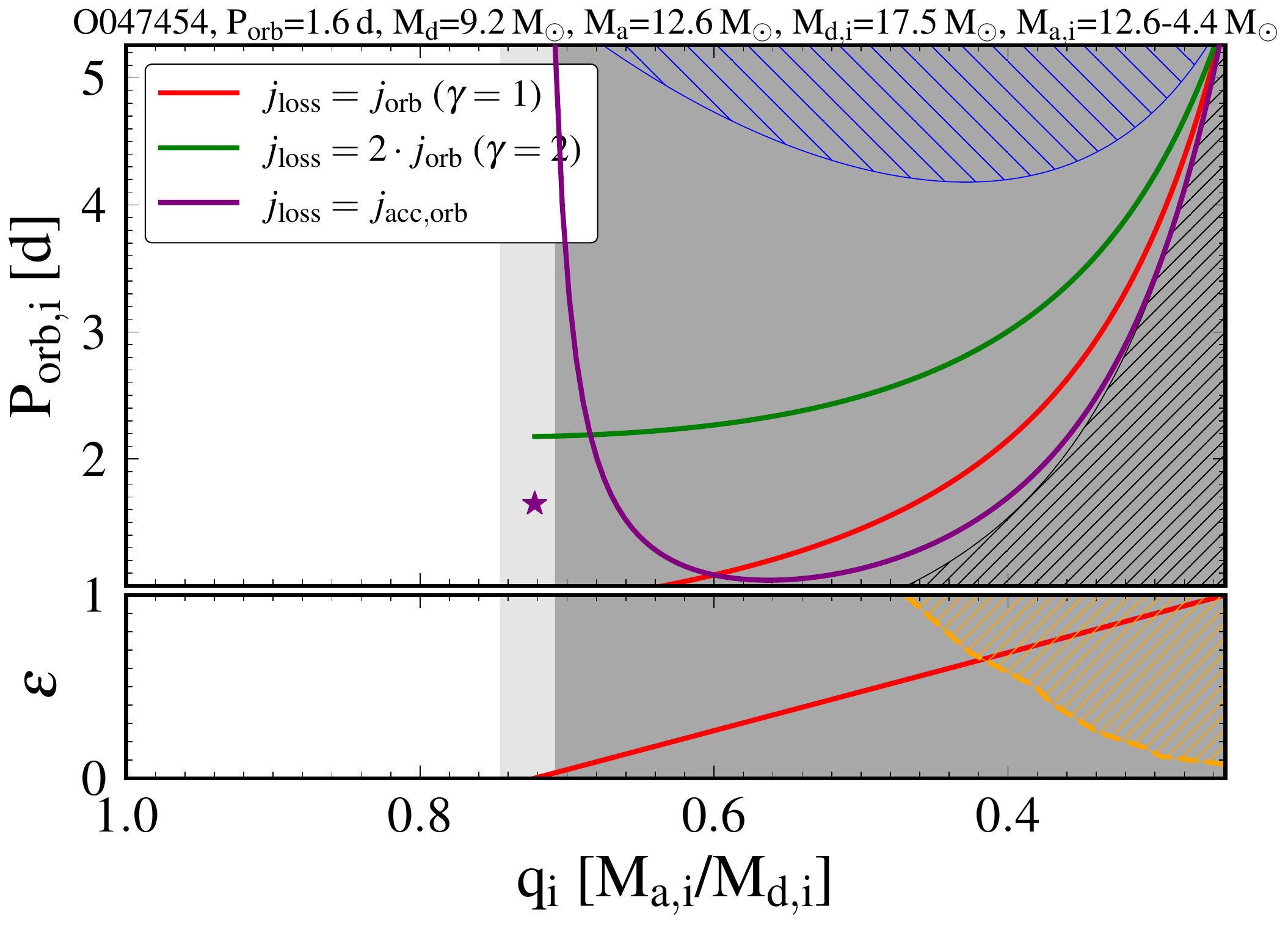}
    \includegraphics[width=0.24\linewidth]{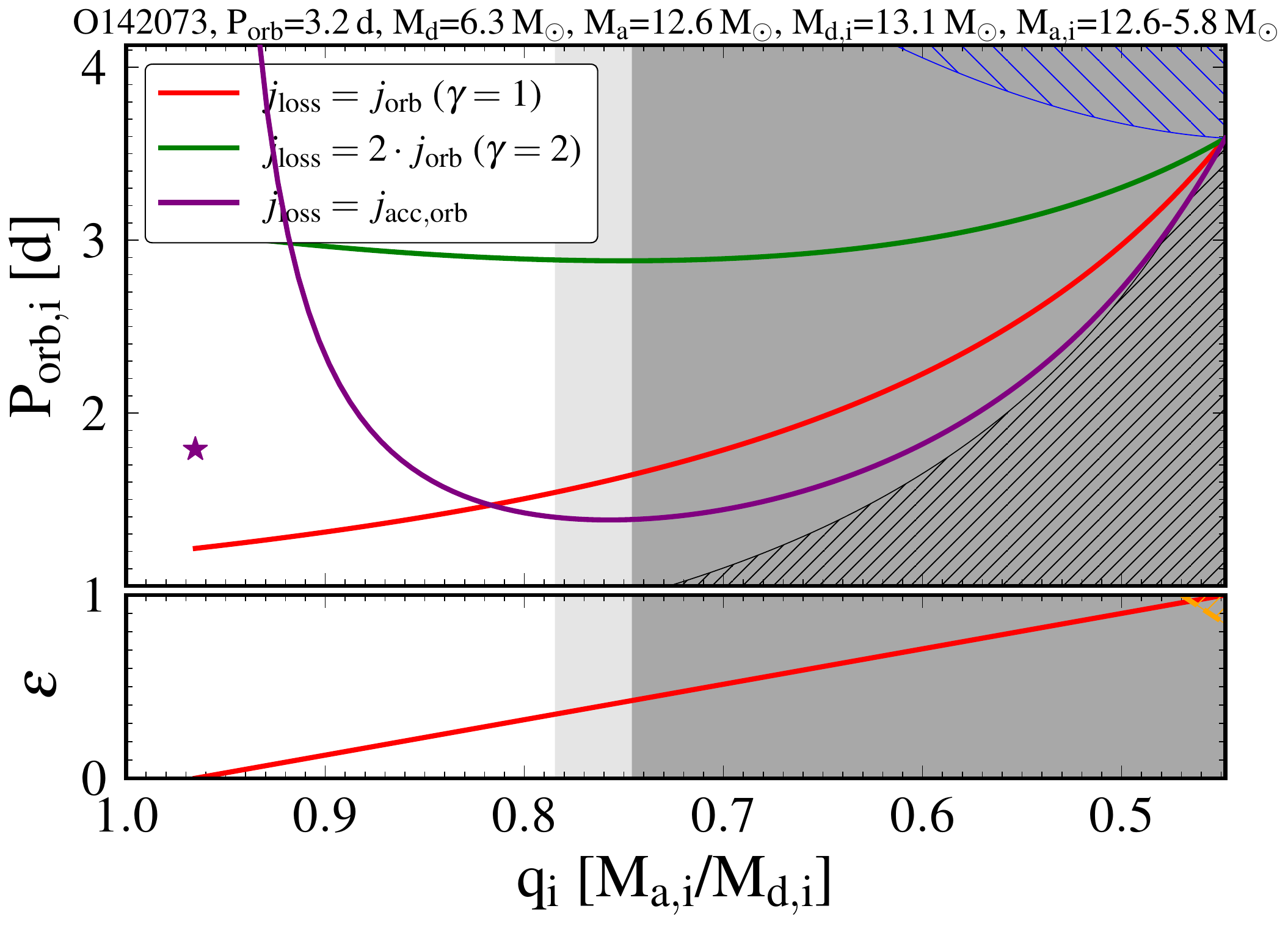}
    \includegraphics[width=0.24\linewidth]{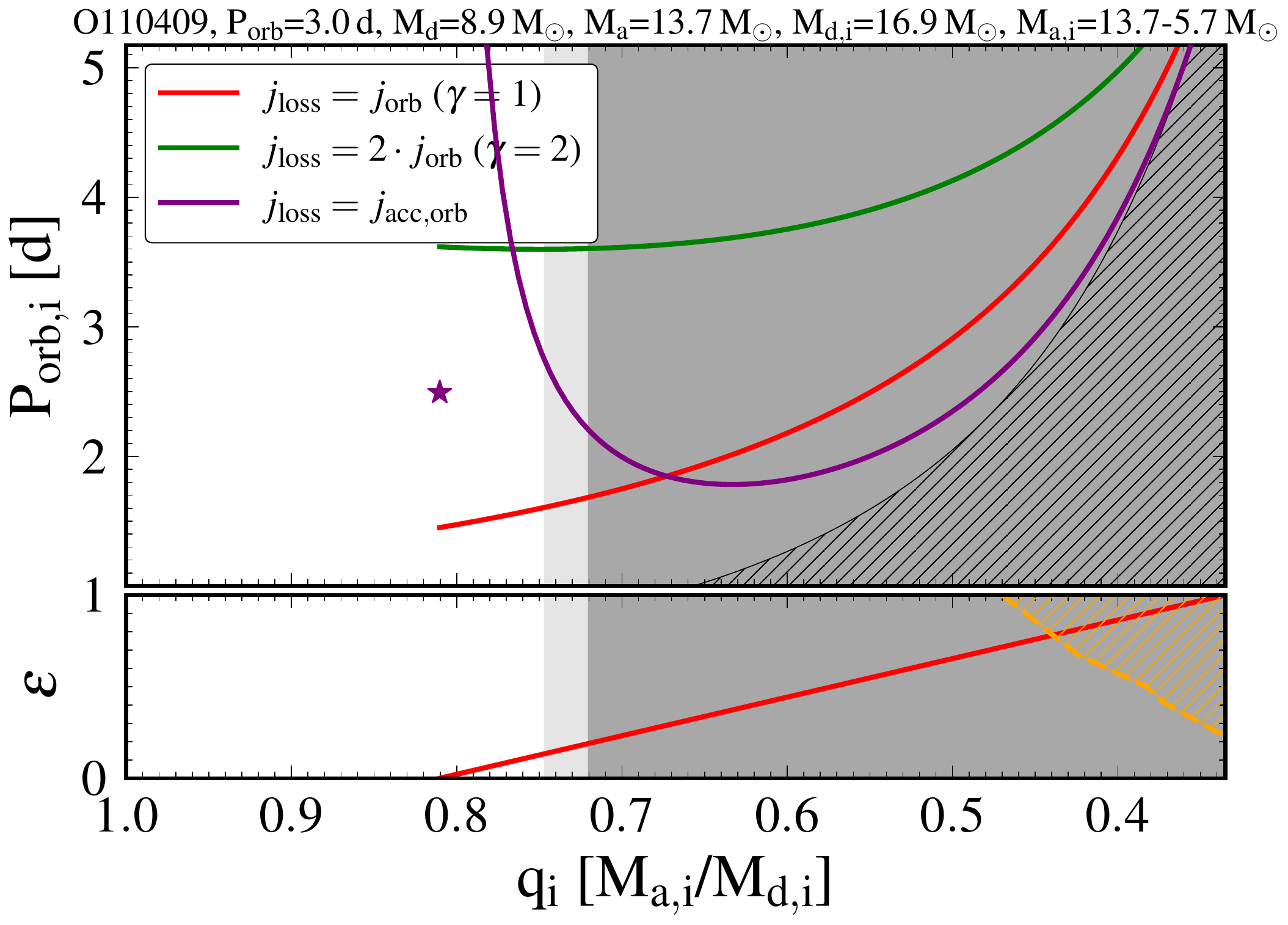}
    \includegraphics[width=0.24\linewidth]{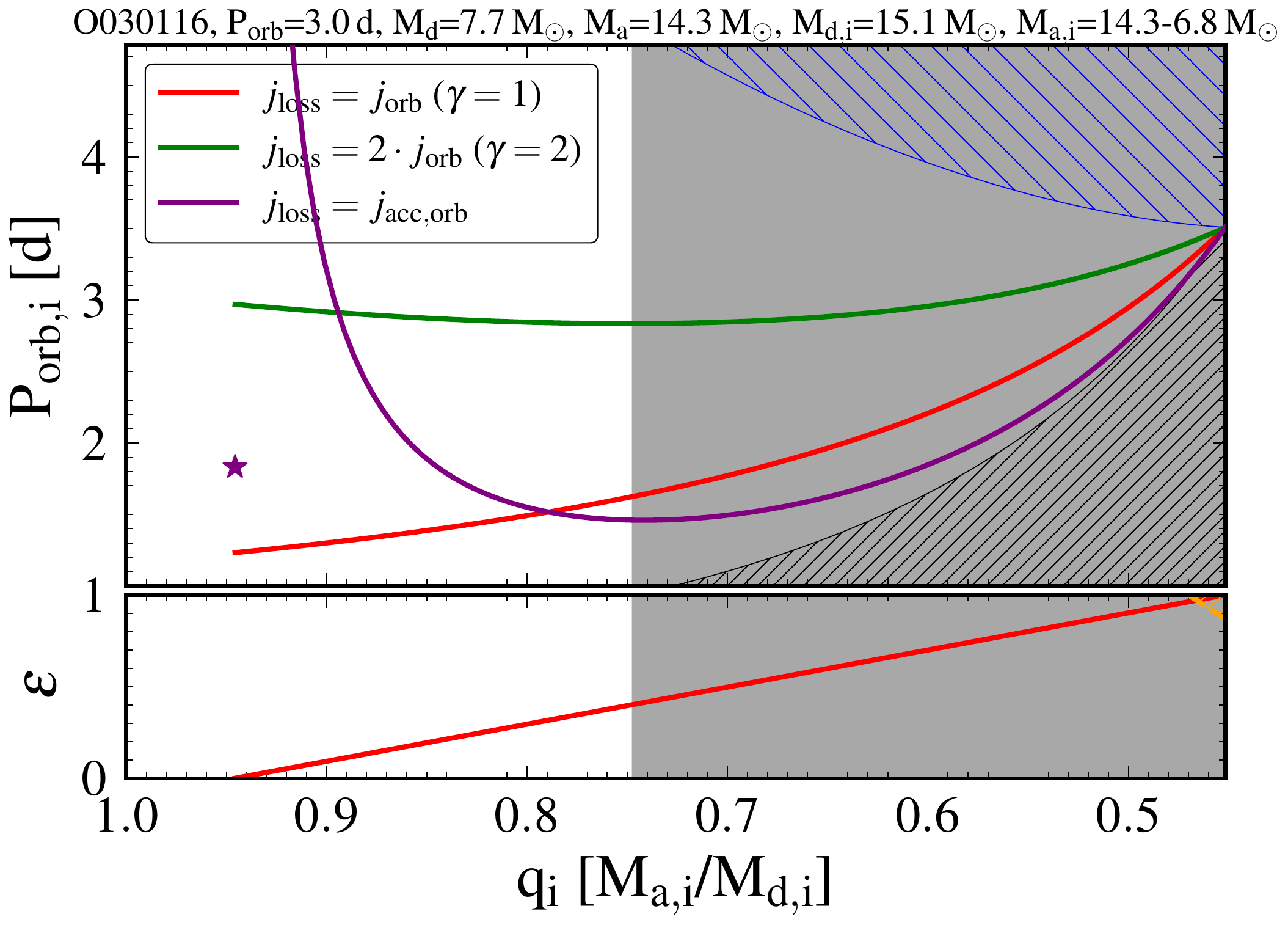}
    \includegraphics[width=0.24\linewidth]{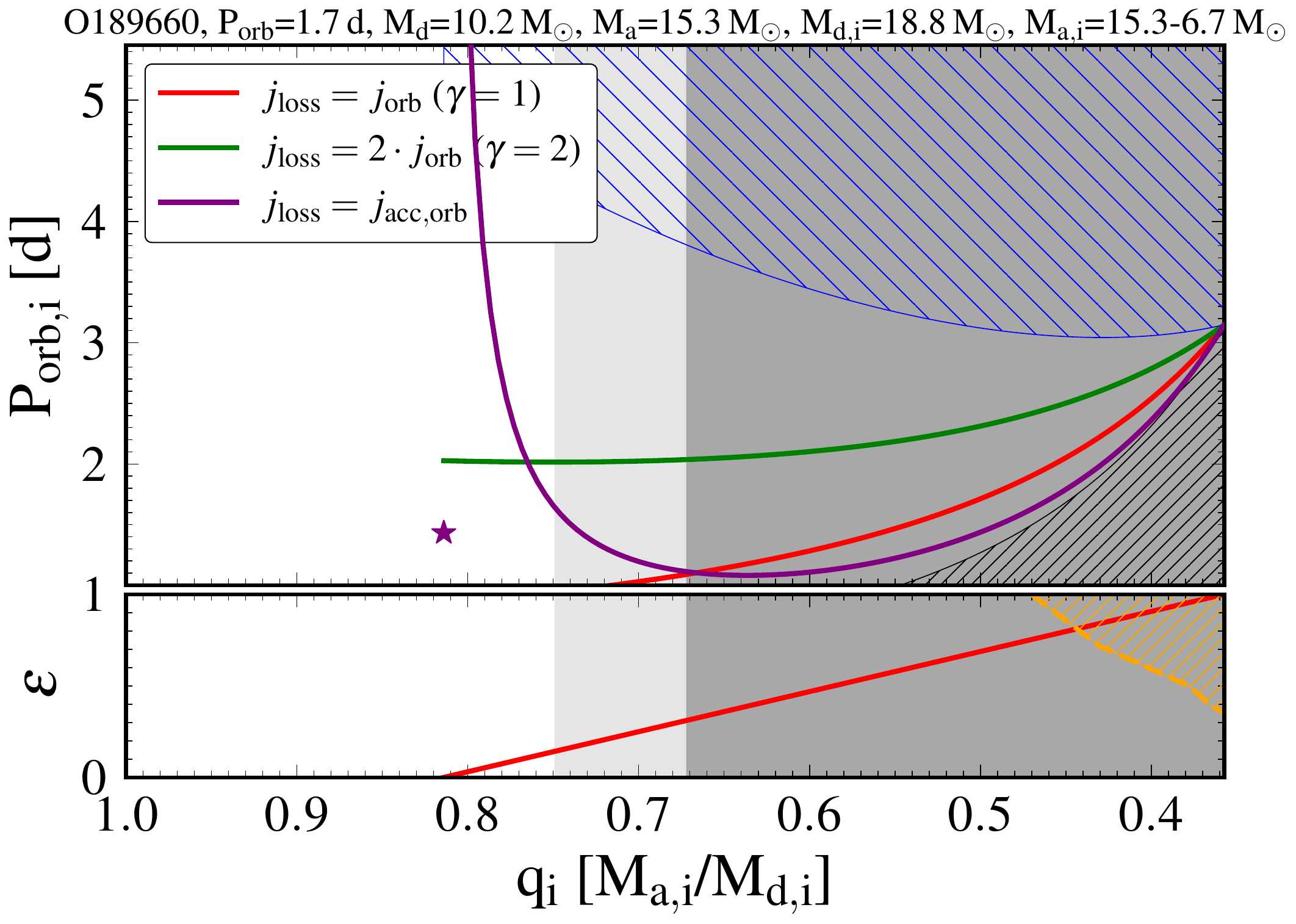}
    \includegraphics[width=0.24\linewidth]{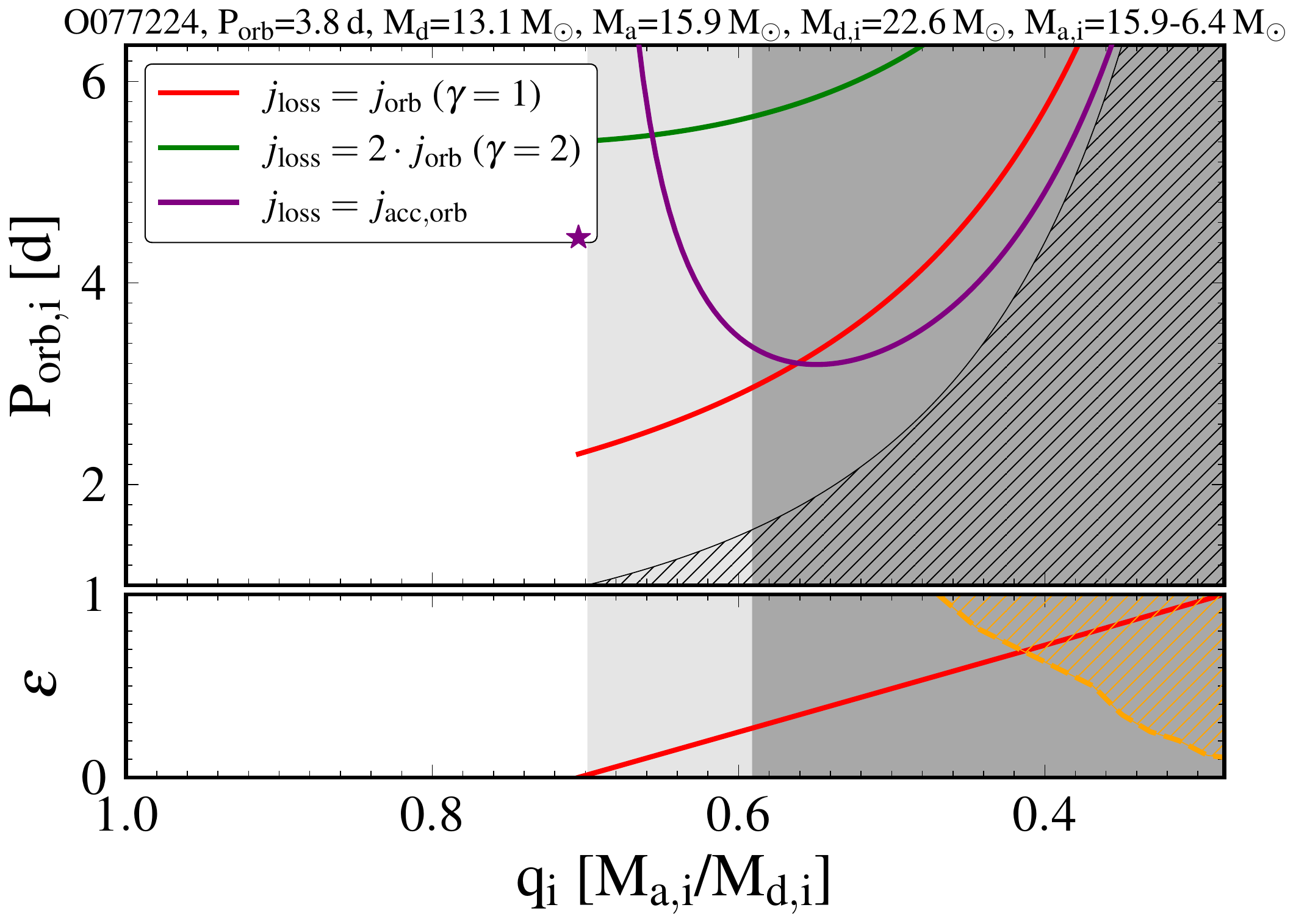}
    \includegraphics[width=0.24\linewidth]{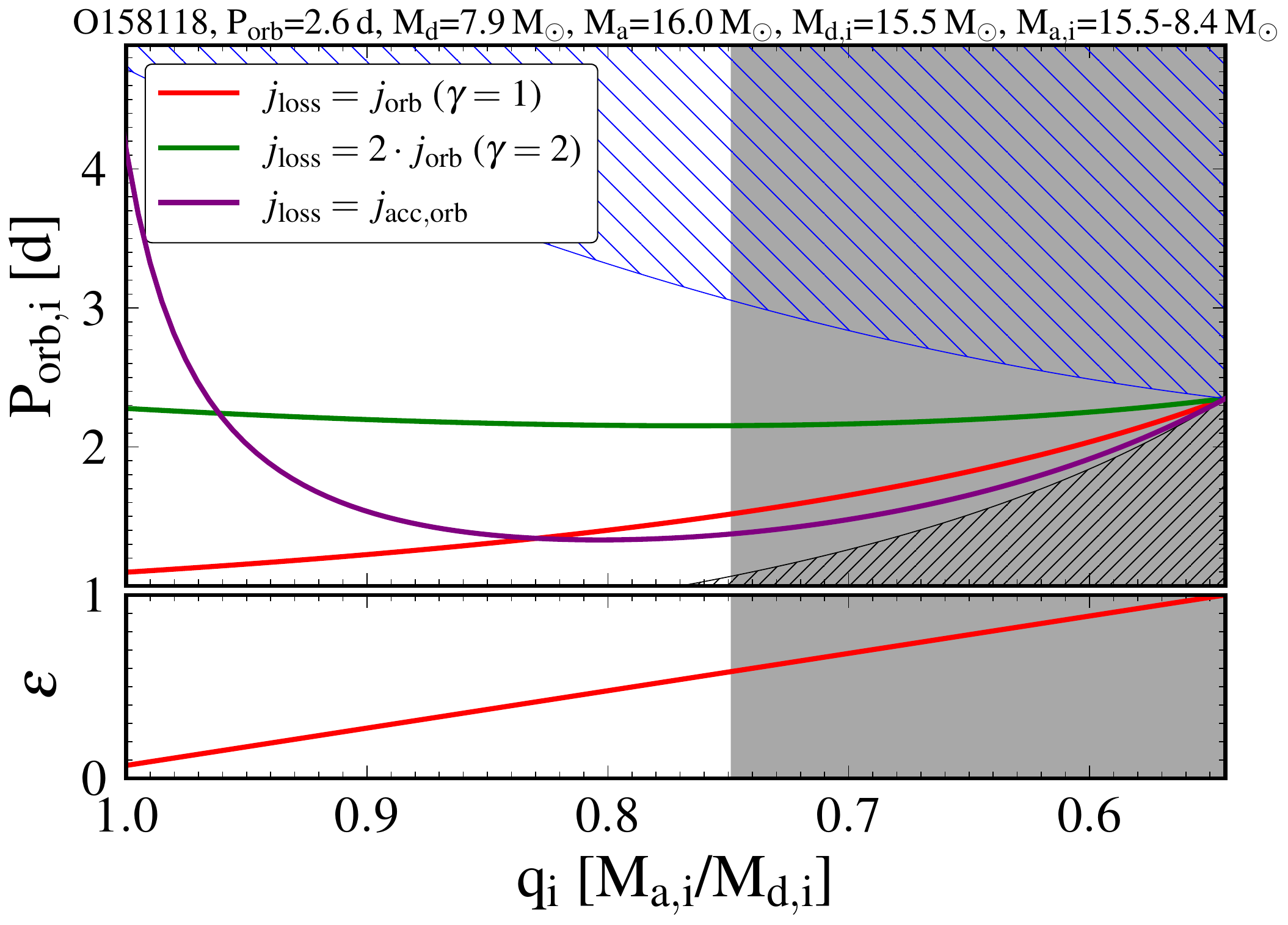}
    \includegraphics[width=0.24\linewidth]{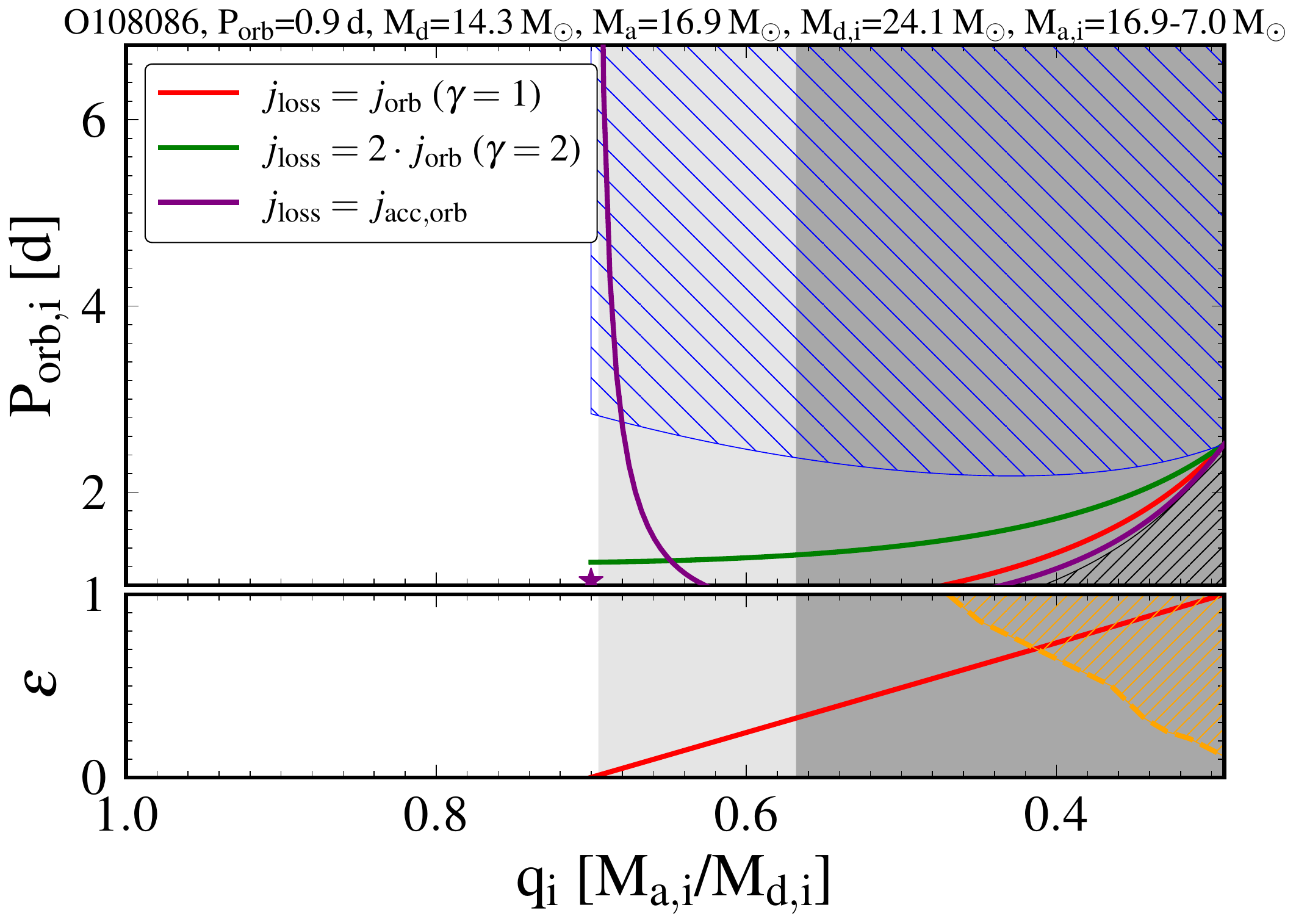}
    \includegraphics[width=0.24\linewidth]{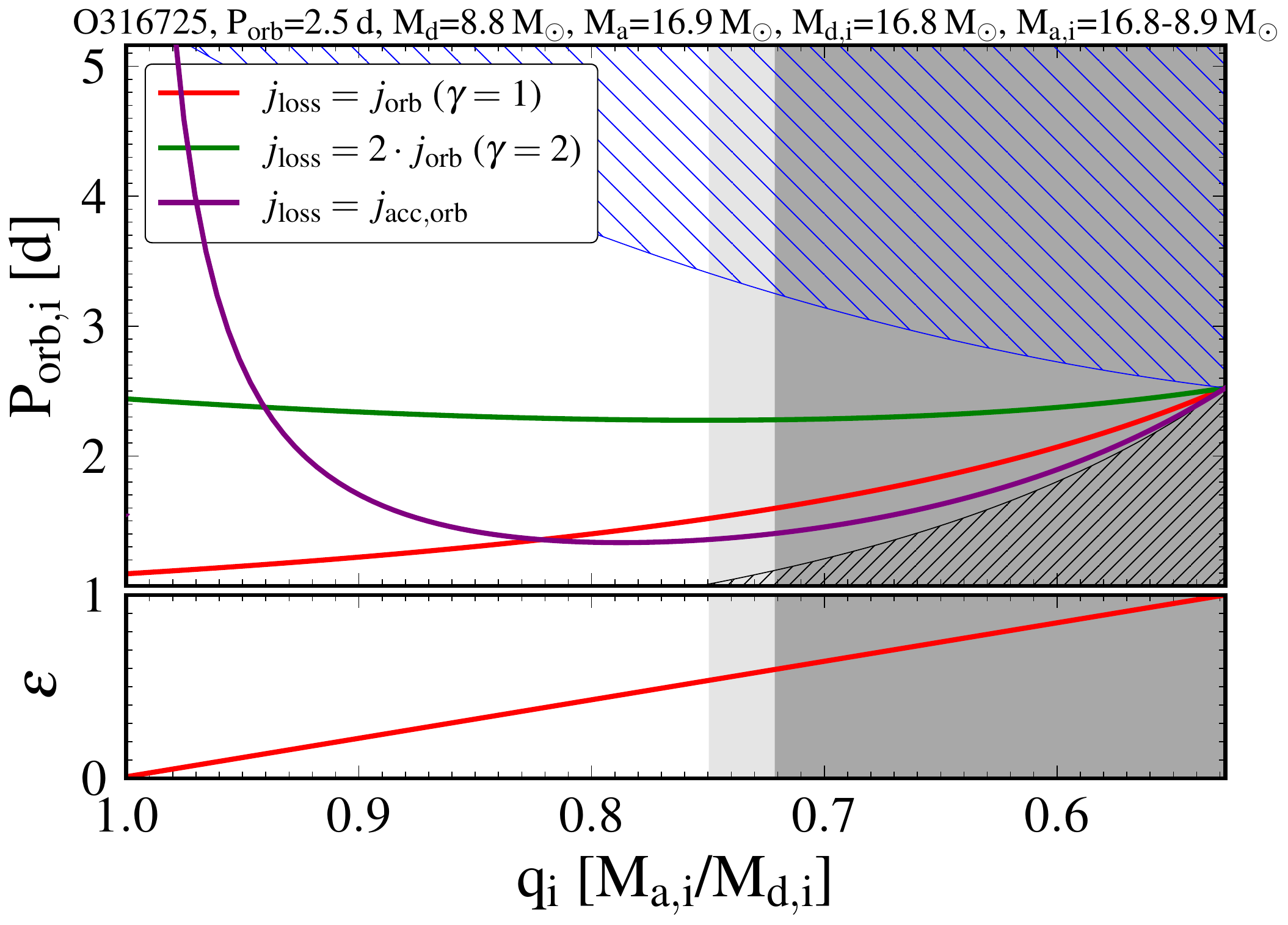}
    \includegraphics[width=0.24\linewidth]{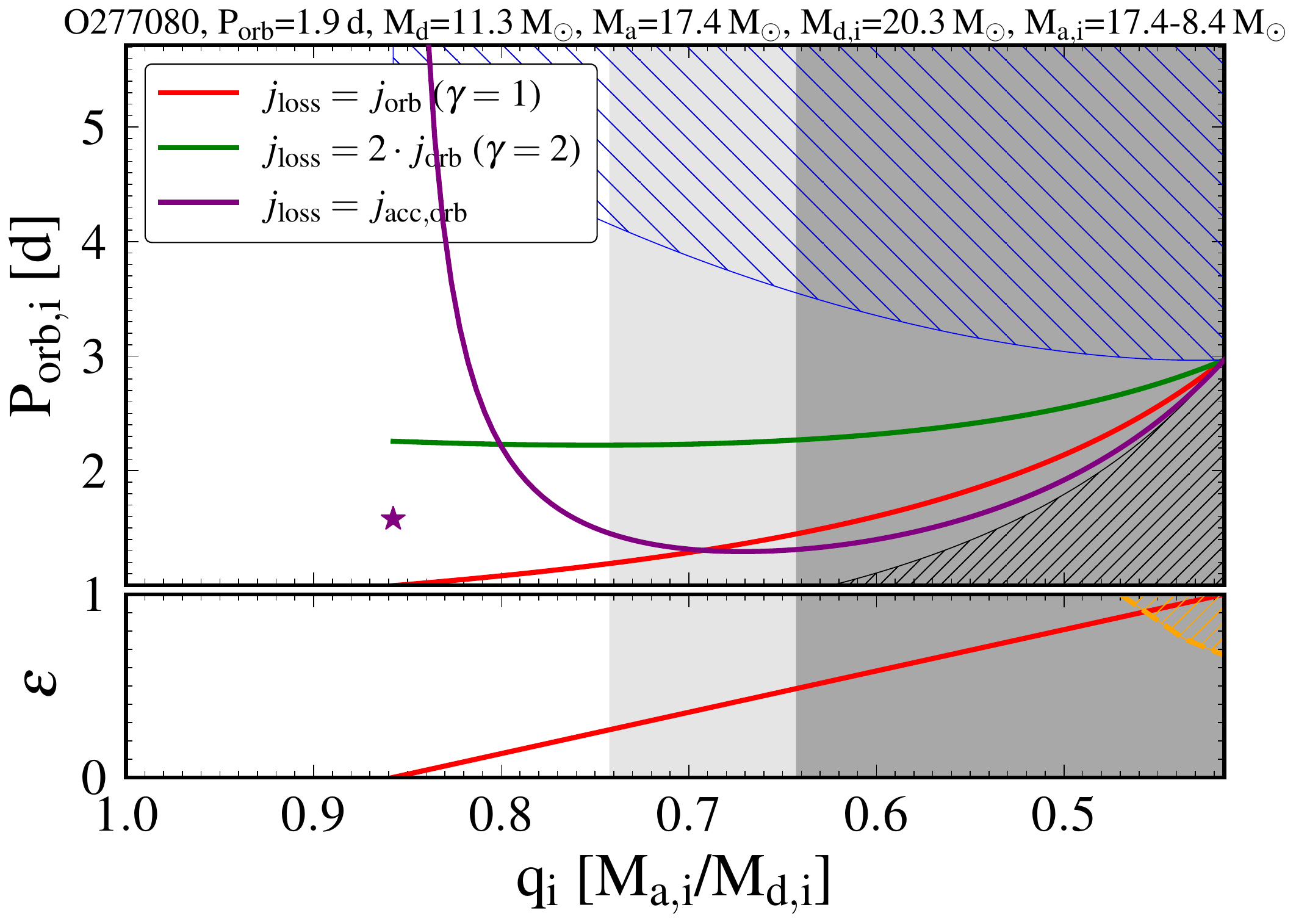}
    \includegraphics[width=0.24\linewidth]{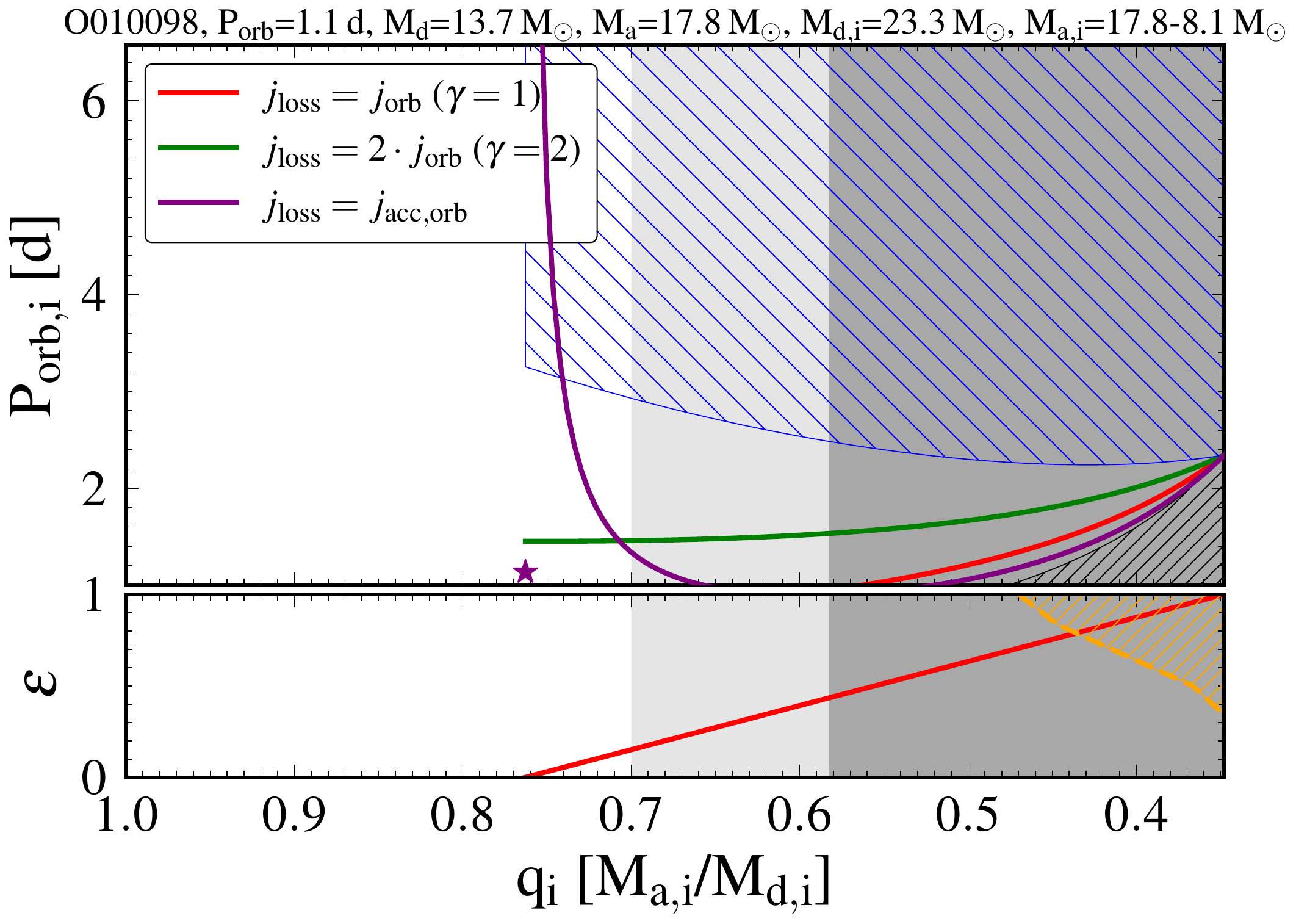}
    \includegraphics[width=0.24\linewidth]{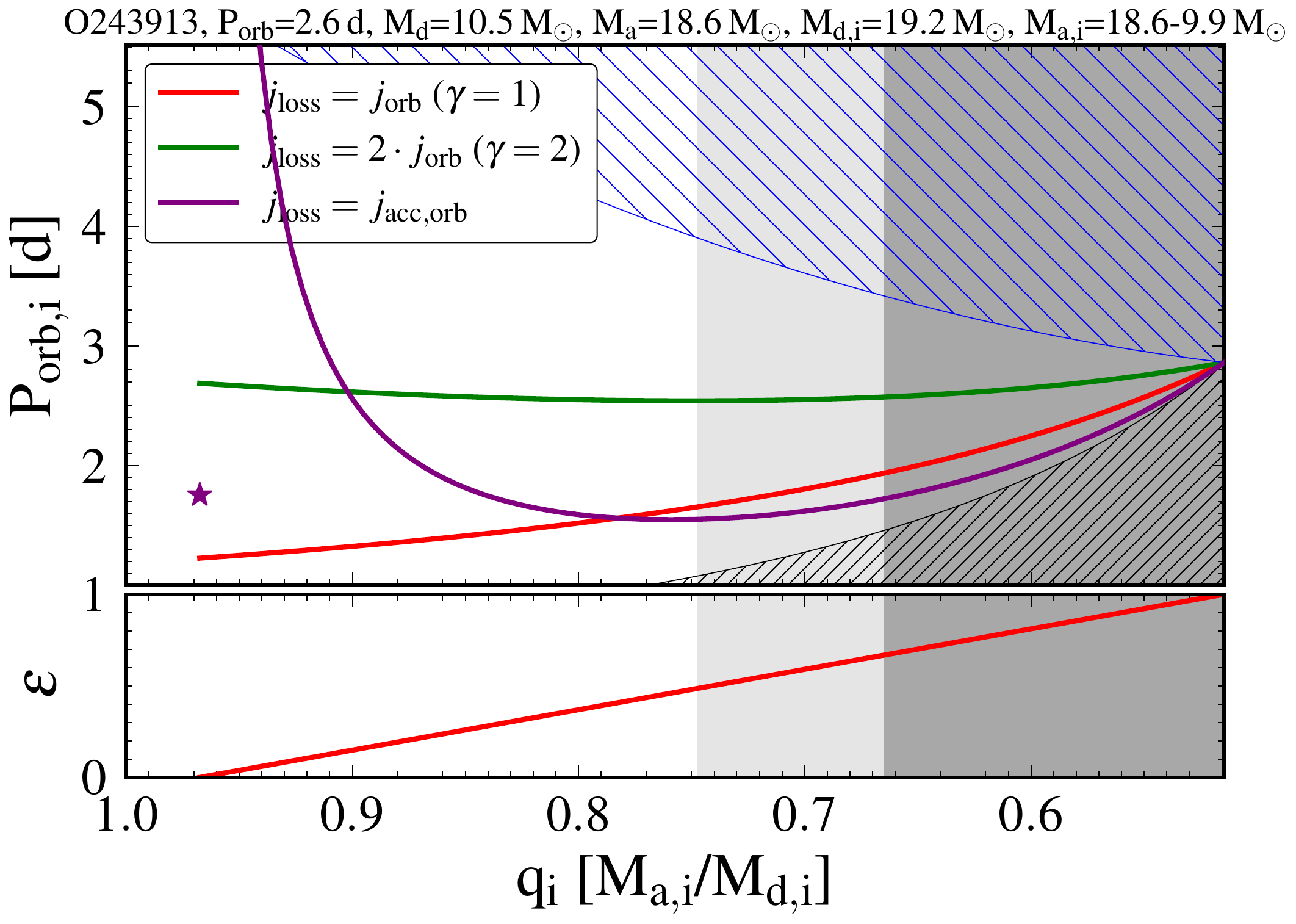}
    \includegraphics[width=0.24\linewidth]{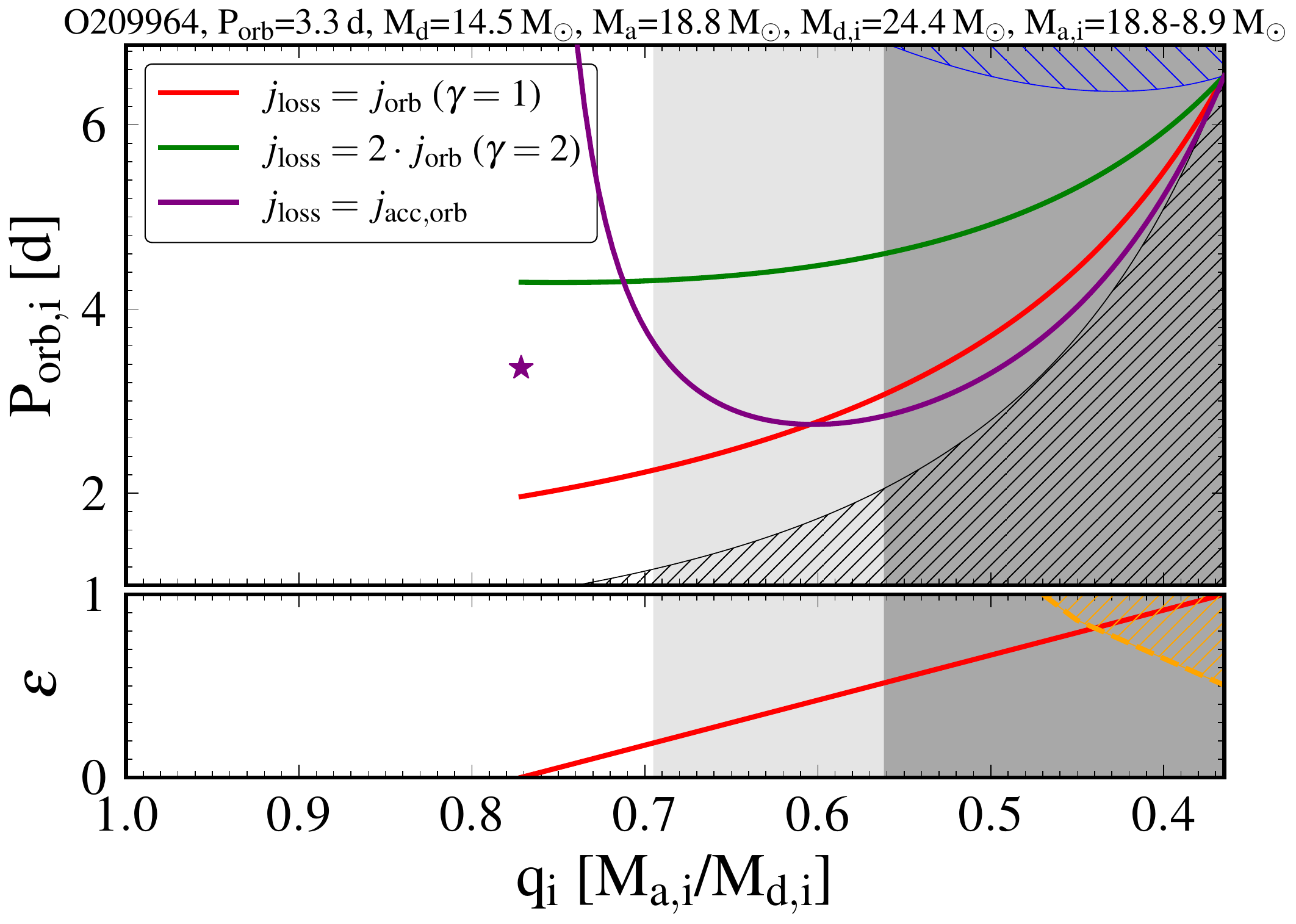}
    \includegraphics[width=0.24\linewidth]{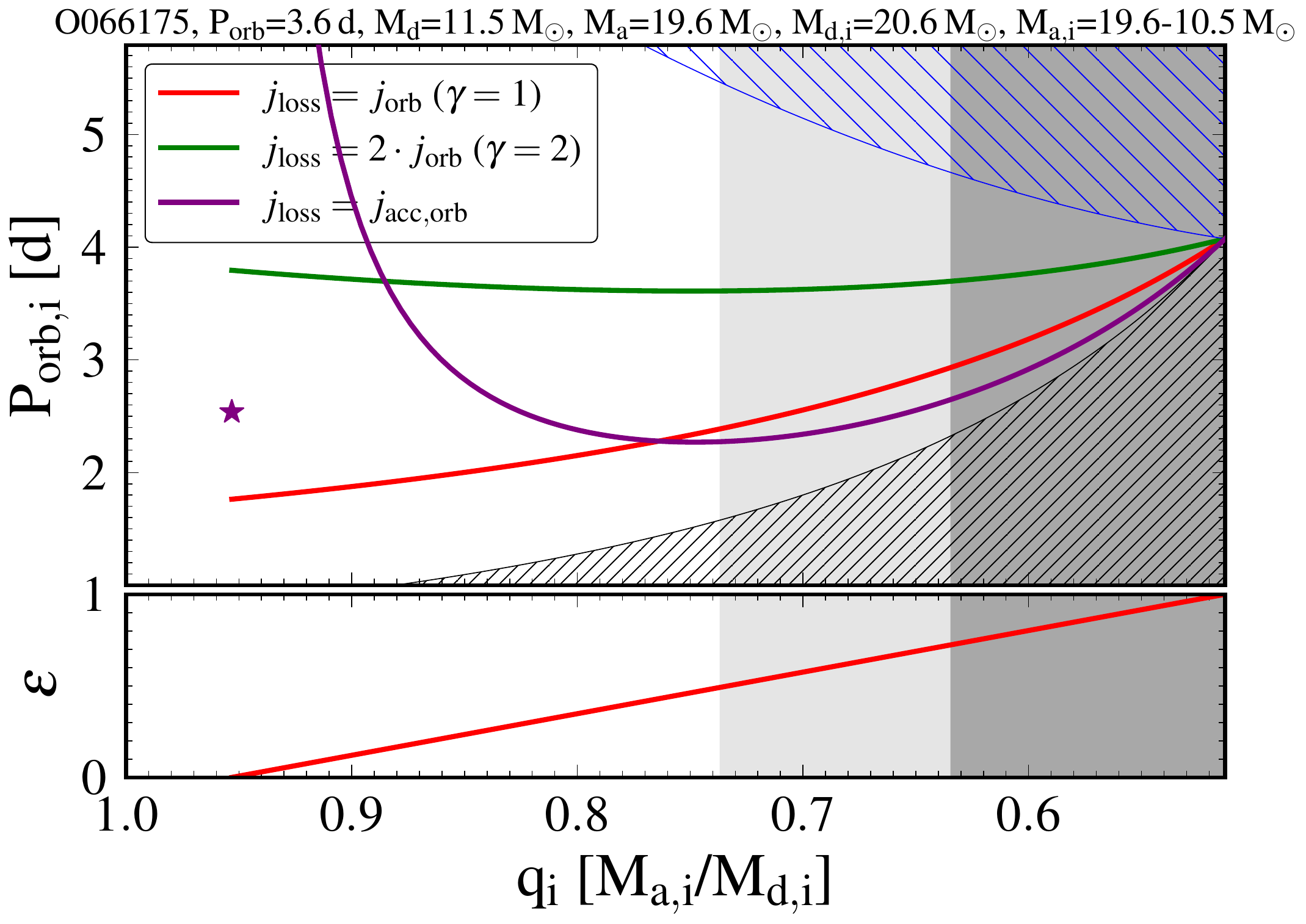}
    \includegraphics[width=0.24\linewidth]{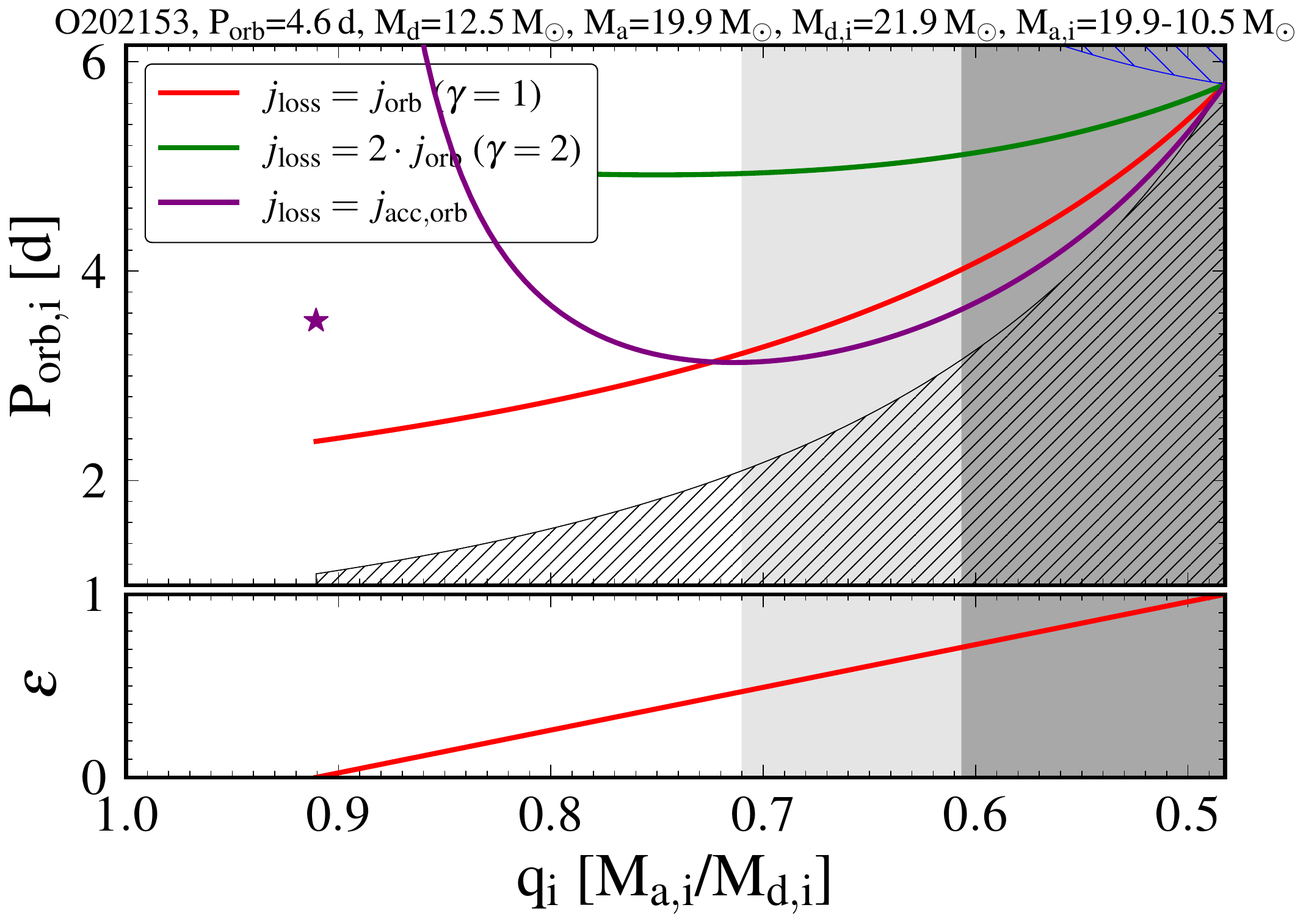}
    \includegraphics[width=0.24\linewidth]{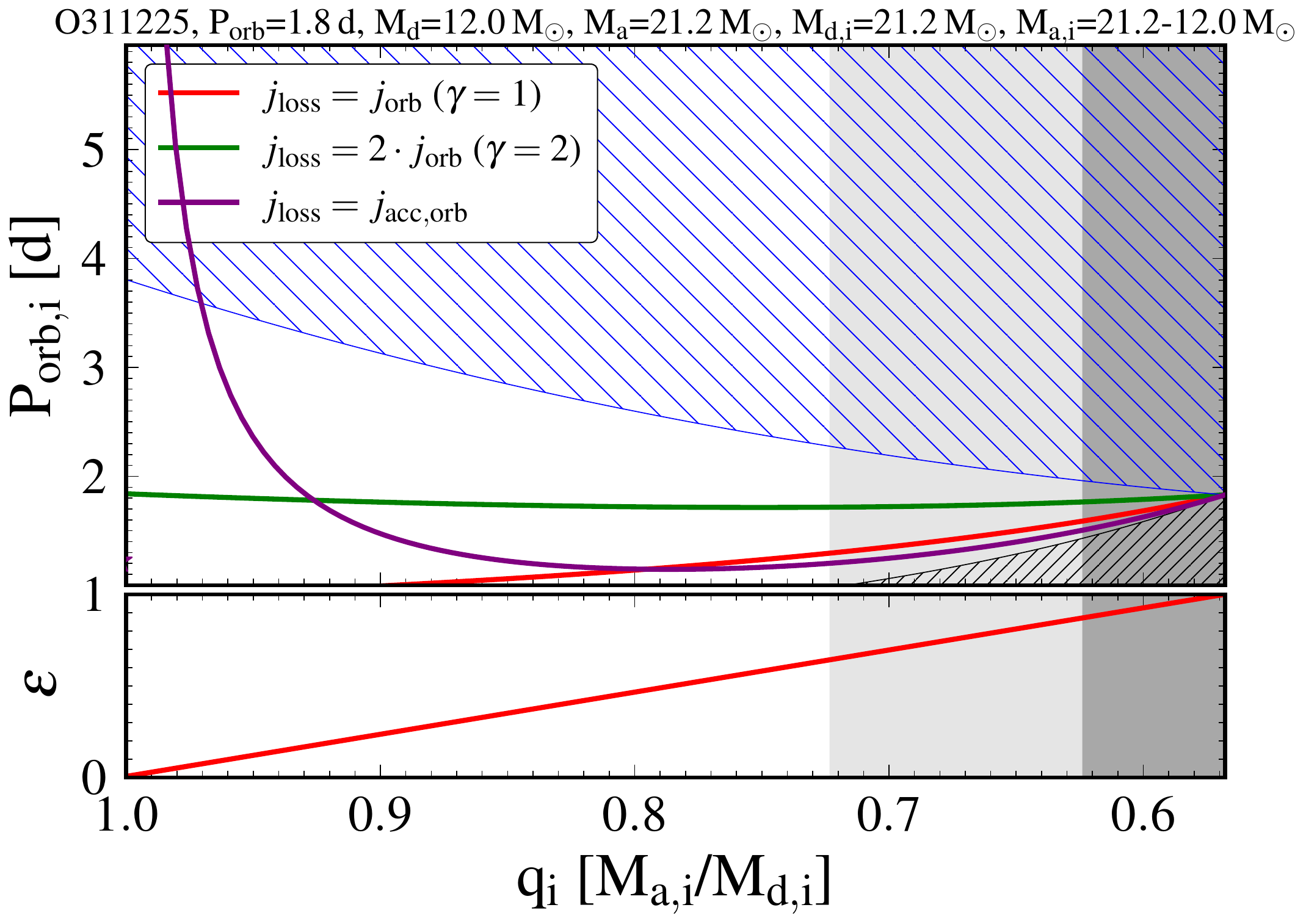}
    \includegraphics[width=0.24\linewidth]{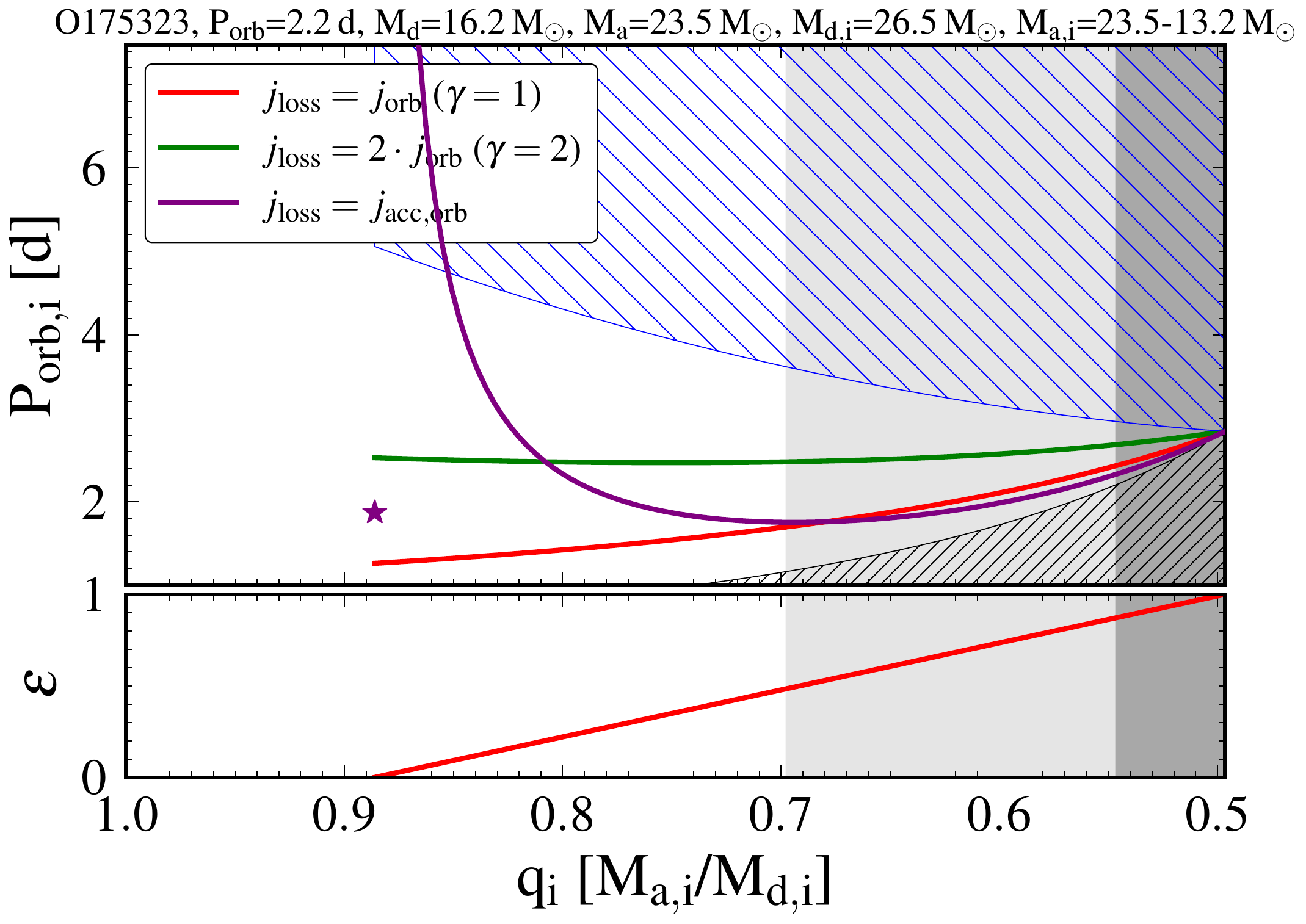}
    \includegraphics[width=0.24\linewidth]{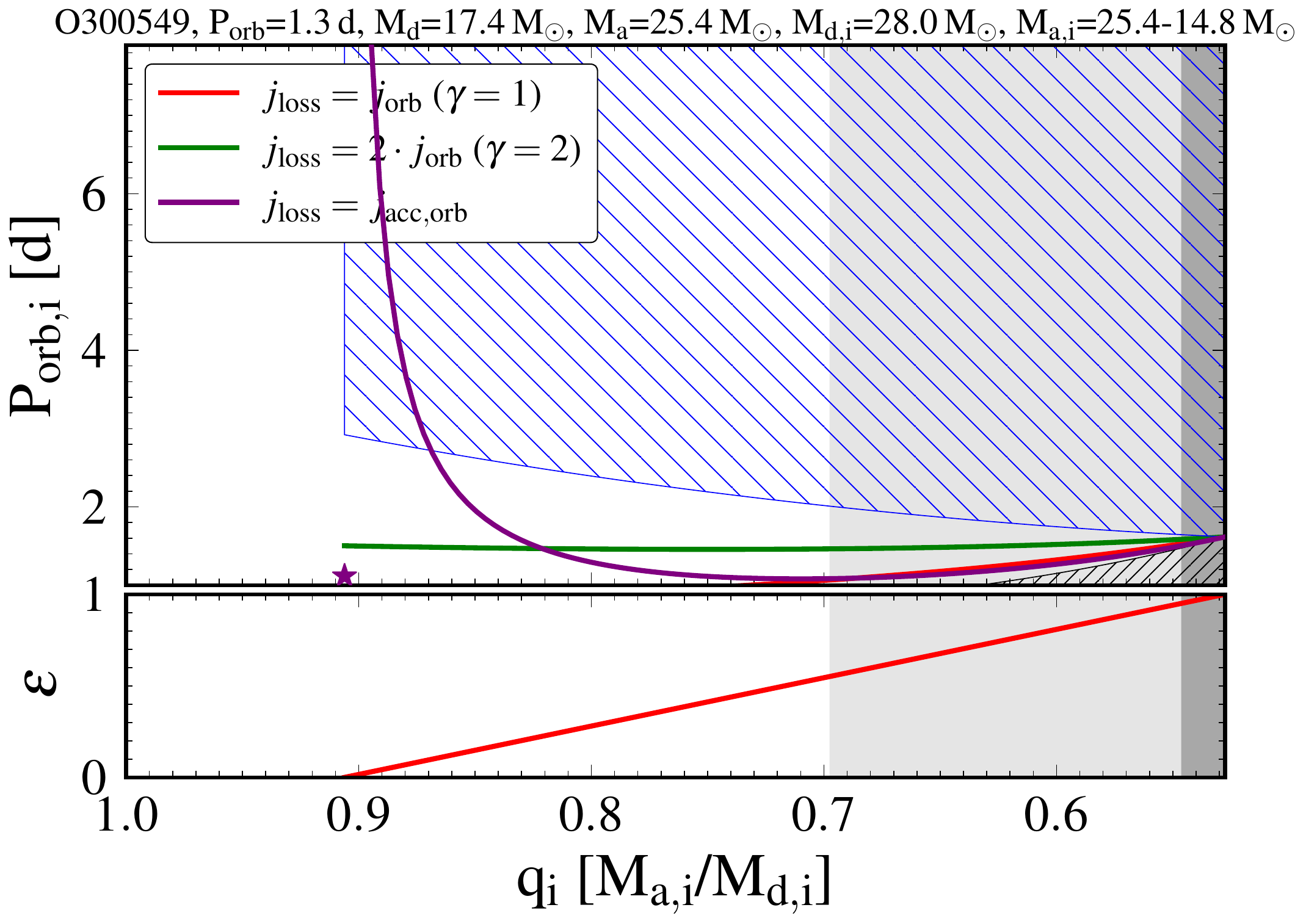}
    \includegraphics[width=0.24\linewidth]{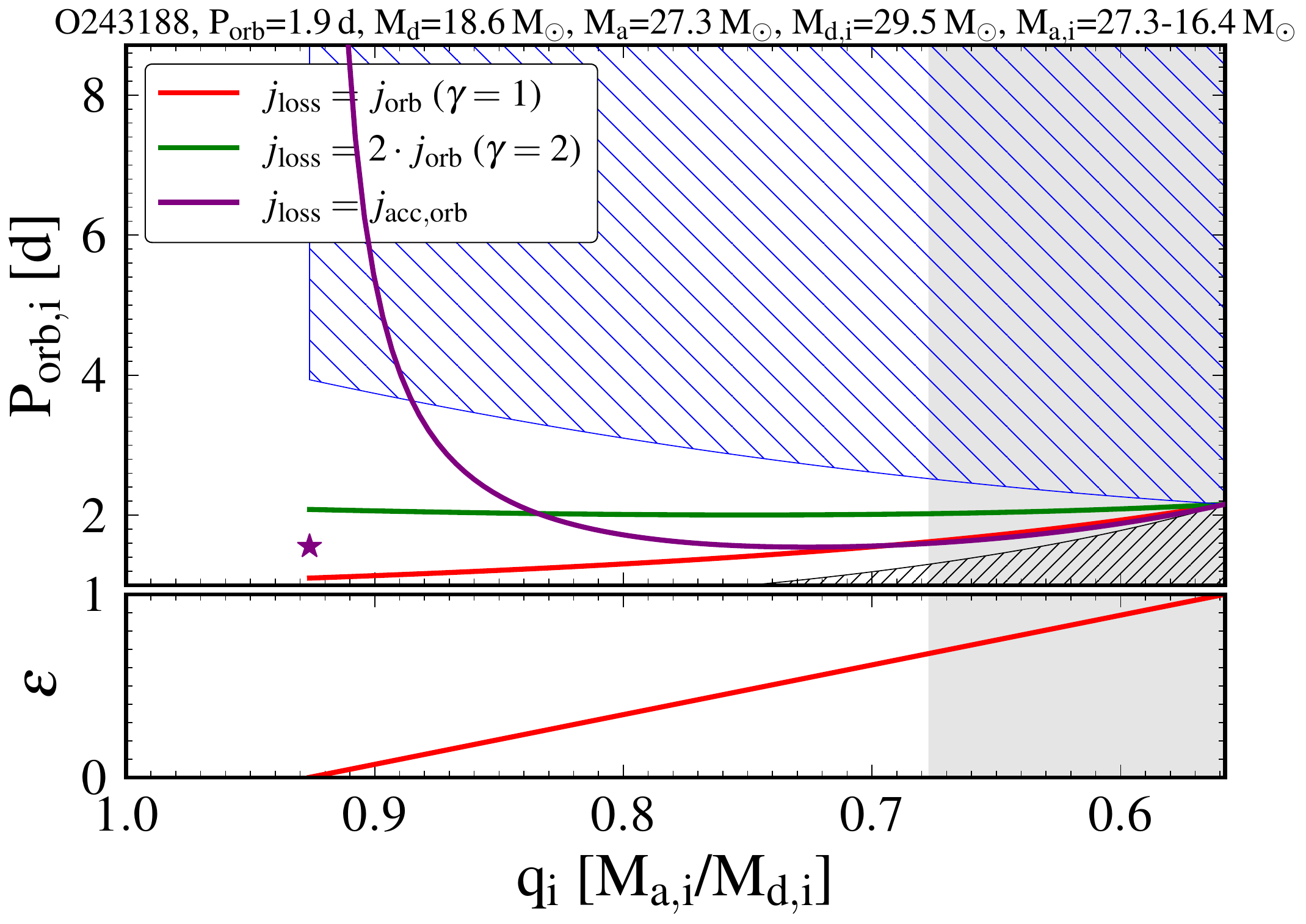}
    \caption{Same as Fig.\,\ref{fig:individual_examples}, for all massive Algols in the SMC. }
    \label{fig:appendix_smc}
\end{figure*}

\begin{figure*}
    \centering
    \includegraphics[width=0.23\linewidth]{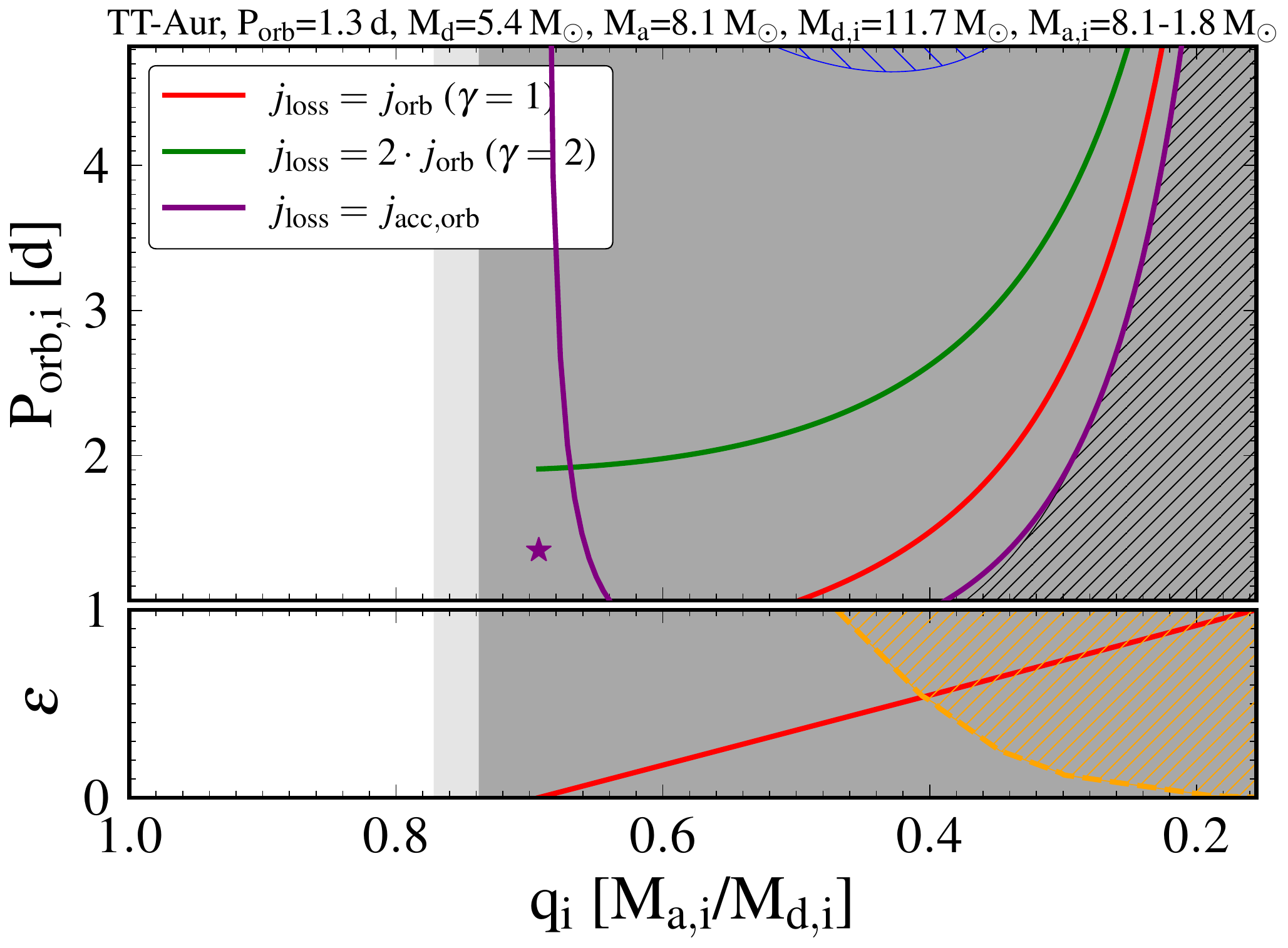}
    \includegraphics[width=0.23\linewidth]{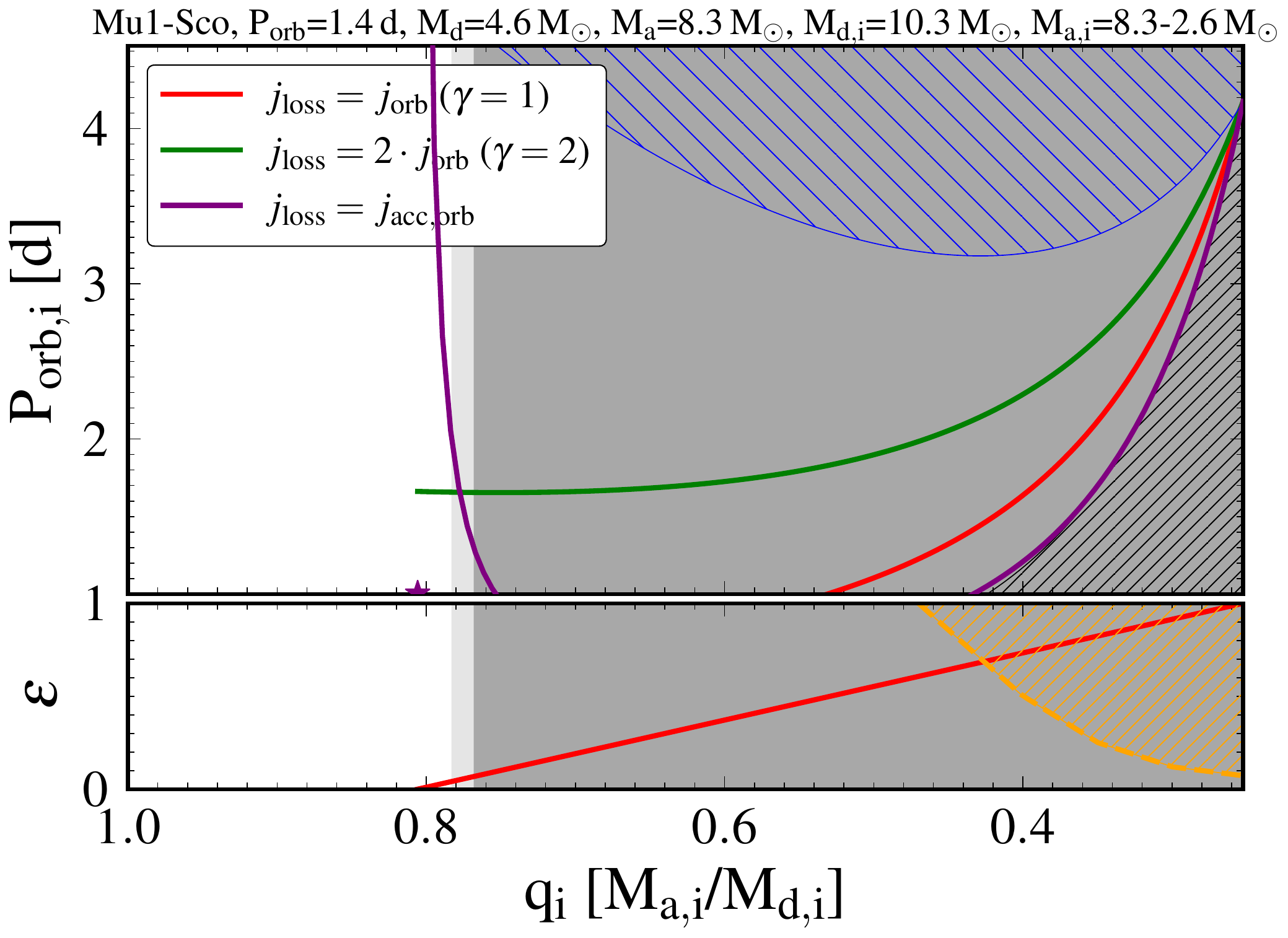}
    \includegraphics[width=0.23\linewidth]{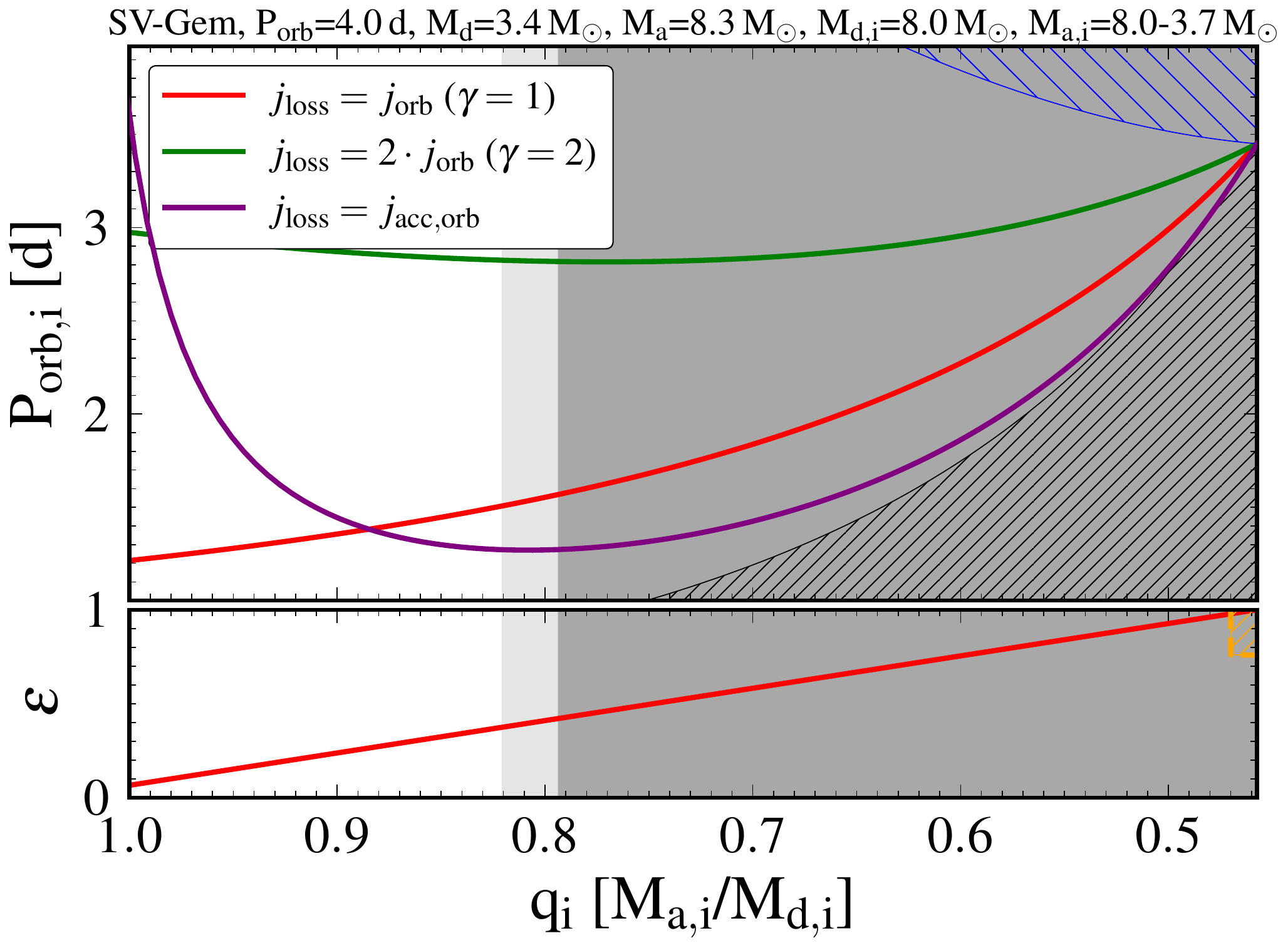}
    \includegraphics[width=0.23\linewidth]{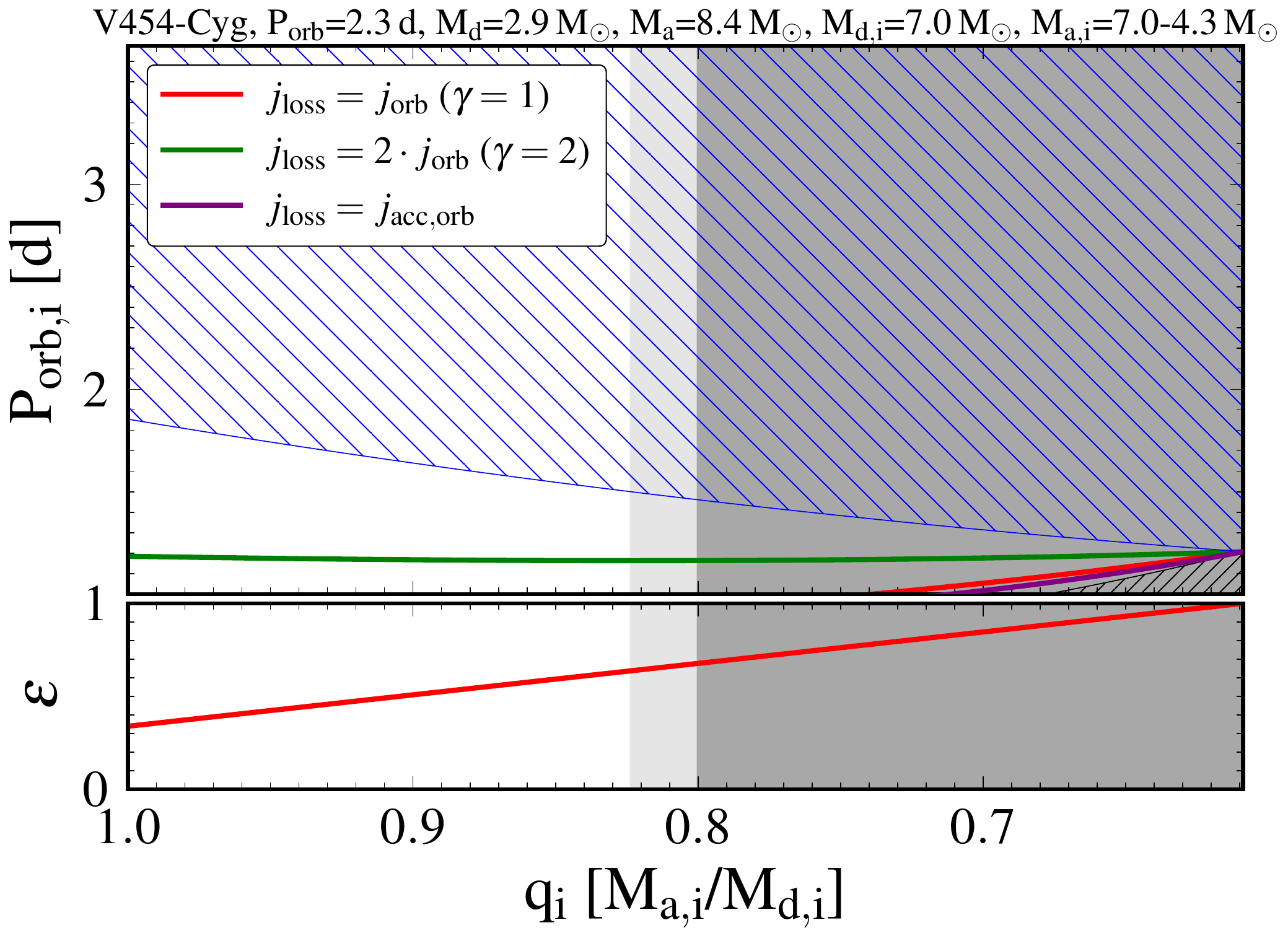}
    \includegraphics[width=0.23\linewidth]{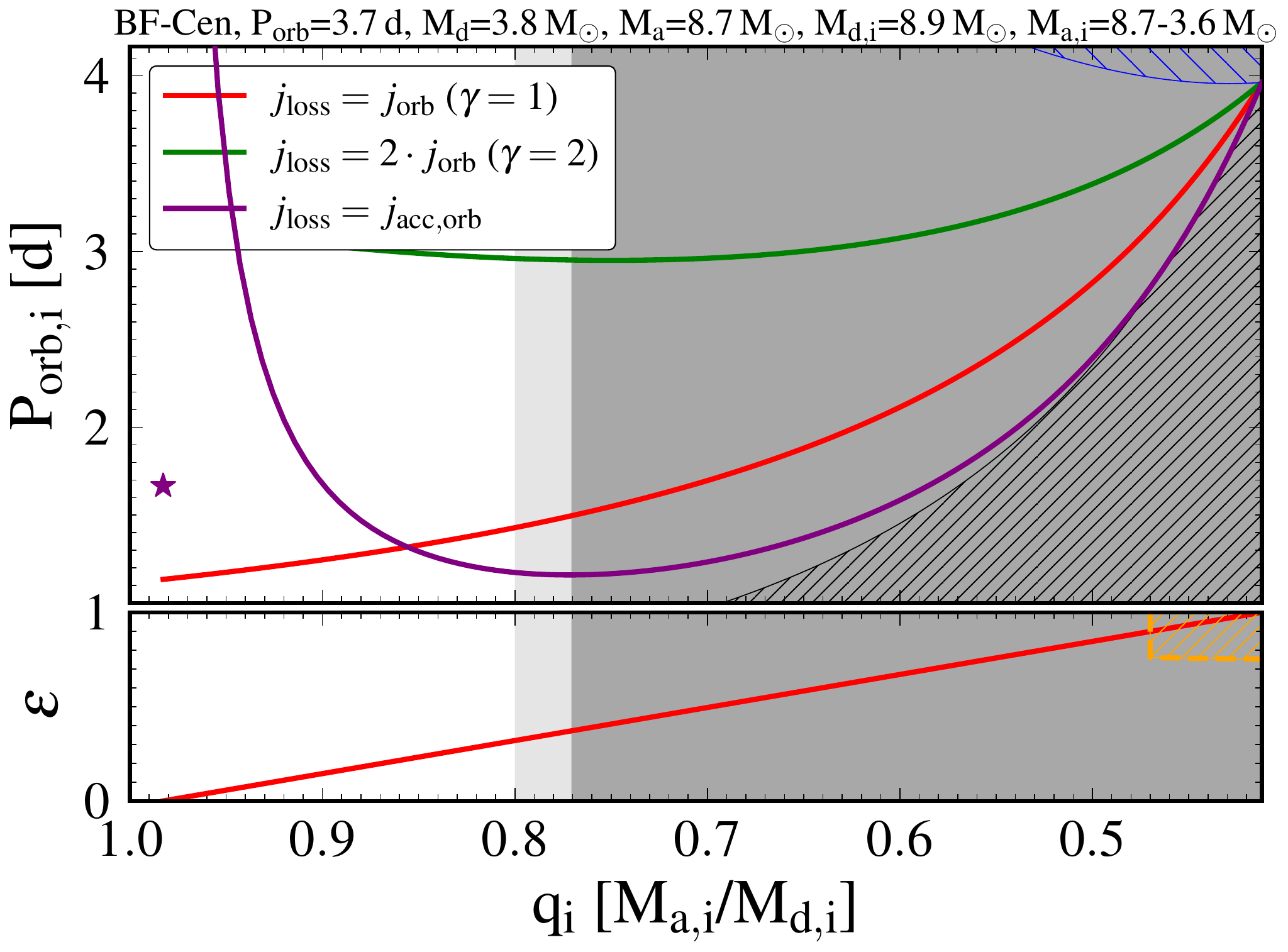}
    \includegraphics[width=0.23\linewidth]{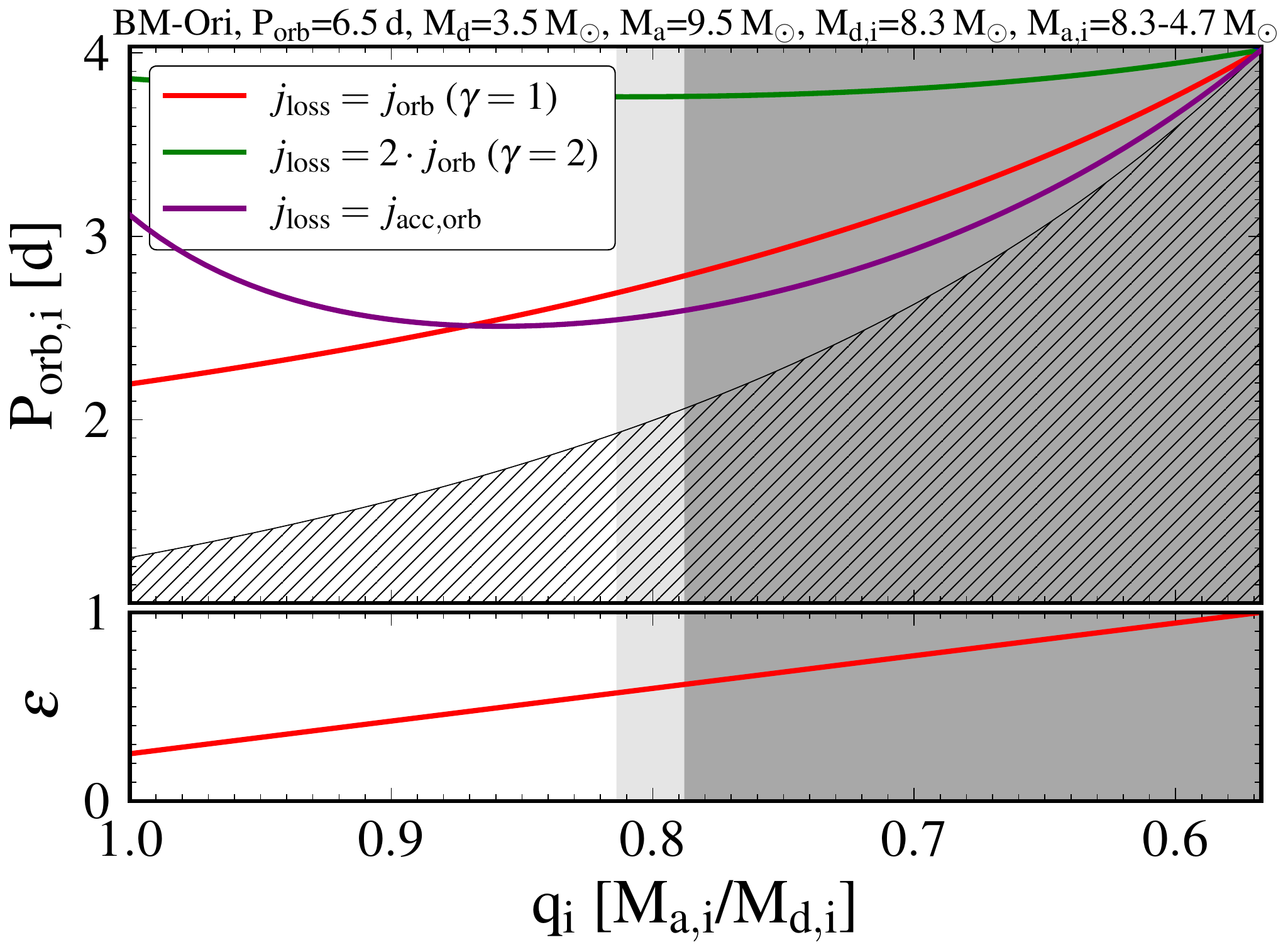}
    \includegraphics[width=0.23\linewidth]{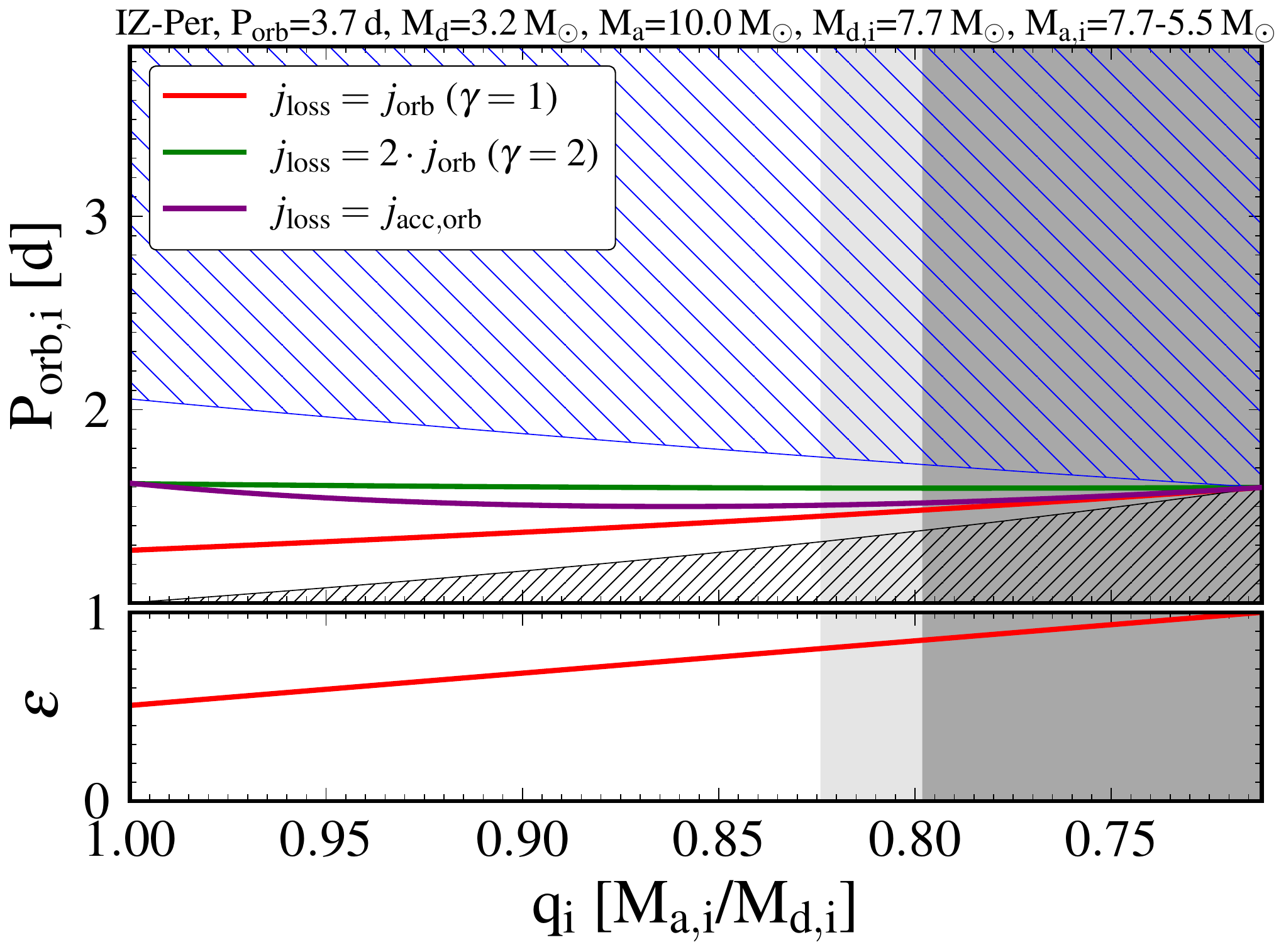}
    \includegraphics[width=0.23\linewidth]{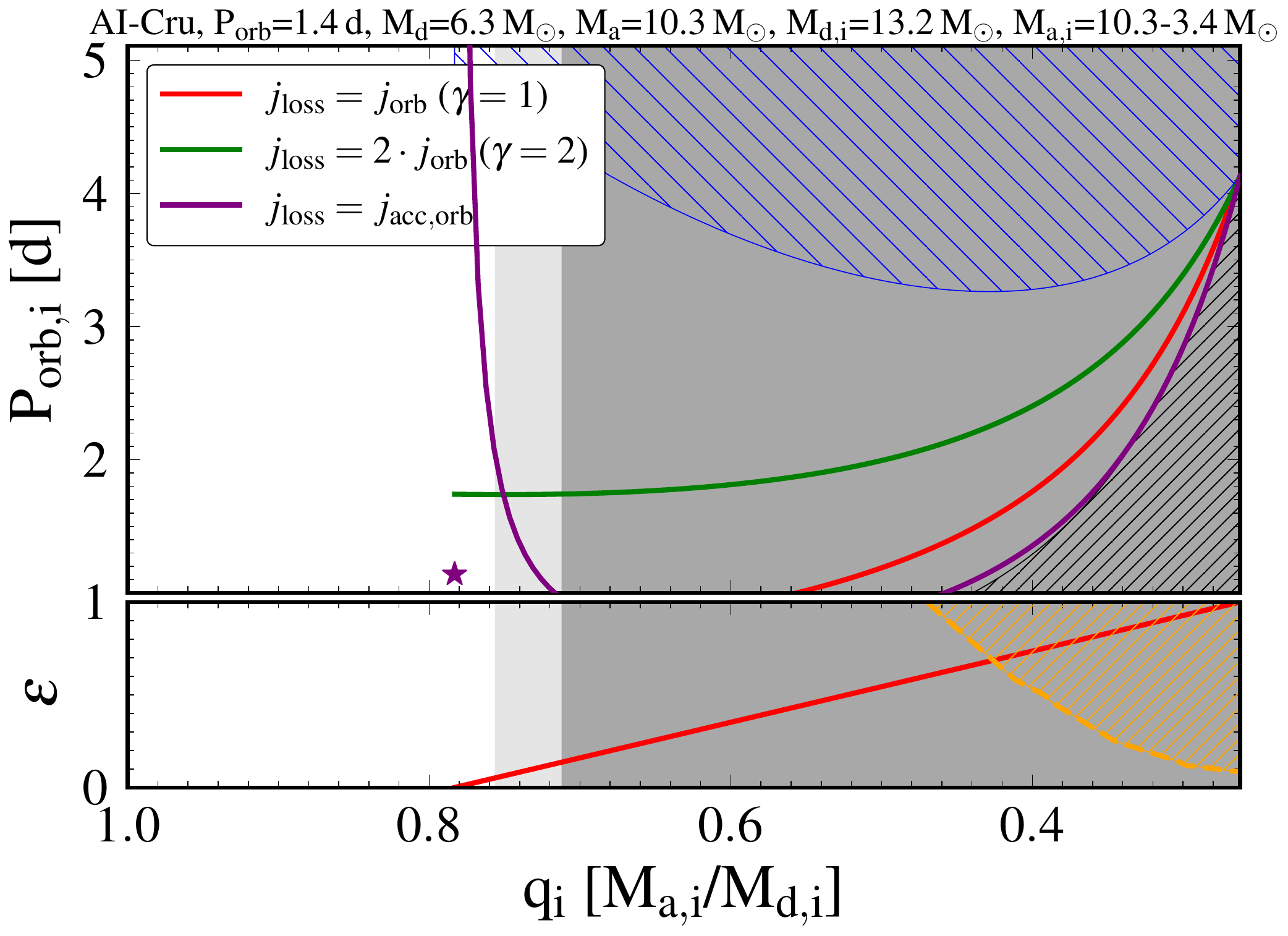}
    \includegraphics[width=0.23\linewidth]{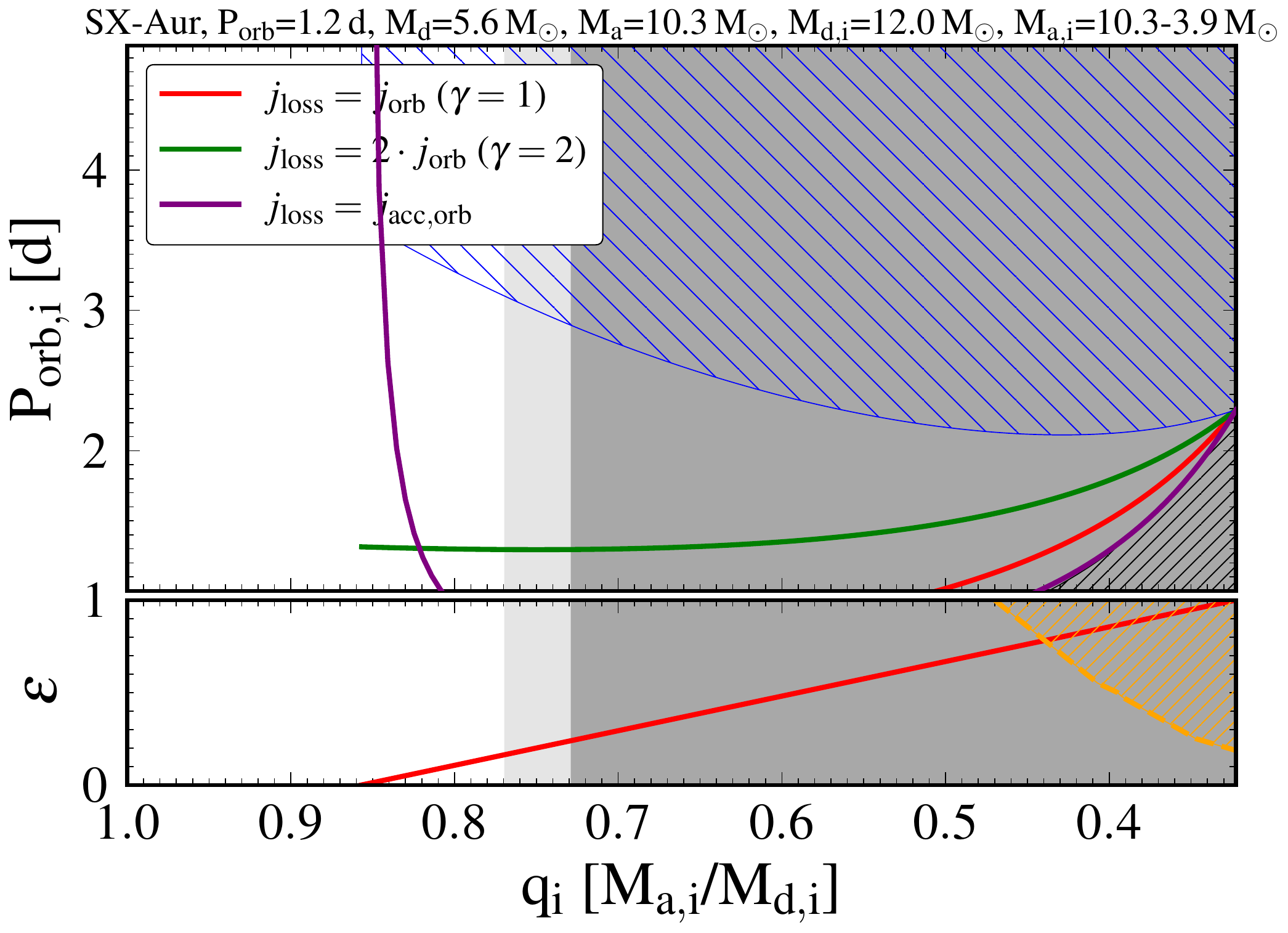}
    \includegraphics[width=0.23\linewidth]{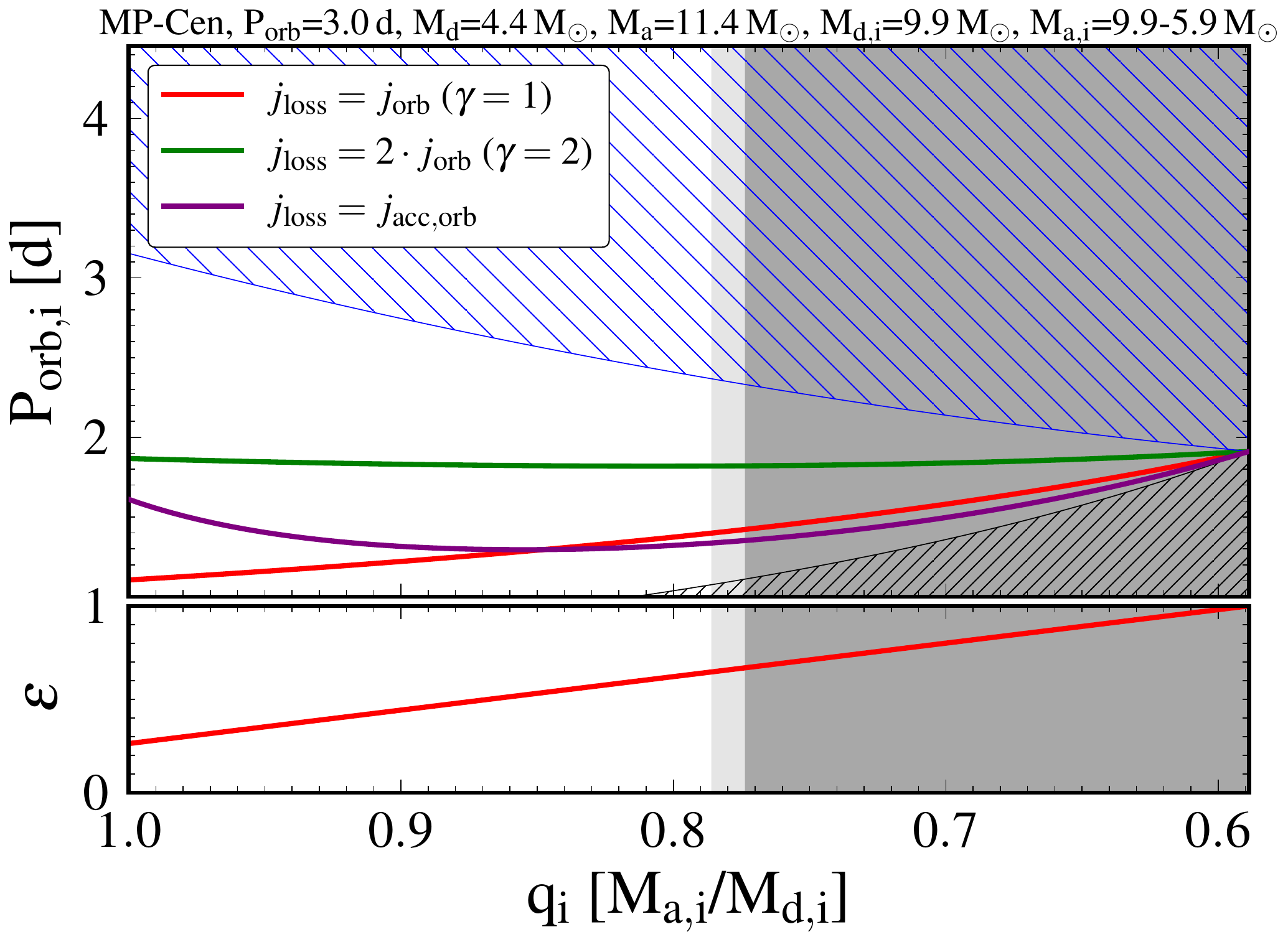}
    \includegraphics[width=0.23\linewidth]{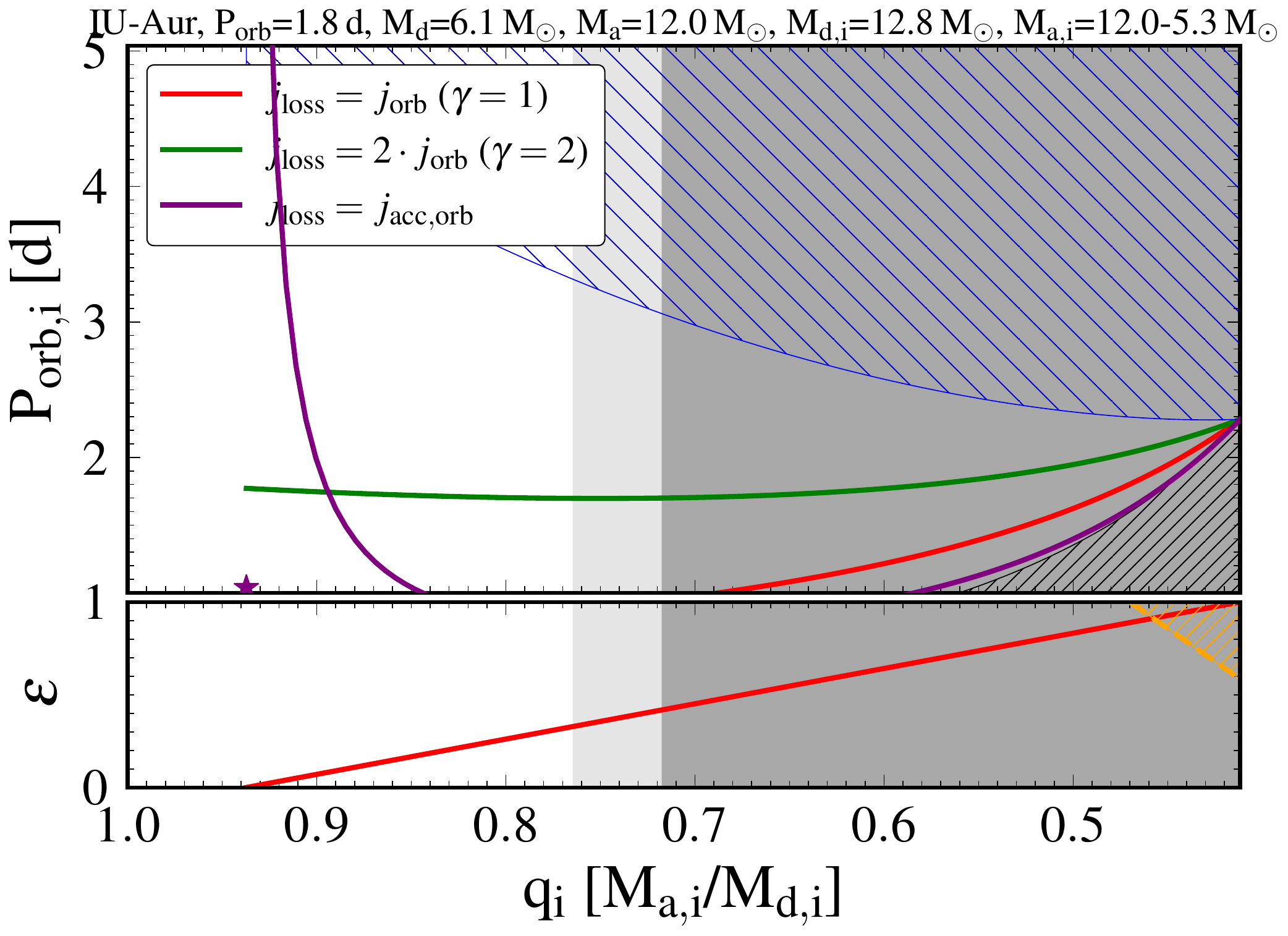}
    \includegraphics[width=0.23\linewidth]{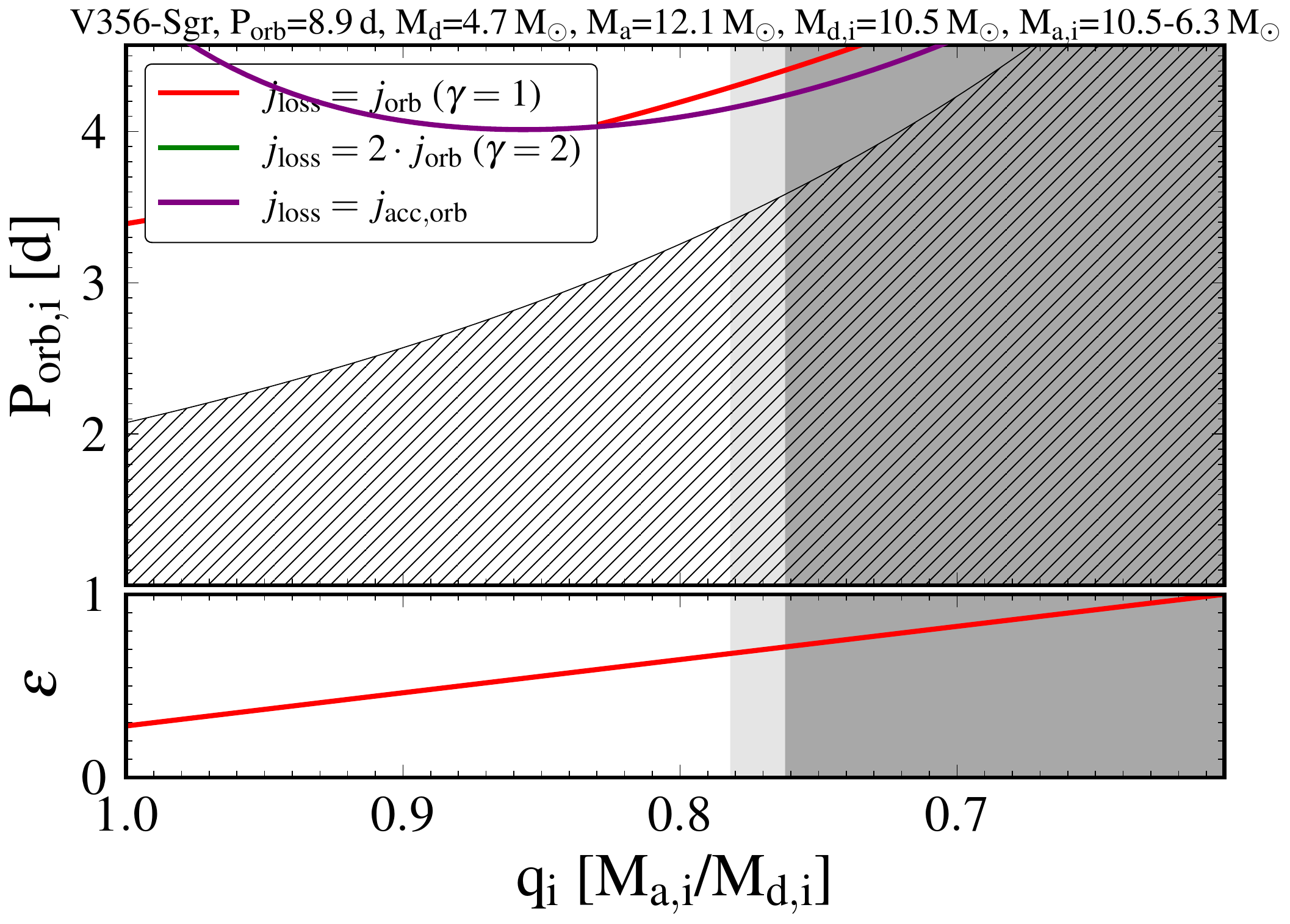}
    \includegraphics[width=0.23\linewidth]{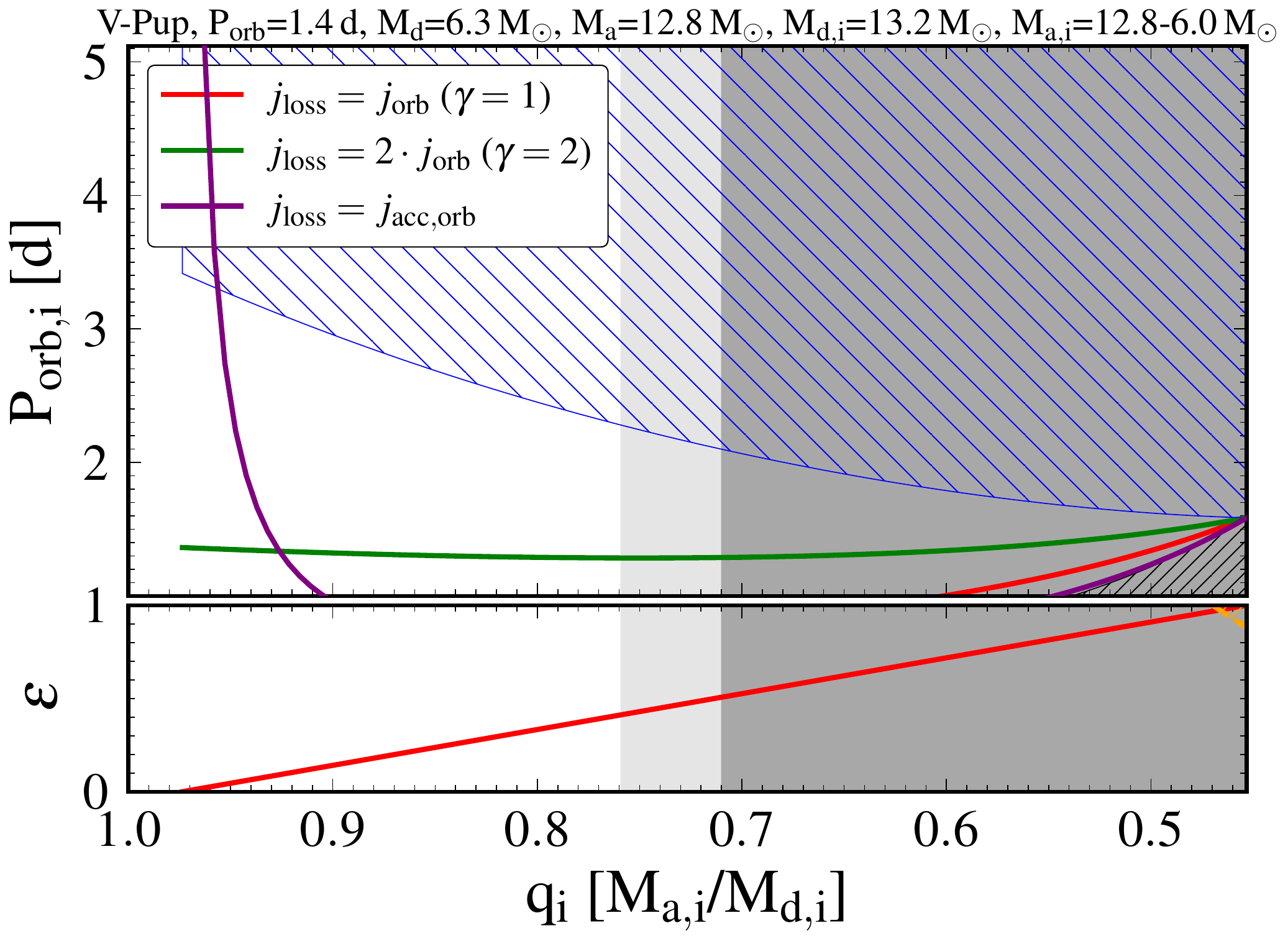}
    \includegraphics[width=0.23\linewidth]{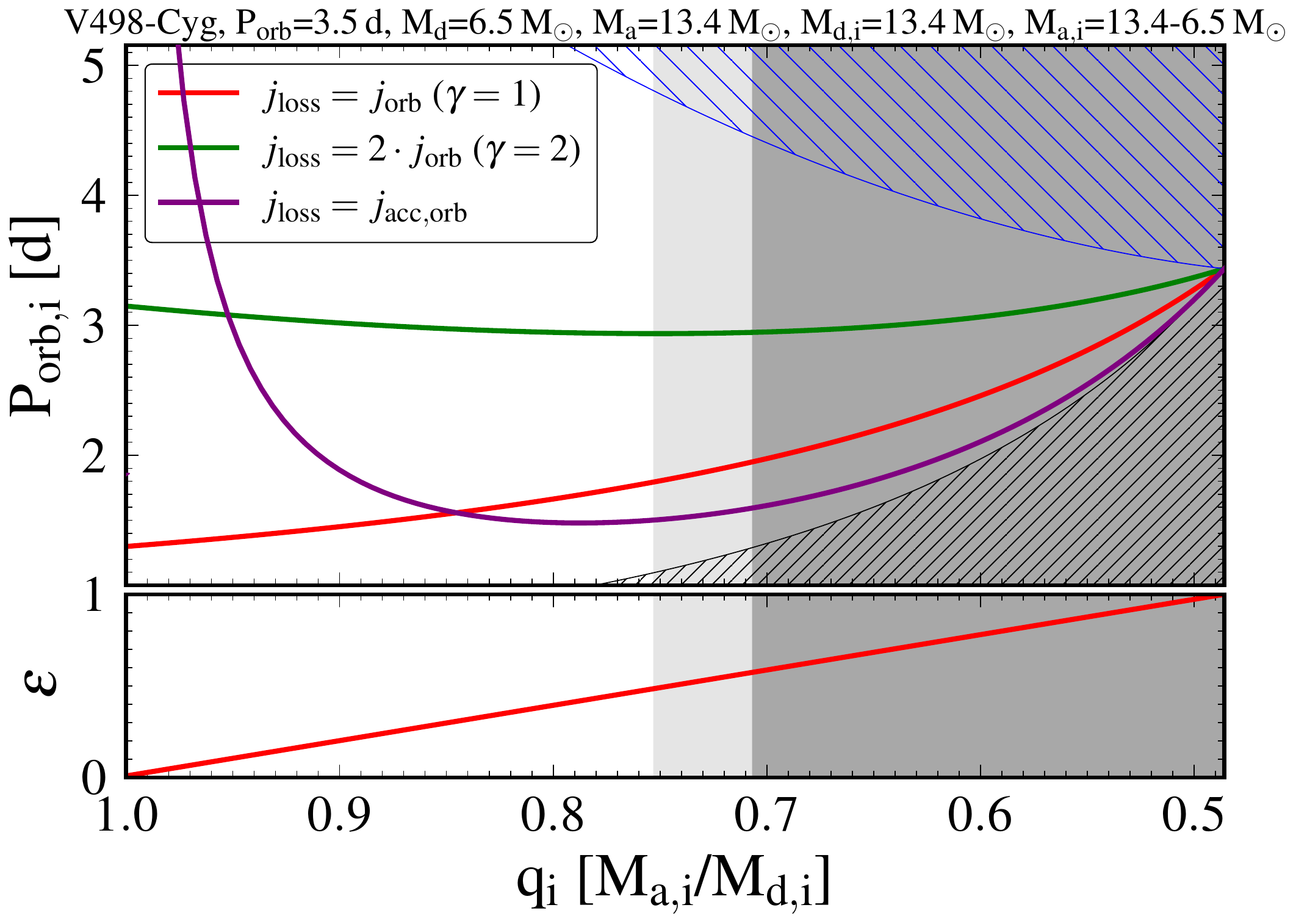}
    \includegraphics[width=0.23\linewidth]{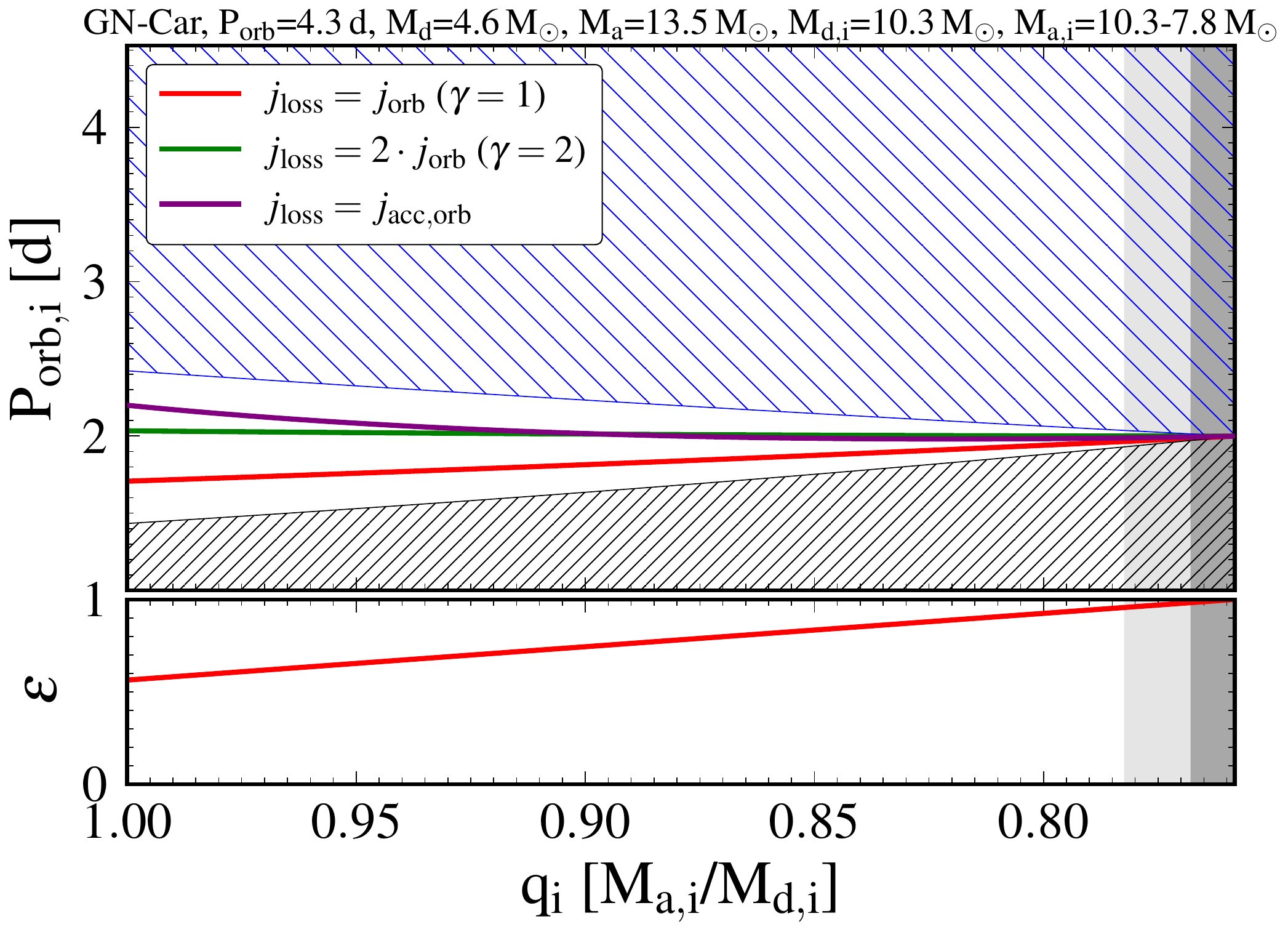}
    \includegraphics[width=0.23\linewidth]{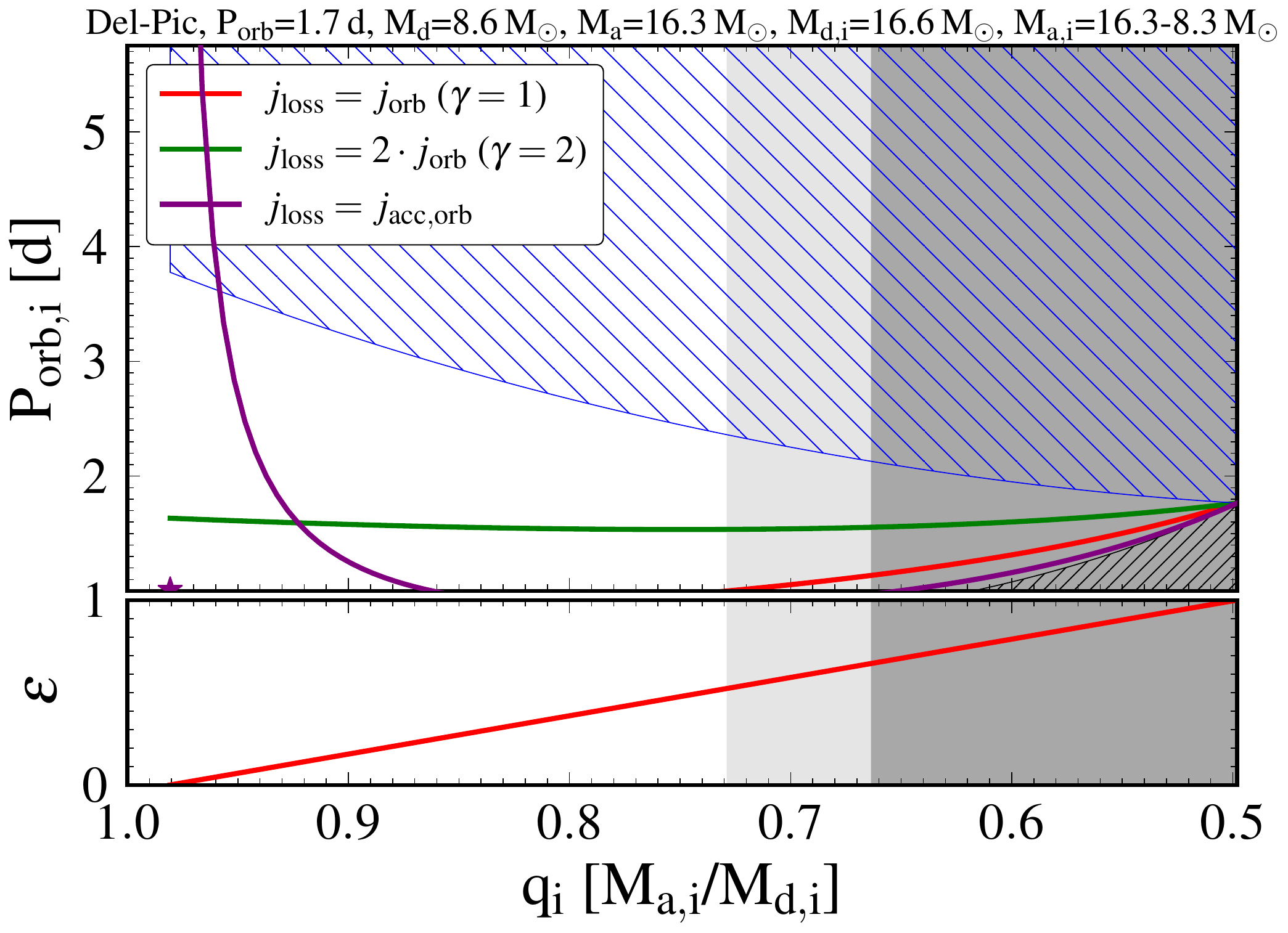}
    \includegraphics[width=0.23\linewidth]{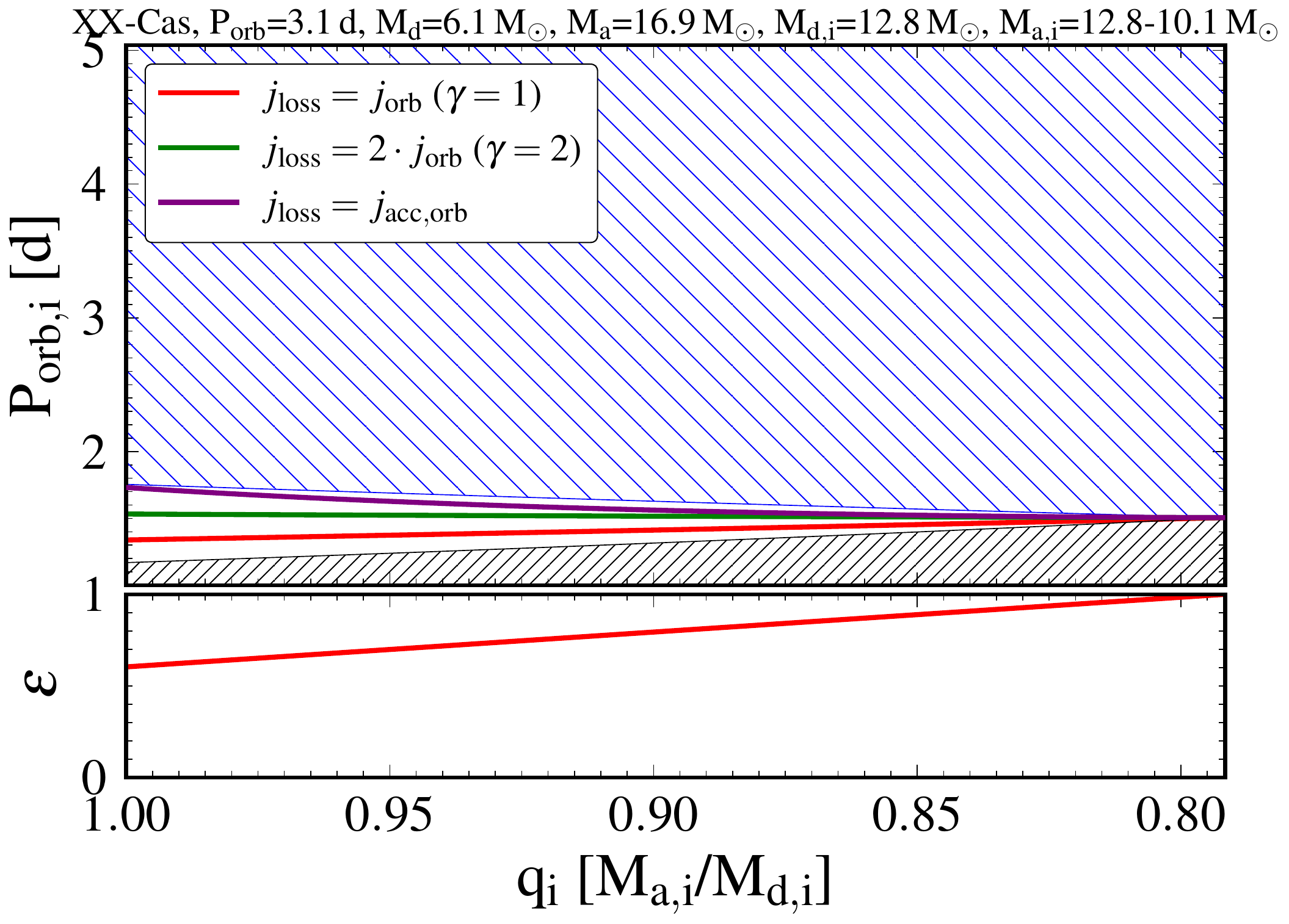}
    \includegraphics[width=0.23\linewidth]{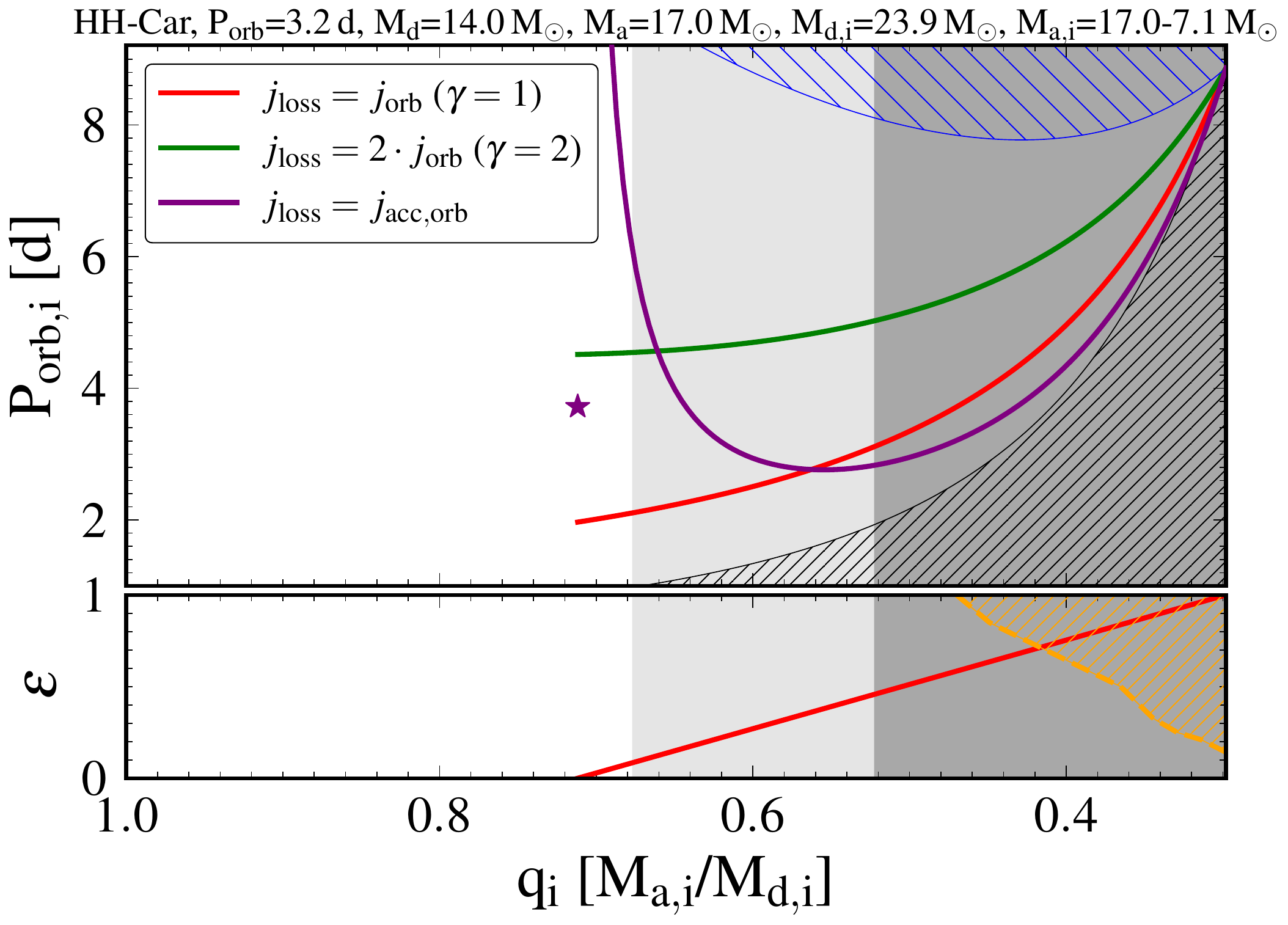}
    \includegraphics[width=0.23\linewidth]{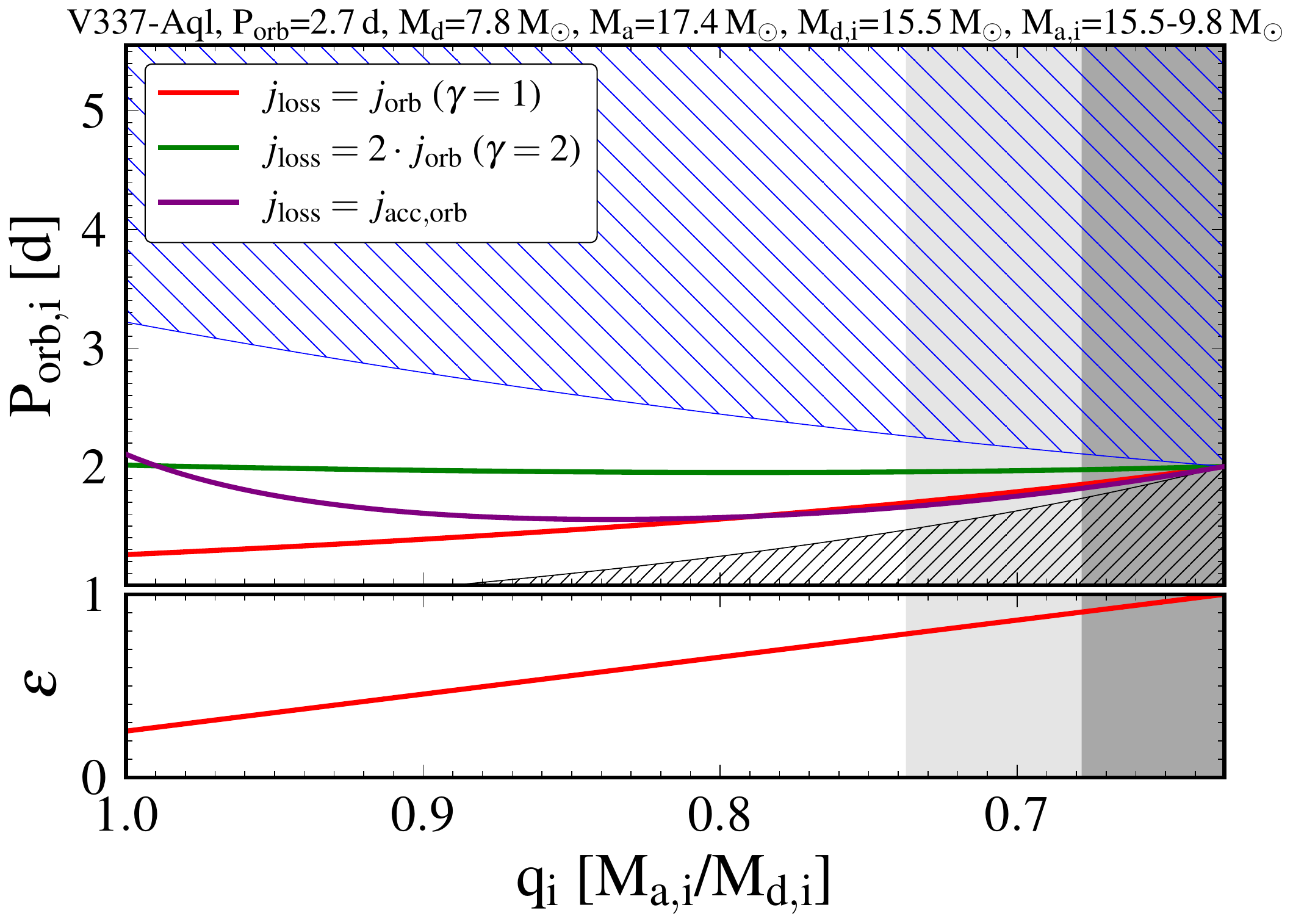}
    \includegraphics[width=0.23\linewidth]{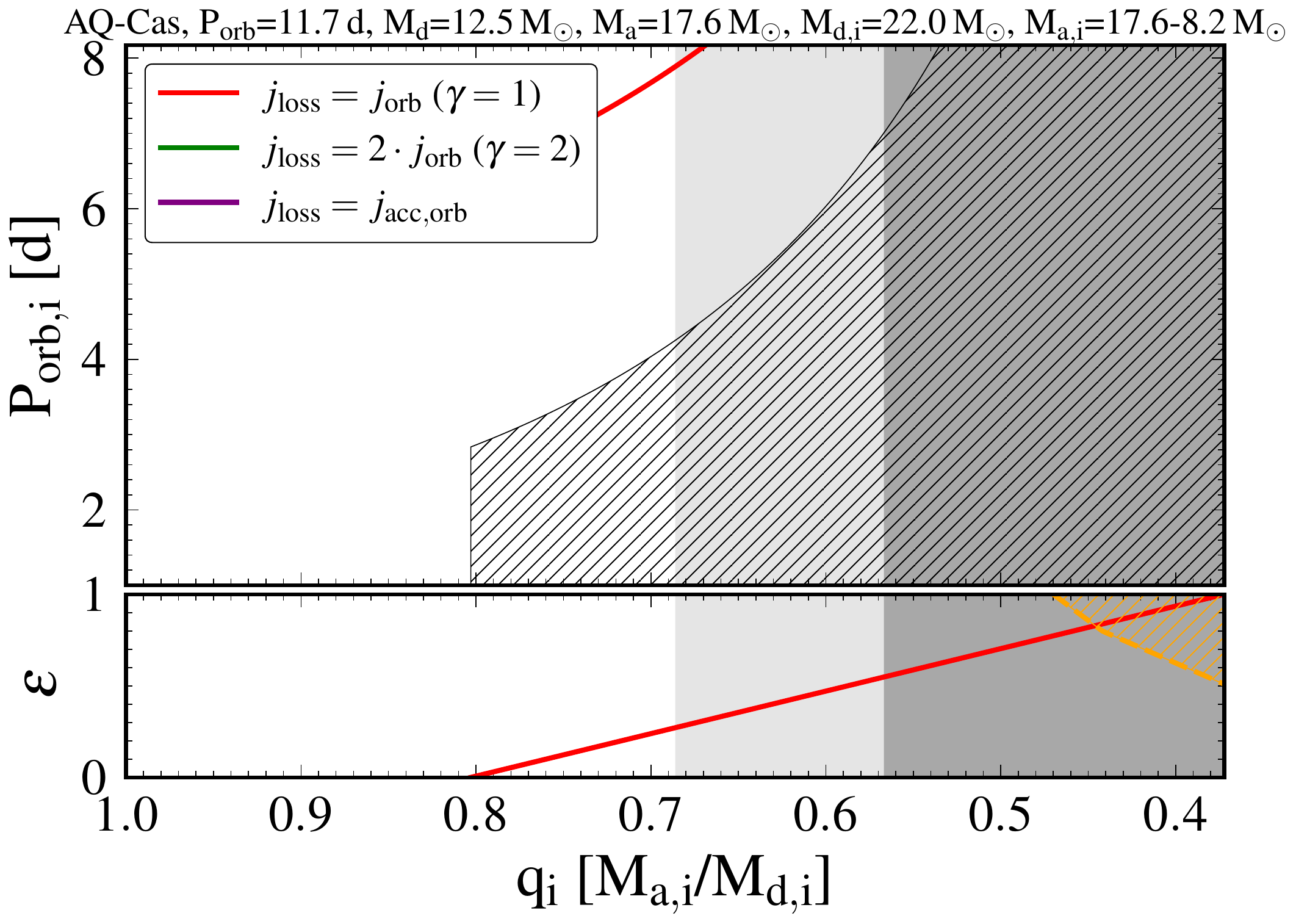}
    \includegraphics[width=0.23\linewidth]{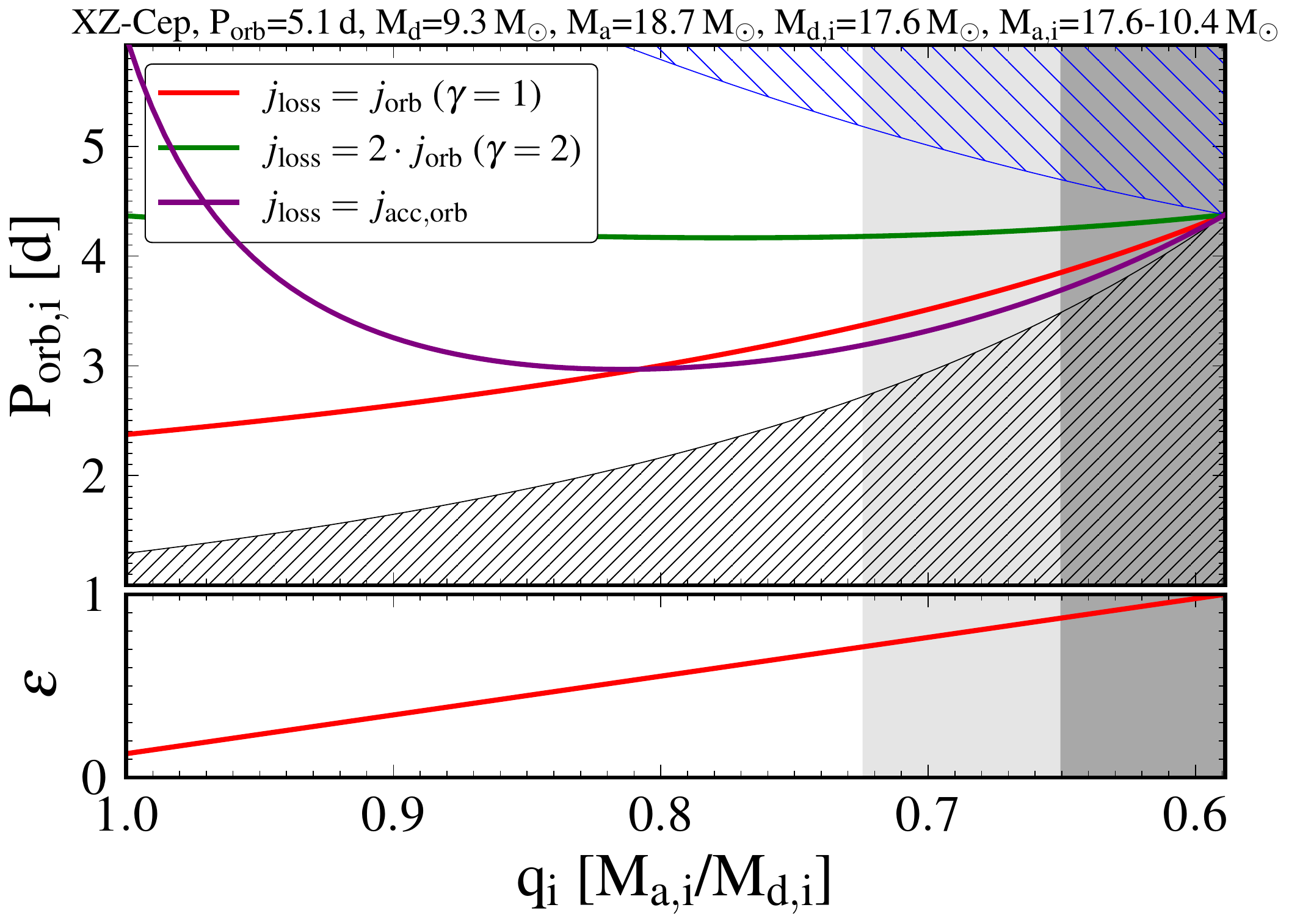}
    \includegraphics[width=0.23\linewidth]{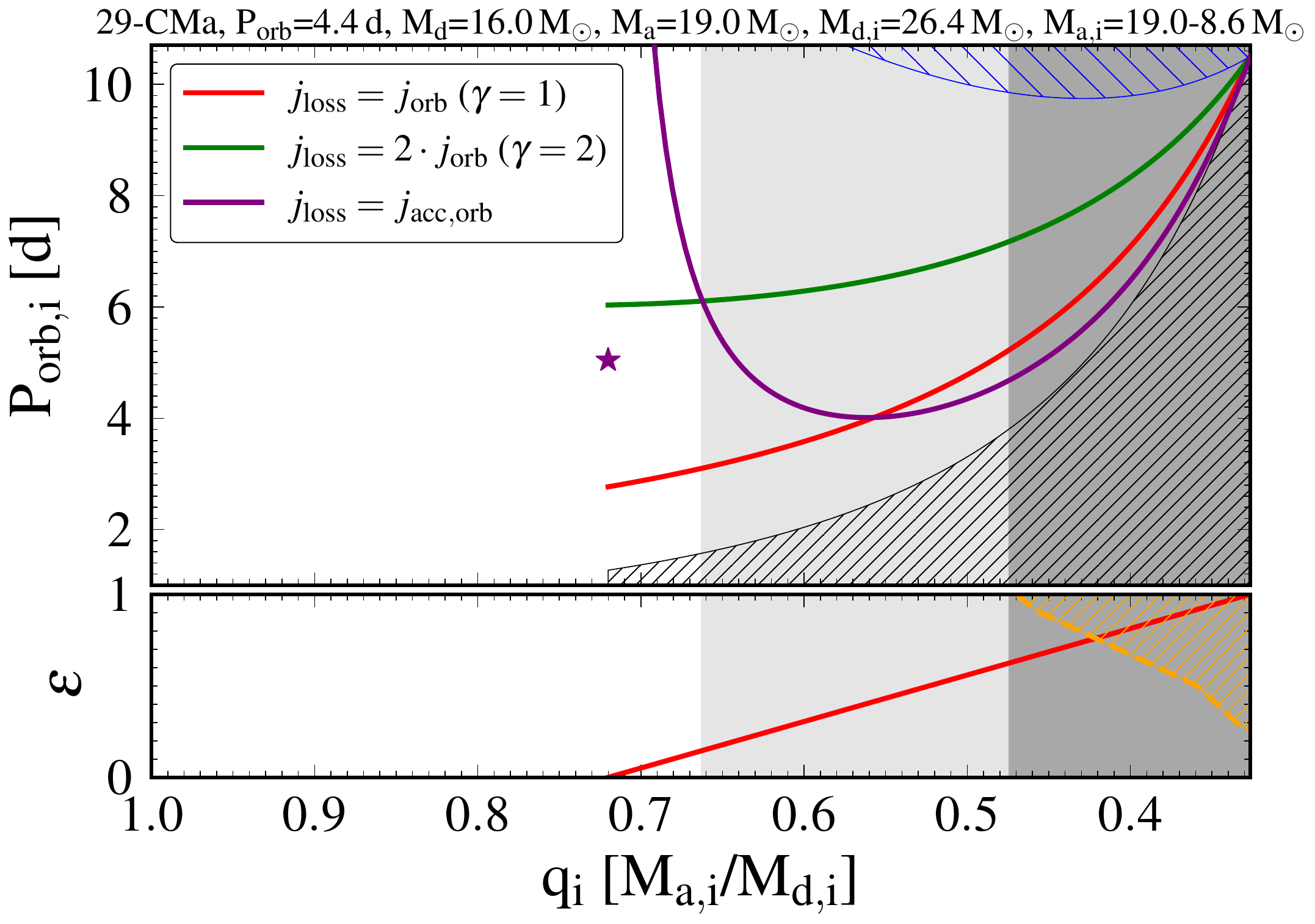}
    \includegraphics[width=0.23\linewidth]{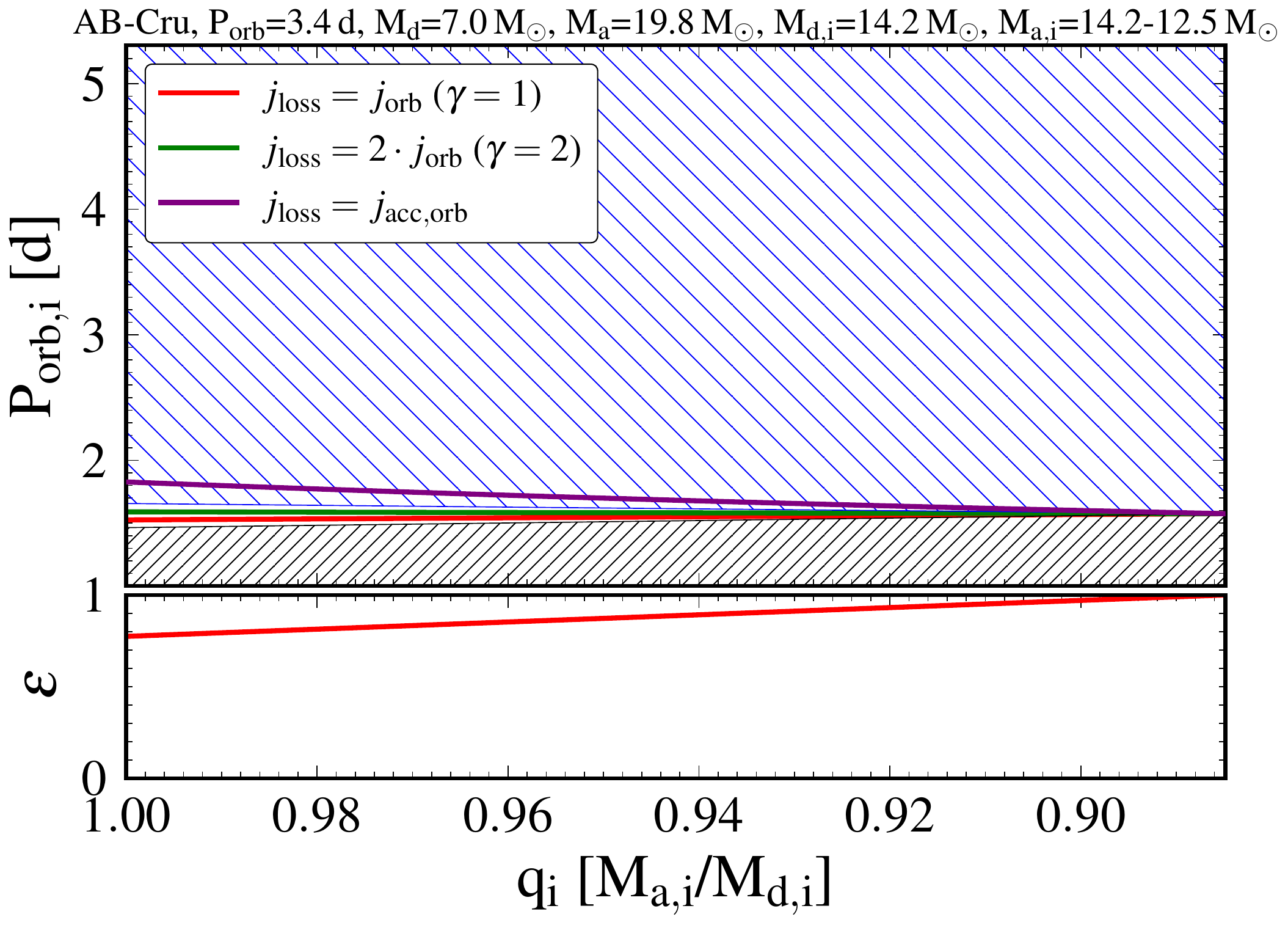}
    \includegraphics[width=0.23\linewidth]{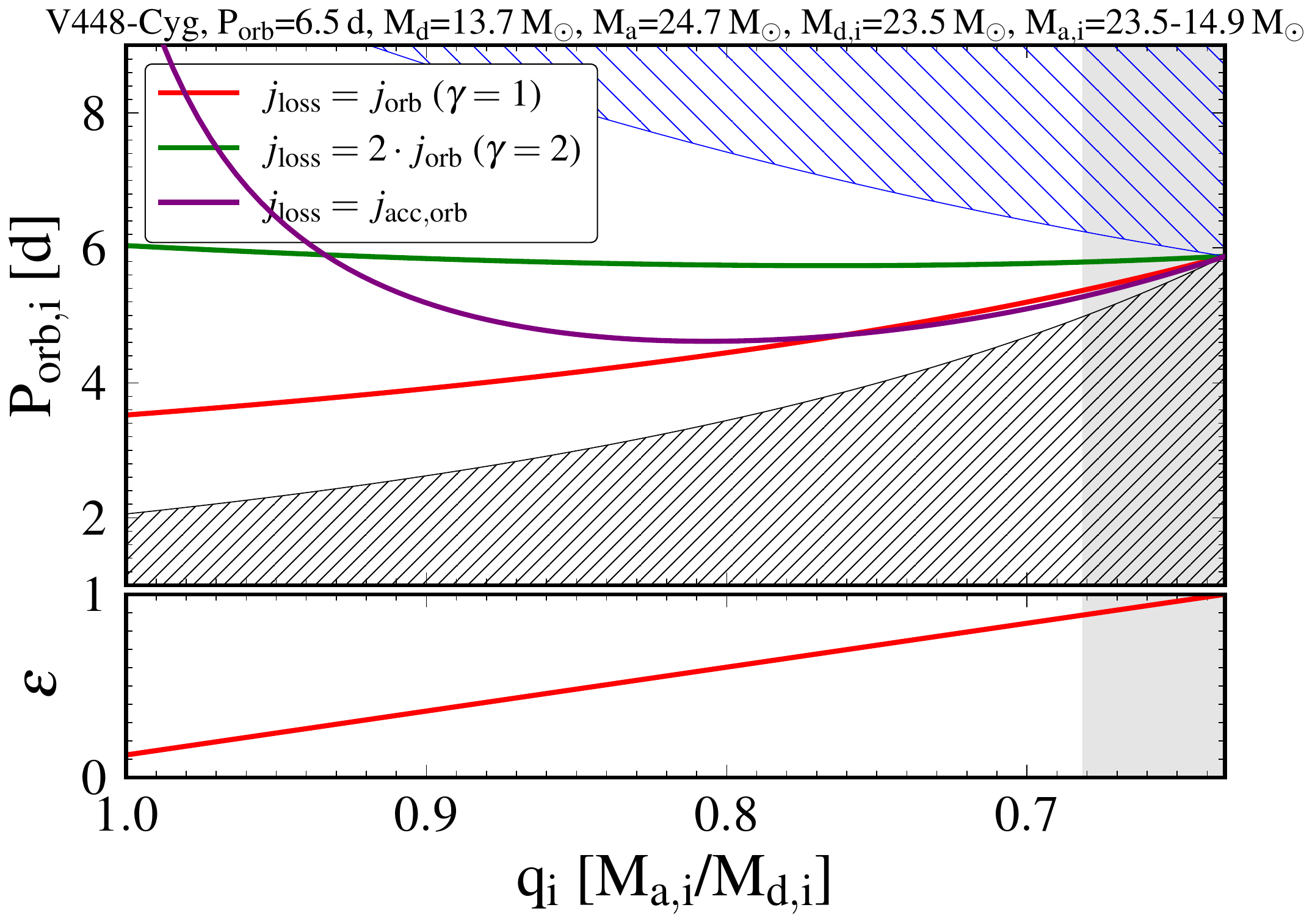}
    \includegraphics[width=0.23\linewidth]{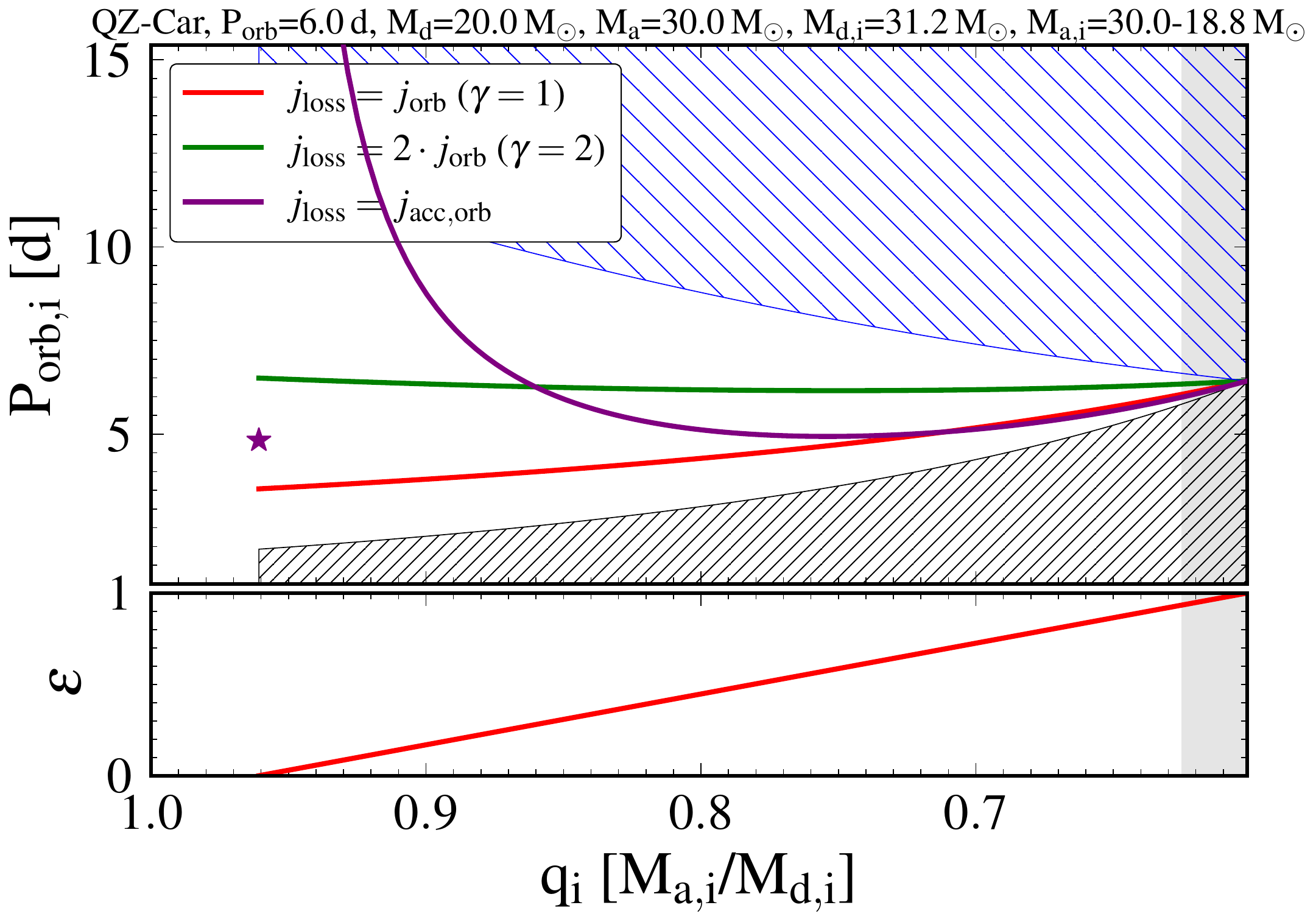}
    \includegraphics[width=0.23\linewidth]{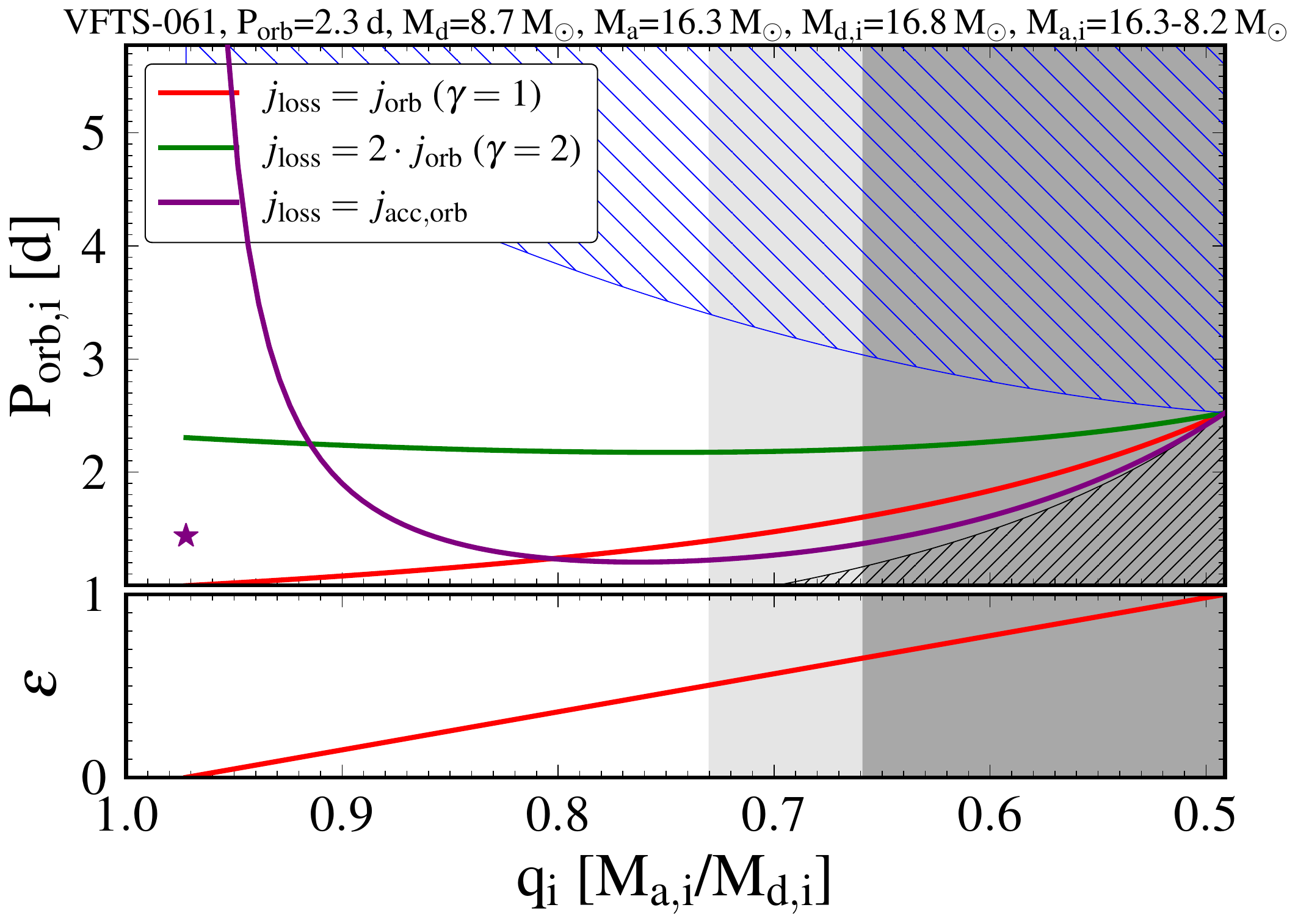}
    \includegraphics[width=0.23\linewidth]{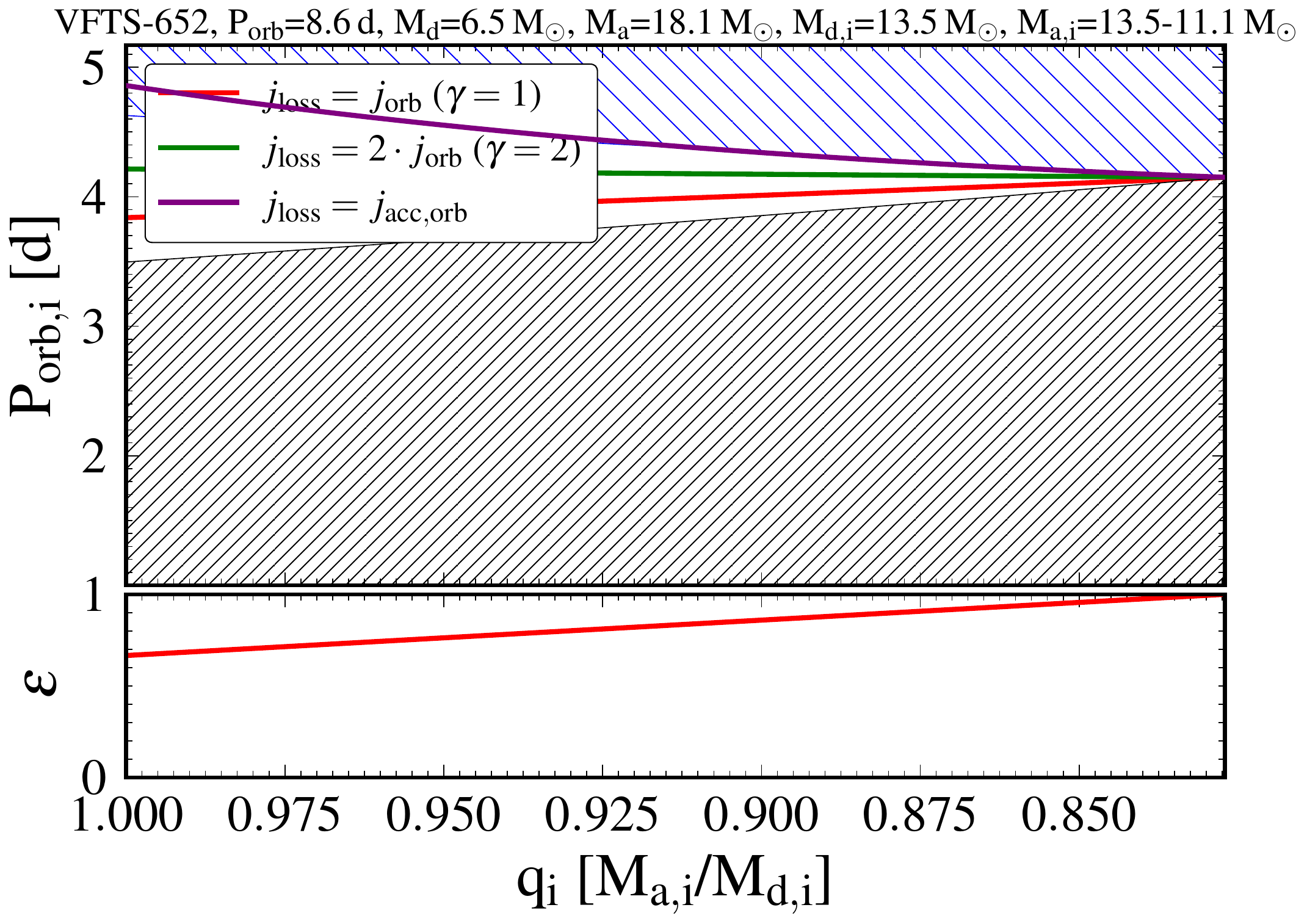}
    \includegraphics[width=0.23\linewidth]{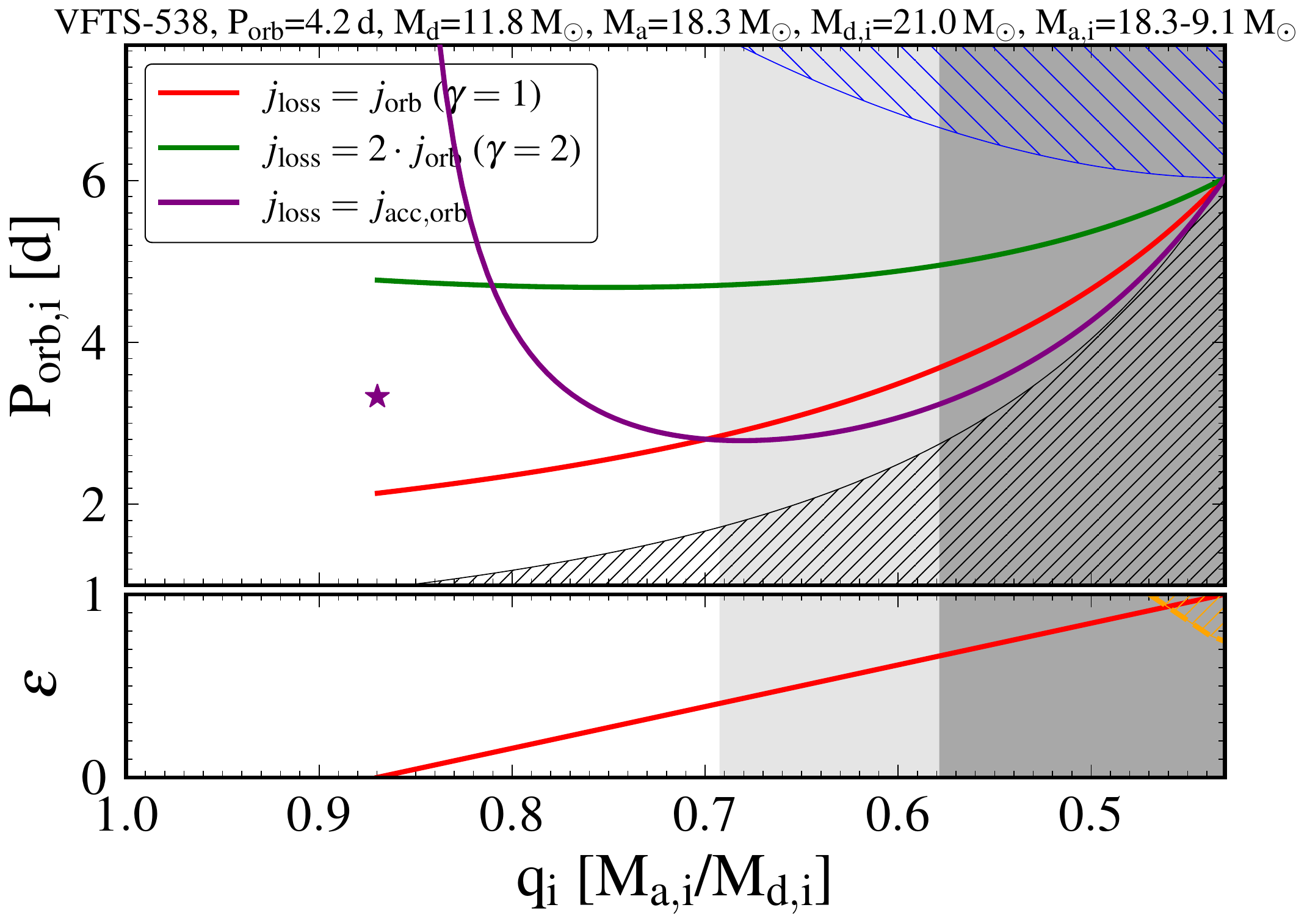}
    \includegraphics[width=0.23\linewidth]{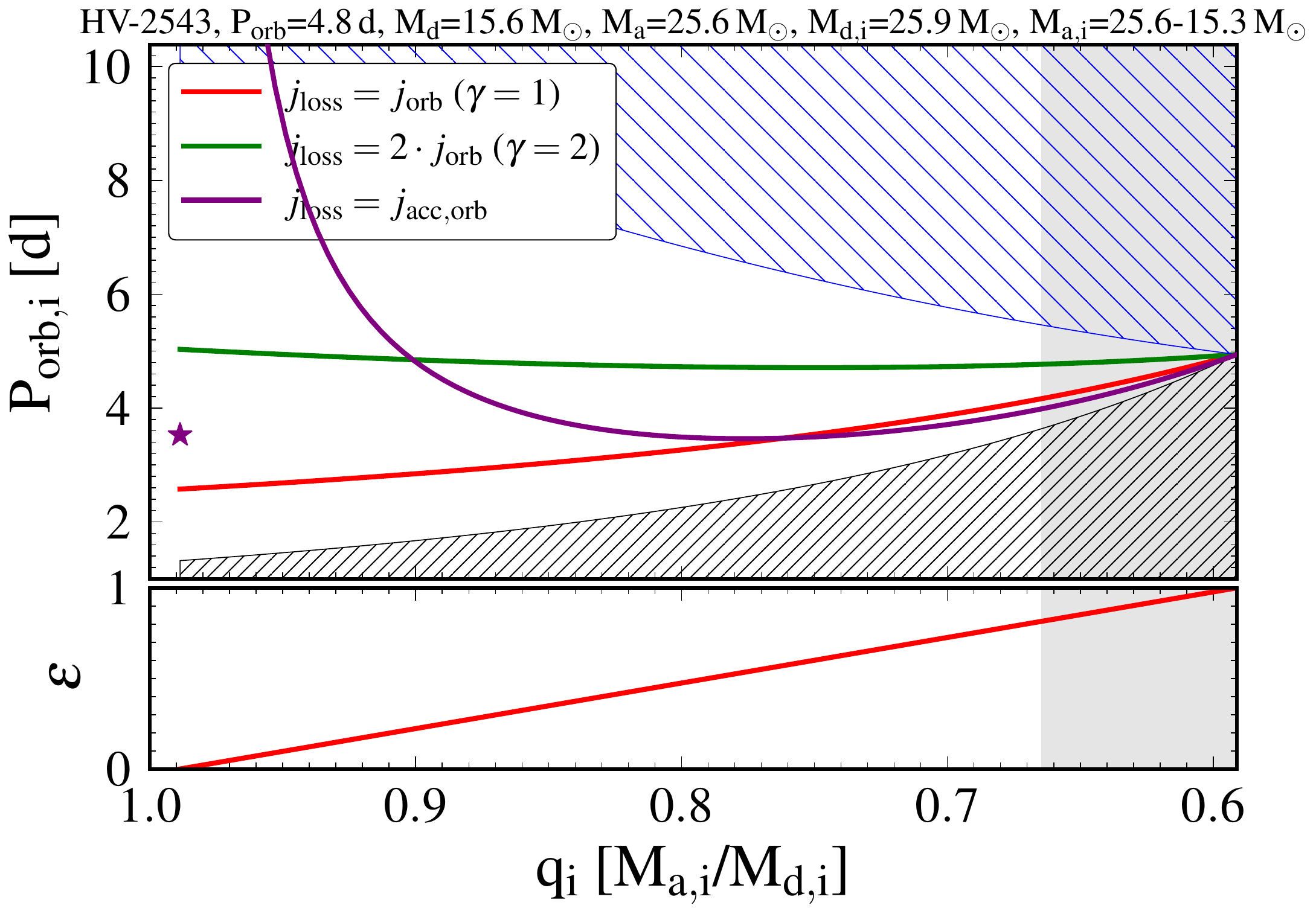}
    \includegraphics[width=0.23\linewidth]{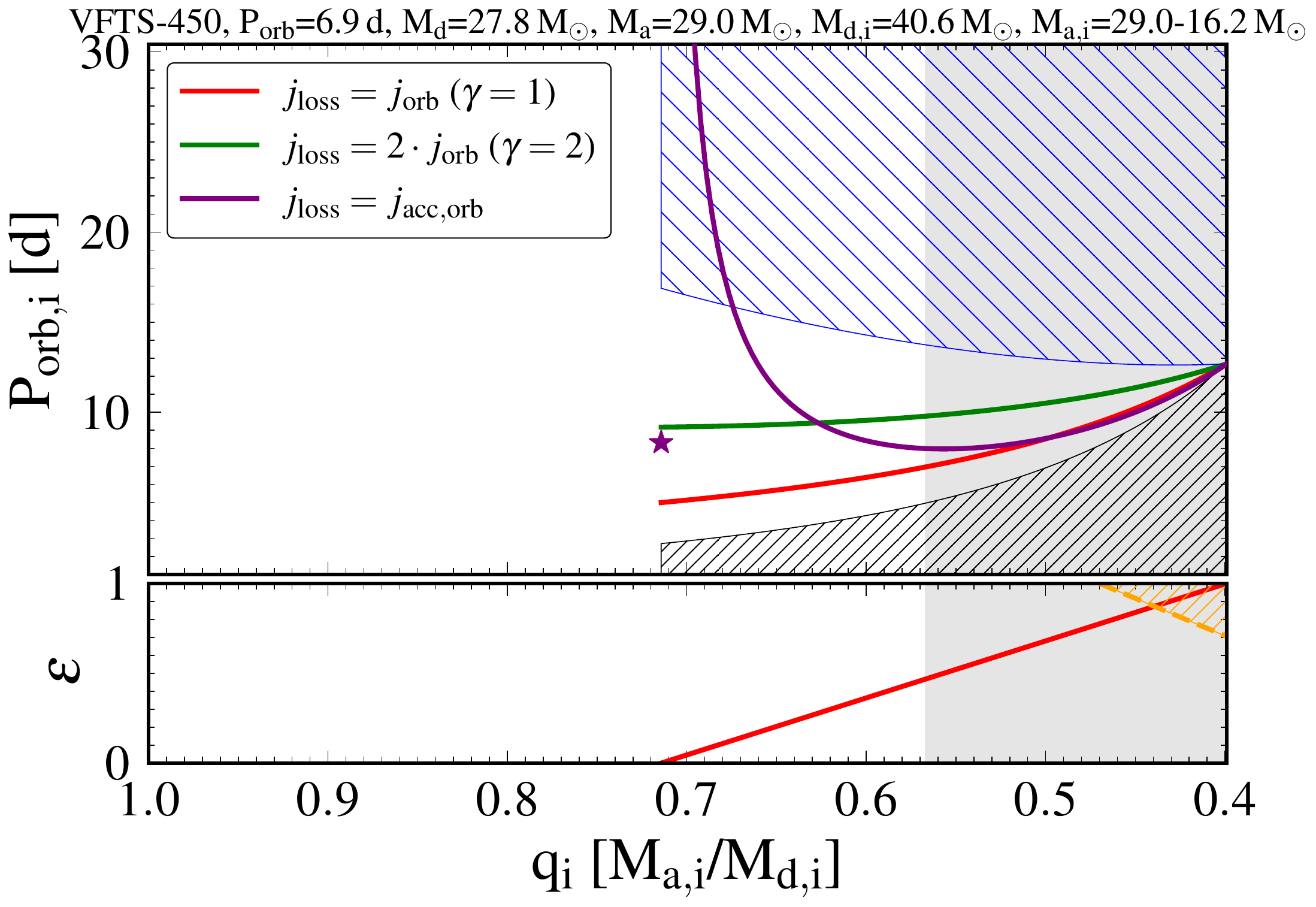}
    \includegraphics[width=0.23\linewidth]{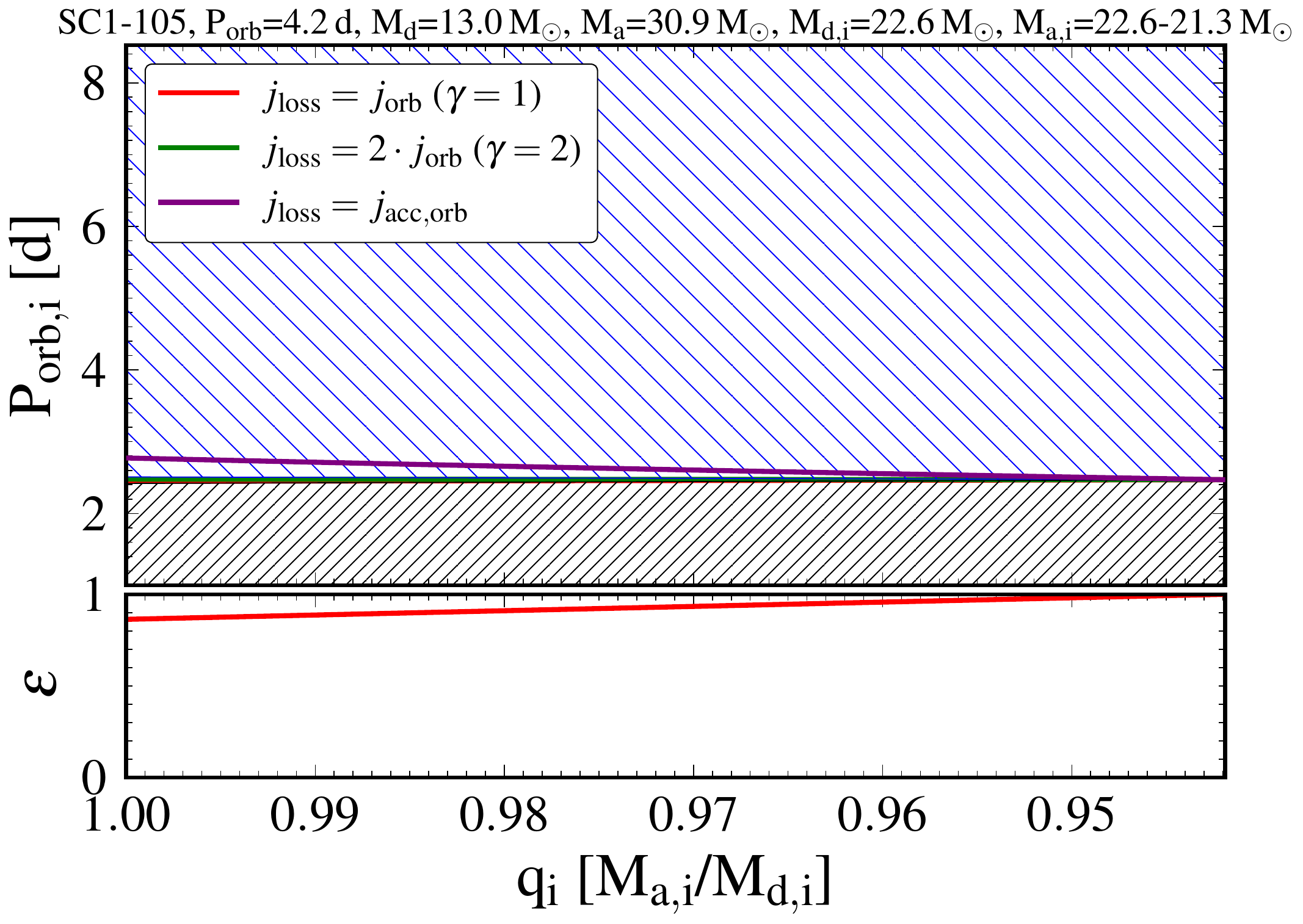}
    \includegraphics[width=0.23\linewidth]{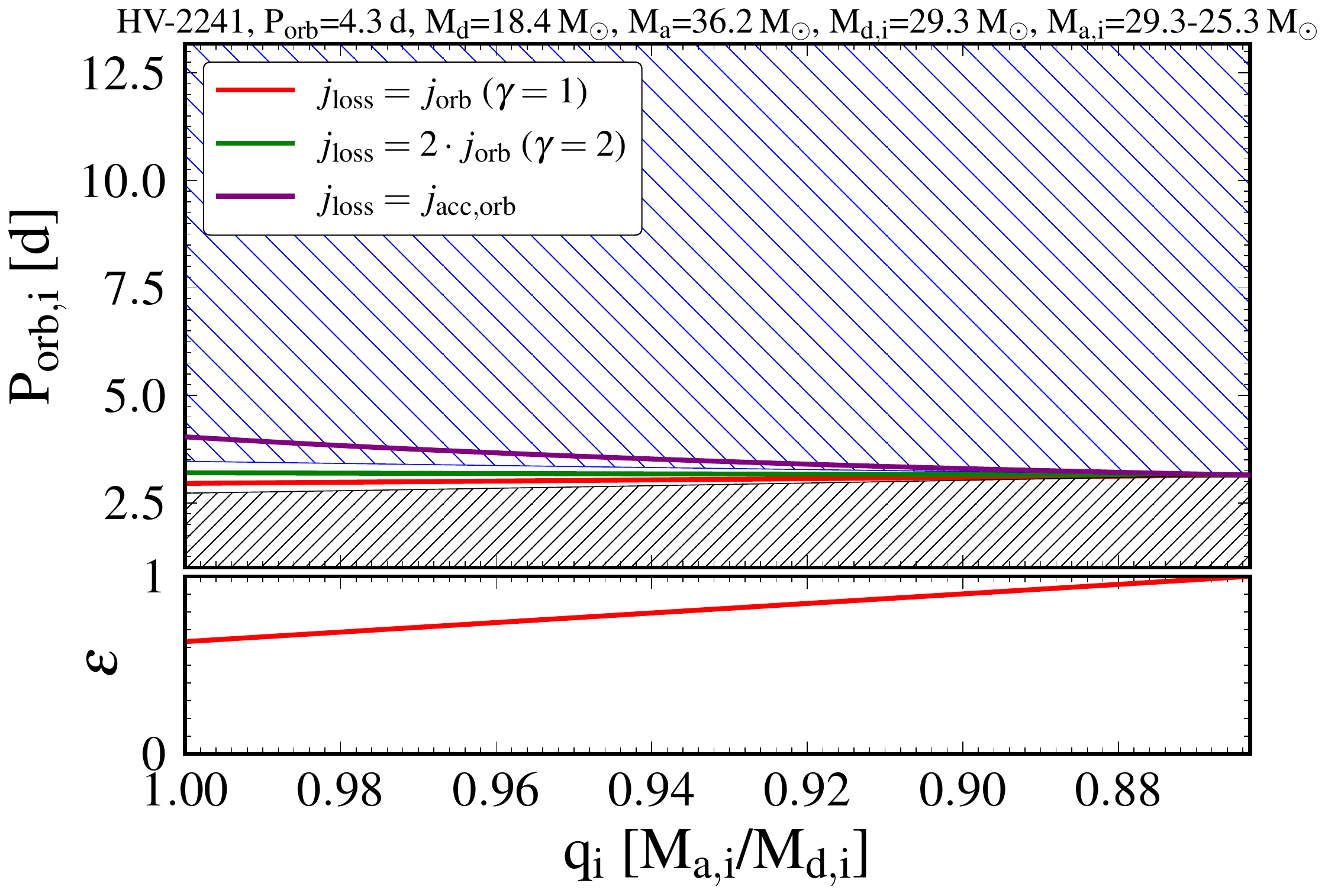}

    \caption{Same as Fig.\,\ref{fig:individual_examples}, for all massive Algols in the Milky Way and LMC. }
    \label{fig:appendix_lmc_mw}
\end{figure*}

\begin{table*}[h!]
\caption{Possible range of mass transfer efficiencies and angular momentum loss for massive Algols in the SMC. }
\centering
\begin{tabular}{lrrrrrrrrrrrr}
\hline\hline
System & $P_{\rm orb}$ & $M_{\rm d}$ & $M_{\rm a}$ & $P_{\rm orb,i,max}$ & $M_{\rm d,i}$ & $M_{\rm a,i,min}$ & $M_{\rm a,i,max}$ & $\varepsilon_{\rm min}$ & $\varepsilon_{\rm qmax}$ & $\varepsilon_{\rm max}^{\rm disk}$ & $\gamma_{\rm min,qmax}$ & $\gamma_{\rm max,qmax}$ \\
\hline
O251047 & 2.51 & 5.50 & 8.10 & 3.72 & 11.75 & 1.85 & 8.10 & 0.00 & 0.00 & 0.54 & 0.00 & 0.00 \\
O064498 & 2.63 & 2.71 & 8.40 & 2.71 & 6.63 & 4.48 & 8.40 & 0.45 & 0.75 & 1.00 & 1.35 & 3.00 \\
O208049 & 3.03 & 4.78 & 10.00 & 3.44 & 10.52 & 4.26 & 10.00 & 0.00 & 0.28 & 0.90 & 0.66 & 2.32 \\
O319960 & 4.05 & 6.75 & 10.60 & 4.35 & 13.75 & 3.60 & 10.60 & 0.00 & 0.02 & 0.68 & 0.37 & 1.83 \\
O060548 & 3.63 & 8.72 & 10.80 & 5.14 & 16.69 & 2.83 & 10.80 & 0.00 & 0.00 & 0.53 & 0.00 & 0.00 \\
O099121 & 2.45 & 6.64 & 11.30 & 4.30 & 13.58 & 4.36 & 11.30 & 0.00 & 0.13 & 0.77 & 0.82 & 2.49 \\
O026631 & 1.41 & 11.26 & 11.50 & 5.70 & 20.24 & 2.52 & 11.50 & 0.00 & 0.00 & 0.45 & 0.00 & 0.00 \\
O193779 & 1.67 & 5.90 & 11.60 & 3.93 & 12.42 & 5.08 & 11.60 & 0.00 & 0.26 & 0.91 & 1.38 & 3.00 \\
O094559 & 1.75 & 9.93 & 12.00 & 5.40 & 18.41 & 3.52 & 12.00 & 0.00 & 0.00 & 0.56 & 0.00 & 0.00 \\
O152981 & 2.00 & 8.15 & 12.50 & 5.01 & 15.88 & 4.77 & 12.50 & 0.00 & 0.08 & 0.73 & 0.97 & 2.78 \\
O142073 & 3.19 & 6.30 & 12.60 & 4.13 & 13.05 & 5.85 & 12.60 & 0.00 & 0.35 & 0.97 & 0.31 & 2.57 \\
O047454 & 1.57 & 9.25 & 12.60 & 5.26 & 17.45 & 4.40 & 12.60 & 0.00 & 0.00 & 0.65 & 0.00 & 0.00 \\
O110409 & 2.97 & 8.87 & 13.70 & 5.17 & 16.91 & 5.66 & 13.70 & 0.00 & 0.13 & 0.78 & 0.42 & 2.45 \\
O030116 & 2.95 & 7.65 & 14.30 & 4.78 & 15.12 & 6.83 & 14.30 & 0.00 & 0.40 & 0.97 & 0.14 & 2.93 \\
O189660 & 1.66 & 10.21 & 15.30 & 5.46 & 18.80 & 6.71 & 15.30 & 0.00 & 0.15 & 0.81 & 1.08 & 3.00 \\
O077224 & 3.82 & 13.05 & 15.90 & 6.36 & 22.57 & 6.38 & 15.90 & 0.00 & 0.02 & 0.69 & 0.00 & 2.19 \\
O158118 & 2.58 & 7.88 & 16.00 & 4.89 & 15.47 & 8.41 & 16.00 & 0.07 & 0.58 & 1.00 & 0.00 & 3.00 \\
O108086 & 0.88 & 14.28 & 16.90 & 6.80 & 24.14 & 7.04 & 16.90 & 0.00 & 0.01 & 0.70 & 1.73 & 3.00 \\
O316725 & 2.55 & 8.82 & 16.90 & 5.16 & 16.84 & 8.88 & 16.90 & 0.01 & 0.54 & 1.00 & 0.00 & 3.00 \\
O010098 & 1.11 & 13.65 & 17.80 & 6.58 & 23.33 & 8.12 & 17.80 & 0.00 & 0.16 & 0.79 & 1.46 & 3.00 \\
O277080 & 1.94 & 11.30 & 17.40 & 5.72 & 20.29 & 8.41 & 17.40 & 0.00 & 0.26 & 0.90 & 0.73 & 3.00 \\
O243913 & 2.63 & 10.52 & 18.60 & 5.52 & 19.23 & 9.89 & 18.60 & 0.00 & 0.49 & 1.00 & 0.00 & 3.00 \\
O066175 & 3.62 & 11.51 & 19.60 & 5.79 & 20.56 & 10.55 & 19.60 & 0.00 & 0.49 & 1.00 & 0.00 & 3.00 \\
O209964 & 3.31 & 14.46 & 18.80 & 6.87 & 24.37 & 8.89 & 18.80 & 0.00 & 0.18 & 0.82 & 0.00 & 2.71 \\
O202153 & 4.60 & 12.50 & 19.90 & 6.16 & 21.86 & 10.54 & 19.90 & 0.00 & 0.47 & 1.00 & 0.00 & 2.52 \\
O311225 & 1.84 & 11.96 & 21.20 & 5.96 & 21.15 & 12.01 & 21.20 & 0.01 & 0.64 & 1.00 & 0.12 & 3.00 \\
O175323 & 2.20 & 16.19 & 23.50 & 7.47 & 26.52 & 13.17 & 23.50 & 0.00 & 0.48 & 1.00 & 0.00 & 3.00 \\
O300549 & 1.33 & 17.42 & 25.40 & 7.90 & 28.03 & 14.79 & 25.40 & 0.00 & 0.55 & 1.00 & 0.81 & 3.00 \\
O243188 & 1.87 & 18.62 & 27.30 & 8.72 & 29.47 & 16.45 & 27.30 & 0.00 & 0.67 & 1.00 & 0.00 & 3.00 \\
\hline
\end{tabular}
\label{table:all_algols_smc}
\end{table*}

\begin{table*}[h!]
\caption{Possible range of mass transfer efficiencies and angular momentum loss for massive Algols in the Milky Way and LMC. }
\centering
\begin{tabular}{lrrrrrrrrrrrr}
\hline\hline
System & $P_{\rm orb}$ & $M_{\rm d}$ & $M_{\rm a}$ & $P_{\rm orb,i,max}$ & $M_{\rm d,i}$ & $M_{\rm a,i,min}$ & $M_{\rm a,i,max}$ & $\varepsilon_{\rm min}$ & $\varepsilon_{\rm qmax}$ & $\varepsilon_{\rm max}^{\rm disk}$ & $\gamma_{\rm min,qmax}$ & $\gamma_{\rm max,qmax}$ \\
\hline
TT-Aur   & 1.33 & 5.40  &  8.10 &  4.82 & 11.69 &  1.81 &  8.10 & 0.00 & 0.00 & 0.54 & 0.00 & 0.00 \\
Mu1-Sco  & 1.44 & 4.60  &  8.30 &  4.53 & 10.30 &  2.60 &  8.30 & 0.00 & 0.04 & 0.69 & 1.52 & 2.95 \\
SV-Gem   & 4.00 & 3.37  &  8.34 &  3.97 &  8.04 &  3.67 &  8.34 & 0.07 & 0.38 & 0.98 & 0.45 & 2.50 \\
V454-Cyg & 2.31 & 2.86  &  8.40 &  3.68 &  7.00 &  4.26 &  8.40 & 0.34 & 0.64 & 1.00 & 1.58 & 3.00 \\
BF-Cen   & 3.69 & 3.80  &  8.70 &  4.17 &  8.85 &  3.65 &  8.70 & 0.00 & 0.32 & 0.90 & 0.51 & 2.47 \\
BM-Ori   & 6.47 & 3.51  &  9.50 &  4.04 &  8.30 &  4.71 &  9.50 & 0.26 & 0.57 & 1.00 & 0.00 & 2.15 \\
IZ-Per   & 3.68 & 3.20  &  9.97 &  3.88 &  7.69 &  5.48 &  9.97 & 0.51 & 0.81 & 1.00 & 0.00 & 3.00 \\
AI-Cru   & 1.41 & 6.30  & 10.30 &  5.11 & 13.15 &  3.45 & 10.30 & 0.00 & 0.05 & 0.69 & 1.45 & 3.00 \\
SX-Aur   & 1.21 & 5.60  & 10.30 &  4.89 & 12.02 &  3.88 & 10.30 & 0.00 & 0.17 & 0.78 & 1.70 & 3.00 \\
MP-Cen   & 2.99 & 4.40  & 11.40 &  4.46 &  9.95 &  5.85 & 11.40 & 0.27 & 0.65 & 1.00 & 0.30 & 3.00 \\
IU-Aur   & 1.81 & 6.07  & 11.99 &  5.04 & 12.79 &  5.27 & 11.99 & 0.00 & 0.33 & 0.91 & 1.21 & 3.00 \\
V356-Sgr & 8.89 & 4.70  & 12.10 &  4.57 & 10.48 &  6.32 & 12.10 & 0.29 & 0.68 & 1.00 & 0.00 & 1.47 \\
V-Pup    & 1.45 & 6.33  & 12.85 &  5.12 & 13.20 &  5.98 & 12.85 & 0.00 & 0.41 & 0.97 & 1.56 & 3.00 \\
V498-Cyg & 3.48 & 6.45  & 13.44 &  5.16 & 13.39 &  6.50 & 13.44 & 0.01 & 0.48 & 1.00 & 0.00 & 3.00 \\
GN-Car   & 4.34 & 4.59  & 13.49 &  4.53 & 10.28 &  7.80 & 13.49 & 0.57 & 0.96 & 1.00 & 0.00 & 3.00 \\
LZ-Cep   & 3.07 & 6.50  & 16.00 &  5.17 & 13.46 &  9.04 & 16.00 & 0.37 & 0.84 & 1.00 & 0.00 & 3.00 \\
Del-Pic  & 1.67 & 8.60  & 16.30 &  5.75 & 16.62 &  8.28 & 16.30 & 0.00 & 0.52 & 1.00 & 1.02 & 3.00 \\
XX-Cas   & 3.06 & 6.07  & 16.85 &  5.04 & 12.79 & 10.13 & 16.85 & 0.61 & 1.00 & 1.00 & 0.00 & 3.00 \\
HH-Car   & 3.23 &14.00  & 17.00 &  9.21 & 23.88 &  7.12 & 17.00 & 0.00 & 0.08 & 0.71 & 0.05 & 2.91 \\
V337-Aql & 2.73 & 7.83  & 17.44 &  5.56 & 15.50 &  9.77 & 17.44 & 0.26 & 0.78 & 1.00 & 0.00 & 3.00 \\
AQ-Cas   &11.70 &12.50  & 17.63 &  8.17 & 21.96 &  8.17 & 17.63 & 0.00 & 0.27 & 0.83 & 0.00 & 1.07 \\
XZ-Cep   & 5.09 & 9.30  & 18.70 &  5.92 & 17.62 & 10.38 & 18.70 & 0.13 & 0.71 & 1.00 & 0.00 & 3.00 \\
29-CMa   & 4.39 &16.00  & 19.00 & 10.71 & 26.39 &  8.61 & 19.00 & 0.00 & 0.15 & 0.76 & 0.00 & 2.83 \\
AB-Cru   & 3.41 & 6.95  & 19.75 &  5.31 & 14.17 & 12.53 & 19.75 & 0.78 & 1.00 & 1.00 & 0.00 & 3.00 \\
V448-Cyg & 6.51 &13.70  & 24.70 &  9.01 & 23.50 & 14.90 & 24.70 & 0.13 & 0.89 & 1.00 & 0.00 & 3.00 \\
QZ-Car   & 5.99 &20.00  & 30.00 & 15.39 & 31.22 & 18.78 & 30.00 & 0.00 & 0.93 & 1.00 & 0.00 & 3.00 \\
\hline
VFTS-061 & 2.33 & 8.70  & 16.30 &  5.78 & 16.76 &  8.24 & 16.30 & 0.00 & 0.51 & 1.00 & 0.26 & 3.00 \\
VFTS-652 & 8.59 & 6.50  & 18.10 &  5.17 & 13.46 & 11.14 & 18.10 & 0.67 & 1.00 & 1.00 & 0.00 & 3.00 \\
VFTS-538 & 4.15 &11.80  & 18.30 &  7.67 & 21.04 &  9.06 & 18.30 & 0.00 & 0.40 & 0.93 & 0.00 & 2.97 \\
HV-2543  & 4.83 &15.60  & 25.60 & 10.39 & 25.89 & 15.31 & 25.60 & 0.00 & 0.81 & 1.00 & 0.00 & 3.00 \\
VFTS-450 & 6.89 &27.80  & 29.00 & 30.43 & 40.60 & 16.20 & 29.00 & 0.00 & 0.46 & 0.87 & 0.00 & 3.00 \\
SC1-105  & 4.25 &13.00  & 30.90 &  8.53 & 22.61 & 21.29 & 30.90 & 0.87 & 1.00 & 1.00 & 0.00 & 3.00 \\
HV-2241  & 4.34 &18.40  & 36.20 & 13.18 & 29.31 & 25.29 & 36.20 & 0.64 & 1.00 & 1.00 & 0.00 & 3.00 \\
\hline
\end{tabular}
\label{table:all_algols_lmc_mw}
\end{table*}

\begin{table*}
\caption{Stellar and binary parameters used in this work (see Sect.\,\ref{sec:method}) taken from detailed binary evolution models for the LMC and SMC. }
\centering
\begin{tabular}{rrrrrrrrr}
\hline\hline
$M_{\rm d,i}$ & $M_{\rm ccd,i}$ & $q_{\min}$ & $q_{\max}$ & $P_{\rm orb,i,max}$ & $M_{\rm ccd,i}$ & $q_{\min}$ & $q_{\max}$ & $P_{\rm orb,i,max}$ \\
\hline
 & $<$------ & LMC & models & ---------$>$ & $<$------ & SMC & models & ---------$>$ \\
\hline
  6.31 &  2.52 & 0.800 & 0.825 &    3.48 &  2.55 & 0.825 & 0.825 &   2.64 \\
  7.08 &  2.90 & 0.800 & 0.825 &    3.70 &  2.93 & 0.825 & 0.825 &   2.80 \\
  7.94 &  3.32 & 0.800 & 0.825 &    3.95 &  3.36 & 0.800 & 0.800 &   2.98 \\
  8.91 &  3.83 & 0.775 & 0.800 &    4.18 &  3.88 & 0.800 & 0.800 &   3.16 \\
 10.00 &  4.43 & 0.775 & 0.787 &    4.47 &  4.48 & 0.800 & 0.800 &   3.35 \\
 11.22 &  5.12 & 0.750 & 0.775 &    4.73 &  5.18 & 0.800 & 0.800 &   3.55 \\
 12.59 &  5.94 & 0.725 & 0.767 &    5.00 &  6.00 & 0.750 & 0.800 &   3.98 \\
 14.12 &  6.92 & 0.700 & 0.750 &    5.30 &  6.99 & 0.750 & 0.750 &   4.47 \\
 15.85 &  8.06 & 0.675 & 0.737 &    5.62 &  8.13 & 0.750 & 0.750 &   5.01 \\
 17.78 &  9.41 & 0.650 & 0.725 &    5.95 &  9.48 & 0.700 & 0.750 &   5.31 \\
 19.95 & 10.97 & 0.600 & 0.700 &    7.08 & 11.04 & 0.650 & 0.750 &   5.62 \\
 22.39 & 12.83 & 0.560 & 0.687 &    8.41 & 12.91 & 0.600 & 0.700 &   6.31 \\
 25.12 & 14.97 & 0.500 & 0.675 &    9.88 & 15.05 & 0.550 & 0.700 &   7.08 \\
 28.18 & 17.46 & 0.450 & 0.650 &   11.88 & 17.54 & 0.550 & 0.700 &   7.94 \\
 31.62 & 20.33 & 0.390 & 0.625 &   15.85 & 20.41 & 0.500 & 0.650 &  10.00 \\
 35.48 & 23.70 & 0.325 & 0.600 &   19.95 & 23.80 & 0.450 & 0.650 &  12.59 \\
 39.81 & 27.54 & 0.275 & 0.575 &   26.61 & 27.64 & 0.400 & 0.650 &  15.85 \\
\hline
\end{tabular}
\label{table:1}
\end{table*}


\begin{figure}
    \centering
    \includegraphics[width=0.49\linewidth]{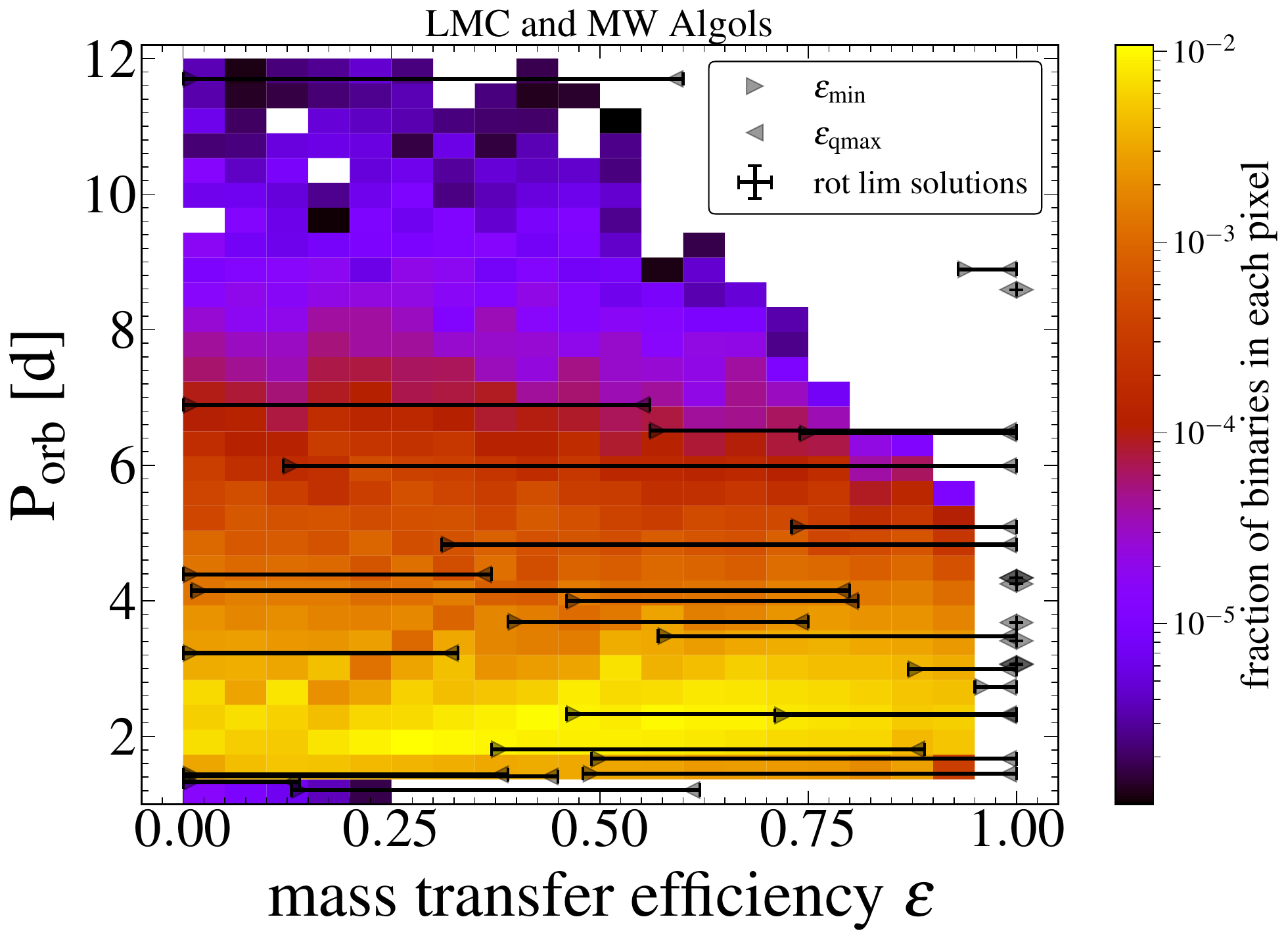}
    \includegraphics[width=0.49\linewidth]{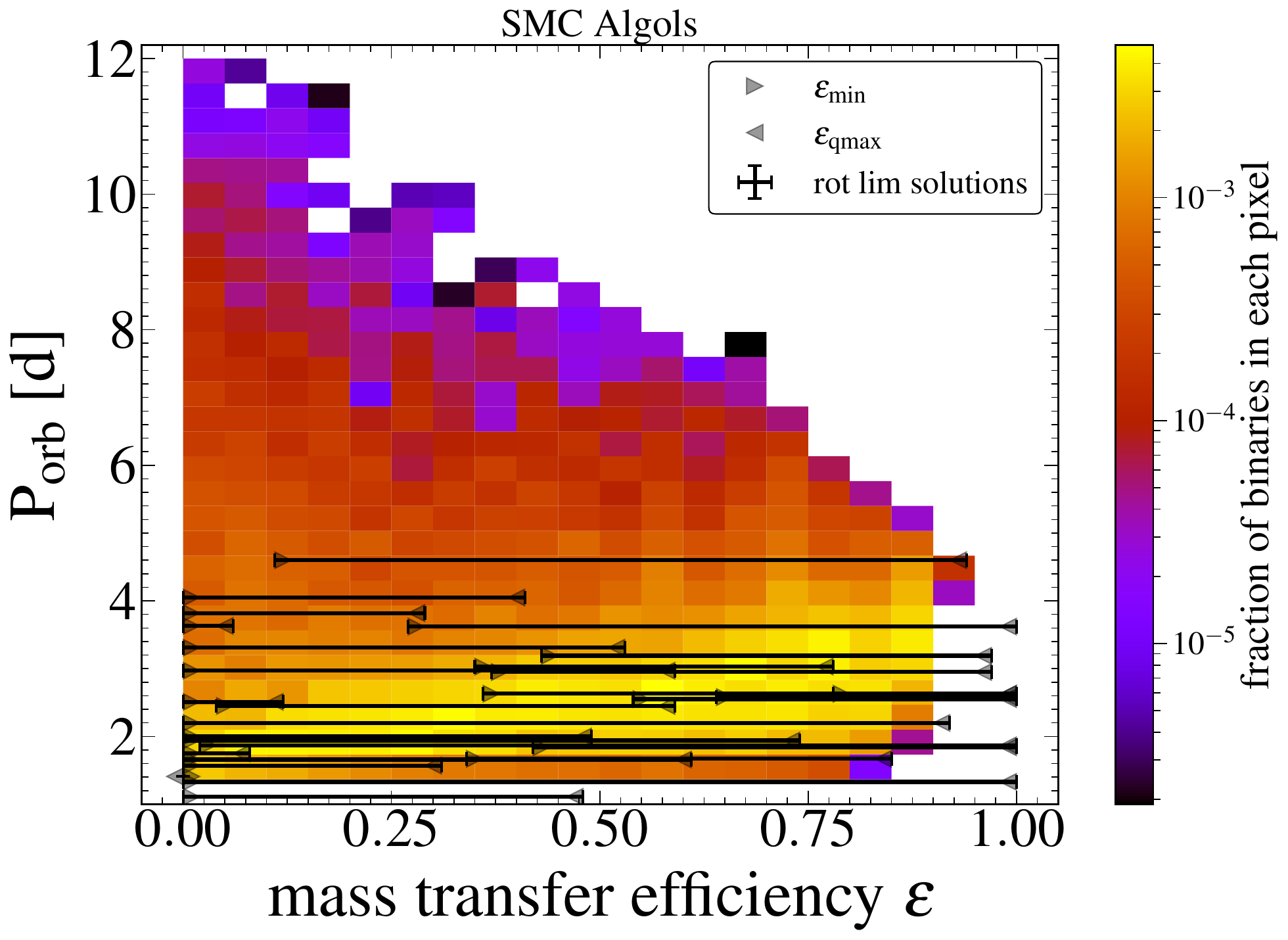}
    \caption{Same as Fig.\,\ref{fig:porb_beta}, but for the shallow stripping assumption (orange curve) in Fig.\,\ref{fig:ccdi}. }
    \label{fig:shallow_stripping_porb_beta}
\end{figure}

\bibliography{sample701}{}
\bibliographystyle{aasjournalv7}



\end{document}